\documentclass[useAMS,usenatbib]{mn2e}

\usepackage{txfonts}
\usepackage{amssymb}
\usepackage{graphicx}
\usepackage{subfigure}
\usepackage{psfrag}
\newcommand{\eqn}[1]
	{(#1)}

\newcommand{\tbl}[1]
	{Table~#1}
\newcommand{\Tbl}[1]
	{Table~#1}
\newcommand{\fig}[1]
	{Figure~#1}
\newcommand{\Fig}[1]
	{Figure~#1}
\newcommand{\sectn}[1]
	{Section~#1}

\newcommand{\appn}[1]
	{Appendix~#1}

\newcommand{\cmb}
	{{CMB}}
\newcommand{\cmbtext}
	{cosmic microwave background}

\newcommand{\cswt}
	{{CSWT}}
\newcommand{\cswttext}
	{continuous spherical wavelet transform}

\newcommand{\dft}
	{{DFT}}
\newcommand{\fft}
	{{FFT}}

\newcommand{\mexhat}
	{Mexican hat}
\newcommand{\Mexhat}
	{Mexican hat}
\newcommand{\morlet}
	{real Morlet}
\newcommand{\Morlet}
	{Real Morlet}

\newcommand{\wmap}
        {{WMAP}}
\newcommand{\wmaptext}
        {Wilkinson Microwave Anisotropy Probe}
\newcommand{\cobe}
	{\mbox{COBE-DMR}}
\newcommand{\cobetext}
        {Cosmic Background Explorer--Differential Microwave Radiometer}
\newcommand{\lcdm}
	{$\Lambda${CDM}}
\newcommand{\lcdmtext}
	{Lambda Cold Dark Matter}
\newcommand{\lambdaarch}
	{{LAMBDA}}
\newcommand{\astroph}
	{astro-ph}

\newcommand{\mnras}
	{\mbox{MNRAS}}

\newcommand{\dthree}
	{{3D}}

\newcommand{\kpzero}
	{{Kp0}}

\newcommand{\healpix}
	{{HEALPix}}
\newcommand{\nside}
	{\ensuremath{{N_{side}}}}

\newcommand{\eg}
	{\mbox{e.g.}}

\newcommand{\etal}
	{\mbox{et al.}}

\newcommand{\ie}
	{\mbox{i.e.}}

\newcommand{\img}
	{\ensuremath{\mathit{i}}}
\newcommand{\dx}
        {\ensuremath{\mathrm{\,d}}}

\newcommand{\lsh}
	{\ensuremath{\ell}}
\newcommand{\sky}
	{\ensuremath{s}}
\newcommand{\skywav}
	{\ensuremath{W}}
\newcommand{\wav}
	{\ensuremath{\psi}}

\newcommand{\dil}
	{\ensuremath{\mathcal{D}}}
\newcommand{\scale}
	{\ensuremath{a}}
\newcommand{\effsize}
	{\ensuremath{\xi}}
\newcommand{\rot}
	{\ensuremath{\mathcal{R}}}
\newcommand{\dmatbig}
	{\ensuremath{D}}
\newcommand{\dmatsmall}
	{\ensuremath{d}}
\newcommand{\eccen}
	{\ensuremath{\epsilon}}

\newcommand{\sa}
	{\ensuremath{\omega}}
\newcommand{\saa}
	{\ensuremath{\theta}}
\newcommand{\sab}
	{\ensuremath{\phi}}
\newcommand{\eulera}
	{\ensuremath{\alpha}}
\newcommand{\eulerb}
	{\ensuremath{\beta}}
\newcommand{\eulerc}
	{\ensuremath{\gamma}}
\newcommand{\eulers}
	{\ensuremath{\eulera, \eulerb, \eulerc}}
\newcommand{\sothree}
        {\ensuremath{\mathrm{SO}(3)}}

\newcommand{\sh}
	{\ensuremath{Y}}
\newcommand{\shcoeff}[1]
	{\ensuremath{\widehat{#1}}}

\newcommand{\conj}
	{\ensuremath{\ast}}
\newcommand{\real}
	{\ensuremath{\mathbb{R}}}
\newcommand{\sphere}
	{\ensuremath{S^2}}
\newcommand{\sigx}
	{\ensuremath{\sigma_x}}
\newcommand{\sigy}
	{\ensuremath{\sigma_y}}

\newcommand{\cswtfftterm}
	{\ensuremath{t}}

\newcommand{\ind}
	{\ensuremath{n}}
\newcommand{\num}
	{\ensuremath{N}}
\newcommand{\p}
	{\ensuremath{^\prime}}
\newcommand{\pp}
	{\ensuremath{^{\prime\prime}}}
\newcommand{\complexity}
	{\ensuremath{\mathcal{O}}}

\newcommand{\cmbtemp}
	{\ensuremath{T}}
\newcommand{\mean}
	{\ensuremath{\mu}}
\newcommand{\skewness}
	{\ensuremath{\zeta}}
\newcommand{\kurtosis}
	{\ensuremath{\kappa}}
\newcommand{\neff}
	{\ensuremath{N_{\rm eff}}}

\newcommand{\spcend}
	{\ensuremath{\:}}

\newcommand{\nstd}
	{\ensuremath{\num_\sigma}}
\newcommand{\ndev}
	{\ensuremath{\num_{\rm dev}}}

\newcommand{\conflevel}
	{\ensuremath{\delta}}

\newcommand{\cov}
        {\ensuremath{C}}
\newcommand{\tstat}
        {\ensuremath{\tau}}

\newcommand{\ngsim}
	{1000}

\newcommand{\nstatmexskew}
	{28}
\newcommand{\nstatmexkurt}
	{47}

\newcommand{\nstdmexskewsgn}
	{\mbox{$-3.38$}}
\newcommand{\nstdmexkurtsgn}
	{\mbox{$3.12$}}
\newcommand{\clmexskew}
	{97.2}
\newcommand{\clmexkurt}
	{95.3}

\newcommand{\nstatmexepskew}
	{39}
\newcommand{\nstatmexepkurt}
	{199}

\newcommand{\nstdmexepskewsgn}
	{\mbox{$-4.10$}}
\newcommand{\nstdmexepkurtsgn}
	{\mbox{$3.01$}}
\newcommand{\clmexepskew}
	{96.1}
\newcommand{\clmexepkurt}
	{80.1}

\newcommand{\nstatmorskew}
	{17}
\newcommand{\nstatmorkurt}
	{642}
\newcommand{\nstdmorskew}
	{\mbox{$5.61$}}
\newcommand{\nstdmorskewteg}  % Additional Tegmark result to quote in text.
	{\mbox{$6.42$}}

\newcommand{\nstdmorskewsgn}
	{\mbox{$-5.61$}}
\newcommand{\nstdmorskewtegsgn}  % Additional Tegmark result to quote in text.
	{\mbox{$-6.42$}}
\newcommand{\nstdmorkurtsgn}
	{\mbox{$2.66$}}
\newcommand{\clmorskew}
	{98.3}
\newcommand{\clmorkurt}
	{35.8}

\newcommand{\ghz}
	{{GHz}}

\title[Non-Gaussianity in the \wmap\ 1-year data]
  {A high-significance detection of non-Gaussianity in the \wmap\ 1-year %
   data using directional spherical wavelets}

\author[J.~D.~McEwen \etal]
  {J.~D.~McEwen,$^1$\thanks{E-mail: mcewen@mrao.cam.ac.uk} 
   M.~P.~Hobson,$^1$ A.~N.~Lasenby$^1$ and D.~J.~Mortlock$^2$\\
  $^1$Astrophysics Group, 
      Cavendish Laboratory, Madingley Road,
      Cambridge CB3 0HE, UK\\
  $^2$Institute of Astronomy, Madingley Road,
      Cambridge CB3 0HA, UK}
\date{1 February 2004}
\pagerange{\pageref{firstpage}--\pageref{lastpage}} 
\pubyear{2004}

\def\LaTeX{L\kern-.36em\raise.3ex\hbox{a}\kern-.15em
    T\kern-.1667em\lower.7ex\hbox{E}\kern-.125emX}

\begin{document}
\label{firstpage}
\maketitle

% -----------------------------------------------------------------------------

\begin{abstract}
A directional spherical wavelet analysis is performed to examine the
Gaussianity of the \wmaptext\ (\wmap) 1-year data.  Such an analysis
is \mbox{facilitated} by the introduction of a fast directional \cswttext.
The directional nature of the \mbox{analysis} allows one to probe
orientated structure in the data.  
Significant deviations from Gaussianity are detected in the skewness
and kurtosis of spherical elliptical \mexhat\ and \morlet\ wavelet
coefficients for both the \wmap\ and \citet{tegmark:2003}
foreground-removed maps. 
The previous non-Gaussianity detection made by \citet{vielva:2003}
using the spherical symmetric \mexhat\ wavelet is confirmed, 
although their detection at the 99.9\% significance
level is only made at the \clmexkurt\% significance level using
our most conservative statistical test.
Furthermore, deviations from Gaussianity in the skewness of
spherical \morlet\ wavelet coefficients 
on a wavelet scale of $550\arcmin$ (corresponding to an effective global
size on the sky of $\sim 26^\circ$ and an internal size of $\sim3^\circ$)
at an azimuthal orientation of $72^\circ$, are made at the
\clmorskew\% significance level, using the same conservative method.
The wavelet analysis inherently allows one to localise on the sky
those regions that introduce skewness and those that introduce
kurtosis.
Preliminary noise analysis indicates that these detected deviation
regions are not atypical and have average noise dispersion.
Further analysis is required to ascertain
whether these detected regions correspond to secondary or instrumental
effects, or whether in fact the non-Gaussianity detected is due to intrinsic
primordial fluctuations in the \cmbtext.
\end{abstract}

% -----------------------------------------------------------------------------


% -----------------------------------------------------------------------------

\section{Introduction}

A range of primordial processes may imprint signatures on the
temperature fluctuations of the \cmbtext\ (\cmb).  The currently
favoured cosmological model is based on the assumption of
initial fluctuations generated by inflation. In the simplest
inflationary models, these result in Gaussian {tempera\-ture} anisotropies
in the \cmb.  Non-standard inflationary models and various cosmic
defect scenarios could, however, lead to non-Gaussian primordial \cmb\
fluctuations.  Non-Gaussianity may also be introduced by secondary
effects, such as the reionisation of the Universe, the integrated
Sachs-Wolfe effect, the Rees-Sciama effect, the Sunyaev-Zel'dovich effect
and gravitational lensing -- in addition to measurement systematics or
foreground contamination.
%Moreover, measurement systematics or
%foreground contamination that may exist in the data may
%also introduce non-Gaussianity.
%
Consequently, probing the microwave sky for non-Gaussianity is of
considerable interest, providing evidence for competing scenarios of
the early Universe and also highlighting important secondary sources
of non-Gaussianity and systematics.

Ideally, one would like to localise any detected non-Gaussian
components on the sky, in particular to determine if they correspond
to secondary effects or systematics.  The ability to probe different
scales is also important to ensure 
non-Gaussian sources present only on certain scales are not
concealed by the predominant Gaussianity of other scales.
Wavelet techniques are thus a perfect candidate for
\cmb\ non-Gaussianity analysis, since they provide both scale and
spatial localisation.  In addition, directional wavelets may provide
further information on orientated structure in the \cmb.

Wavelets have already been used to analyse the Gaussianity of the
\cmb.  For example, \citet{hobson:1999} and \citet{bh:2001}
investigated the use of planar wavelets in detecting and
characterising non-Gaussianity on patches of the \cmb\ sky. This
approach was used by \citet{mhl:2000} to analyse planar faces of the
4-year \cobetext\ (\cobe) data in the QuadCube pixelisation, showing
that the data is consistent with Gaussianity (correcting an earlier claim of
non-Gaussianity by \citealt{pando:1998}).  To consider a full sky \cmb\
map properly, however, wavelet analysis must be extended to spherical
geometry.  A spherical Haar wavelet analysis of the \cobe\ data
was performed by \citet{barreiro:2000}, but no evidence of
non-Gaussianity was found. Employing the approach described by
\citet{antoine:1998} for performing continuous wavelet transforms on a
sphere, \citet{cayon:2001} used the isotropic \mexhat\ wavelet to
analyse the \cobe\ maps; again, no significant deviations from
Gaussianity were detected.  \citet{martinez:2002} subsequently
compared the performance of spherical Haar and \mexhat\ wavelets for
non-Gaussianity detection and found the \Mexhat\ wavelet to be
superior.

Since the release of the \wmaptext\ (\wmap)\ 1-year data, a wide range of
Gaussianity analyses have been performed, calculating measures
such as the bispectrum and Minkowski functionals
\citep{komatsu:2003,mm:2004,lm:2004},
the genus \citep{cg:2003,eriksen:2004}, the 3-point
correlation function \citep{gw:2003}, multipole alignment statistics 
\citep{copi:2004,oliveira:2004,slosar:2004}, phase associations
\citep{chiang:2003,coles:2004}, local curvature 
\citep{hansen:2004,cabella:2004} and hot and cold spot statistics
\citep{larson:2004}. Some
statistics show consistency with Gaussianity, whereas others provide
some evidence for a non-Gaussian signal and/or an asymmetry
between the northern and southern Galactic hemispheres. One of the
highest significance levels for non-Gaussianity yet reported
was obtained by \citet{vielva:2003} using a spherical \mexhat\ wavelet
analysis.  This result has been confirmed by \citet{mw:2004}, who show
it to be robust to different Galactic masks and assumptions regarding
noise properties.
In particular, it was found that the kurtosis of the
wavelet coefficients in the southern hemisphere,
at an approximate size on the sky of $10^\circ$,
lies just outside the $3\sigma$ Gaussian confidence level.

Previous wavelet analyses of the \cmb\ have been restricted
to rotationally symmetric wavelets.
A directional analysis on the full sky has previously been prohibited by
the computational infeasibility of any implementation.  In this paper,
by applying a fast directional \cswttext\ (\cswt), we extend
non-Gaussianity analysis to examine directional structure in the \cmb.

The remainder of this paper is structured as follows.  The directional
\cswt\ and the construction of new directional spherical wavelets is
presented in \sectn{\ref{sec:cswt}}.  In
\sectn{\ref{sec:non_gaussianity}} the procedure followed to analyse
the \wmap\ 1-year data for non-Gaussianity is described.  Results and
further analysis are presented in
\sectn{\ref{sec:results}}.  Concluding remarks are made in
\sectn{\ref{sec:conclusions}}.

% -----------------------------------------------------------------------------

\section{Directional continuous spherical wavelet analysis}
\label{sec:cswt}

To perform a wavelet analysis of full sky maps defined on the
celestial sphere, Euclidean wavelet analysis must be extended to
spherical geometry.
We consider the directional \cswt\ constructed by \citet{antoine:1998}.  
This transform was constructed from group theoretic principles,
however we present here an equivalent construction based on a few
simple operations and norm-preserving properties.

% ---------------------------------------

\subsection{Transform}

A wavelet basis is constructed on the sphere by applying the spherical
extension of Euclidean motions and dilations to mother
wavelets defined on the sphere -- analogous to the construction of a
Euclidean wavelet basis. 

The natural extension of Euclidean motions on the sphere are rotations.
These are characterised by the elements of the rotation group \sothree,
which we parameterise in terms of the three Euler angles
$(\eulers)$.  The rotation of a square-integrable function $f$ on
the 2-sphere \sphere\ (i.e.\ $f \in L^2(\sphere)$) is defined by
\begin{equation}
(\rot_\rho f)(\sa) = f(\rho^{-1} \sa), \; \; \rho \in \sothree 
\spcend ,
\end{equation}
where $\sa$ denotes spherical coordinates (\ie\ $\sa \in \sphere$).

\begin{figure}
\scriptsize
\psfrag{x}{$x$}
\psfrag{y}{$y$}
\psfrag{z}{$z$}
\psfrag{t}{$2\tan(\saa/2)$}
\psfrag{a}{$\saa$}
\centerline{
  \includegraphics[width=56mm]{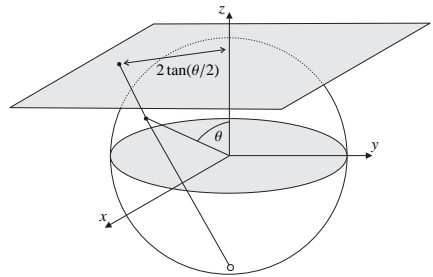}}
\caption[Stereographic projection]{Stereographic projection of the
  sphere onto the plane.}
\label{fig:stereographic_projection}
\end{figure}

Dilations on the sphere are constructed by first
lifting \sphere\ to the plane by a stereographic projection
from the south pole (\fig{\ref{fig:stereographic_projection}}),
followed by the usual Euclidean dilation in the plane, before
re-projecting back onto \sphere.  A spherical dilation is thus
defined by
\begin{equation}
(\dil_\scale f)(\sa) = f_\scale(\sa) 
= \sqrt{\lambda(\scale,\saa)} \: f(\sa_{1/\scale}),
\; \; \scale \in \real^{+}_{\ast}
\spcend ,
\end{equation}
where $\sa_\scale = (\saa_\scale, \sab)$ and $\tan(\saa_\scale/2)
= \scale \tan(\saa/2)$.  The $\lambda(\scale,\saa)$ cocycle term is
introduced to preserve the 2-norm and is defined by
\begin{displaymath}
\lambda(\scale,\saa) = \frac{4 \: \scale^2}
{[(\scale^2-1)\cos{\saa}+(\scale^2+1)]^2}
\spcend .
\end{displaymath}

A wavelet basis on the sphere may now be constructed by rotations and
dilations of an admissible%
\footnote{Candidate mother wavelets must also satisfy certain
  admissibility criteria to qualify as a spherical wavelet
  (see \citealt{antoine:2002} for a definition of the strict
  admissibility criterion and the more practical, necessary but not
  sufficient, zero mean criterion).} 
mother spherical wavelet $\wav \in
L^2(\sphere)$ (described further in \sectn{\ref{sec:mother_wavelets}}).
The corresponding wavelet family 
\mbox{$\{ \wav_{\scale,\rho} \equiv \rot_\rho \dil_\scale \wav,
\; {\rho \in \sothree,} \; {\scale \in \real_{\ast}^{+}} \}$}
provides an
over-complete set of functions in $L^2(\sphere)$.  The \cswt\
of $\sky \in L^2(\sphere)$ is given
by the projection onto each wavelet basis function in the usual
manner,
\begin{equation}
\skywav(\scale, \eulers) =
\int_{\sphere}
(\rot_{\eulers} \wav_\scale)^\conj(\sa) \;
\sky(\sa) \:
\dx \mu(\sa)
\spcend ,
\label{eqn:cswt}
\end{equation}
where the \conj\ denotes complex conjugation and 
\mbox{$\dx\mu(\sa)=\sin(\saa)\dx\saa \dx\sab$} is the usual rotationally
invariant measure on the sphere.  

% -- Removed --
%\citet{holschneider:1996} suggest that any \cswt\ should satisfy the
%Euclidean limit, that is the transform should reduce locally to the
%usual Euclidean Continuous Wavelet Transform (CWT).  By construction,
%the \cswt\ proposed by \citet{antoine:1998} satisfy this criterion (a
%precise mathematical definition may be constructed from group
%theoretic principles \citep{antoine:1998}).
% -- Removed --

The transform is general in the sense that all orientations in the
rotation group \sothree\ are considered, thus directional structure is
naturally incorporated.  It is important to note, however, that only
\emph{local} directions make any sense on \sphere.  There is no global
way of defining directions on the sphere\footnote{There is no
differentiable vector field of constant norm on the sphere and hence
no global way of defining directions.} -- there will always be some
singular point where the definition fails.  
%Directional analysis on the
%sphere is necessarily a small scale (local) operation.

A full directional wavelet analysis on the sphere has previously been
prohibited by the computational infeasibility of any
implementation. We rectify this problem by presenting a fast
algorithm in \appn{\ref{sec:fast_cswt}} to perform the directional
\cswt.

% ---------------------------------------

\subsection{Mother spherical wavelets}
\label{sec:mother_wavelets}

The wavelet basis previously described is constructed from rotations
and dilations of an admissible mother spherical wavelet.  Mother spherical wavelets
are simply constructed by projecting admissible Euclidean planar wavelets onto
the sphere by an inverse stereographic projection,
\begin{equation}
\wav_{\sphere}(\saa, \sab) 
= (\Pi^{-1} \wav_{\real^2}) (\saa, \sab)
\equiv \frac{2}{1+\cos(\saa)} 
\: \wav_{\real^2}(r, \sab)
\spcend ,
\label{eqn:mother_wav_proj}
\end{equation}
where $r = 2 \tan(\saa/2)$.  The modulating term is again introduced
to preserve the 2-norm.

\begin{figure*}
\begin{minipage}{175mm}
\centering
\mbox{
\subfigure[\Mexhat\ $\eccen=0.00$]{\includegraphics[width=50mm]{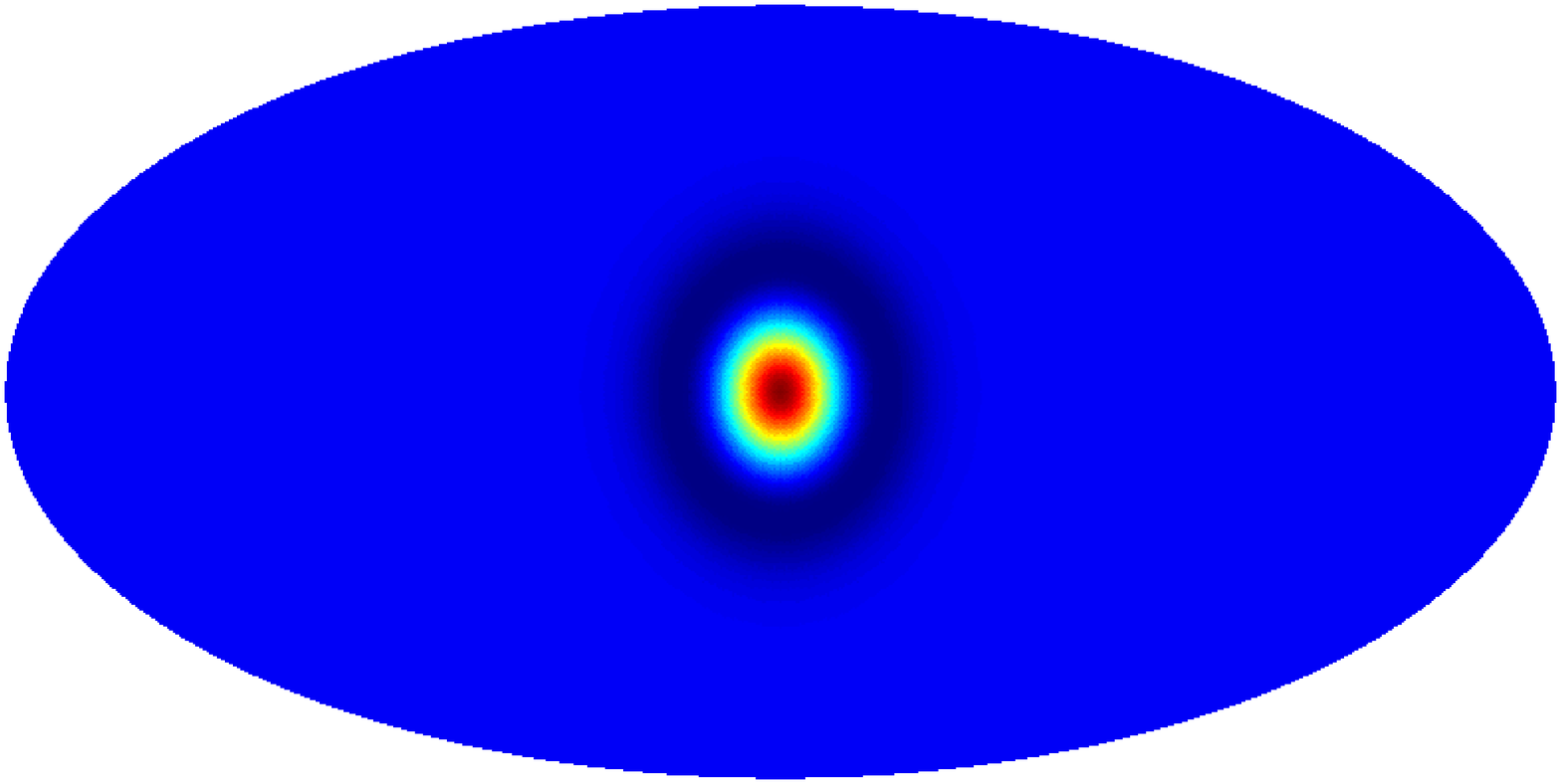}} \quad
\subfigure[\Mexhat\ $\eccen=0.95$]{\includegraphics[width=50mm]{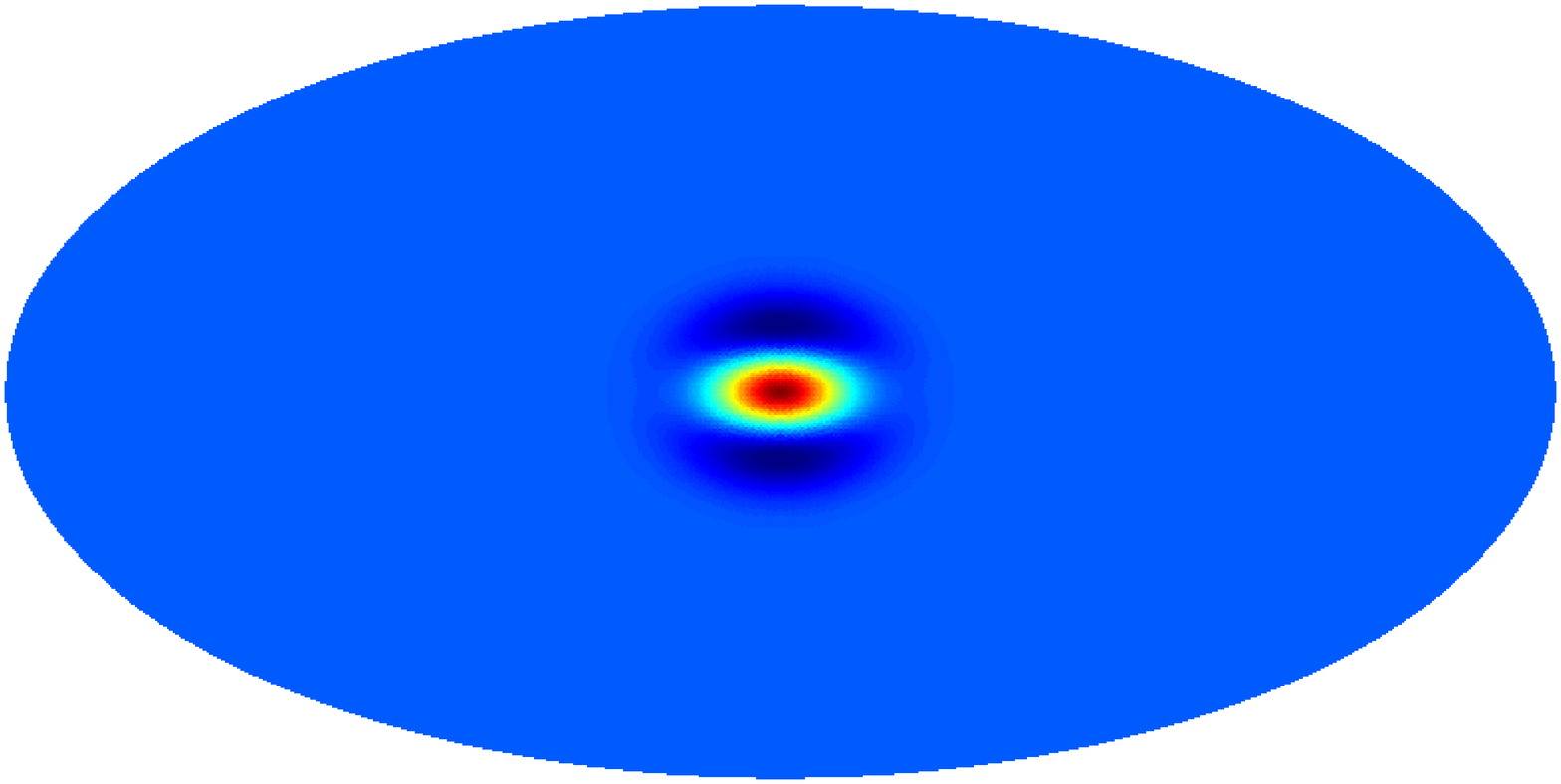}} \quad
\subfigure[\Morlet\ $\bmath{k}=\left( 10, 0 \right)^{T}$]{\includegraphics[width=50mm]{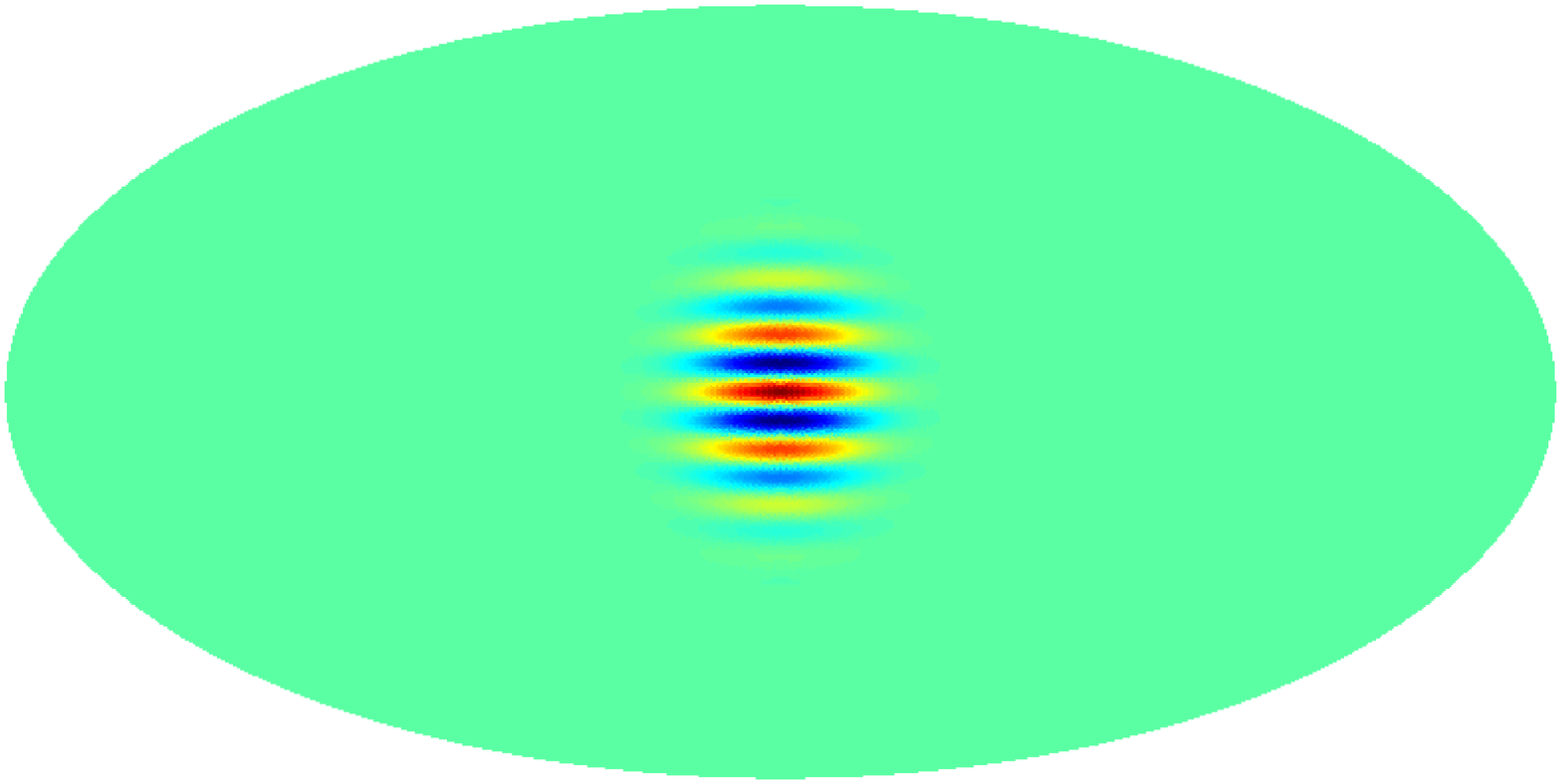}}
}
\caption{Spherical wavelets (dilation $\scale=750\arcmin$; size on sky
  $\effsize_1=2100\arcmin$, $\effsize_2=236\arcmin$).}
\label{fig:mother_wavelets}
\end{minipage}
\end{figure*}

Directional spherical wavelets may be naturally constructed in this
setting -- they are simply the projection of directional Euclidean planar
wavelets onto the sphere.  Two directional planar Euclidean mother wavelets
are defined in the following subsections: the elliptical \mexhat\ and \morlet\
wavelets.  The corresponding spherical wavelets are illustrated in
\fig{\ref{fig:mother_wavelets}}. 
The \mexhat\ wavelet and the \morlet\ wavelet, chosen for its sensitivity to
scanning artifacts, are subsequently applied to the detection of
non-Gaussianity in the \wmap\ 1-year data.  

\subsubsection{Elliptical \mexhat\ wavelet} 

We propose a directional extension of the usual \mexhat\ wavelet.  The
\emph{elliptical \mexhat\ wavelet} is defined as the negative of the Laplacian
of an elliptical 2-dimensional Gaussian,
\begin{eqnarray}
\lefteqn{
\wav_{\real^2}^{\rm Mex}(x,y; \sigx, \sigy) =
\frac{1}{2\pi \sigx^3 \sigy^3} 
\left (
 \sigx^2 + \sigy^2
  - \frac{x^2}{(\sigx / \sigy)^2}
  - \, \frac{y^2}{(\sigy / \sigx)^2} \right )
} \hspace{20mm} \nonumber \\
& & 
\times \:
\exp \left [ - \frac{1}{2}
\left ( \frac{x^2}{\sigx^2} + \frac{y^2}{\sigy^2} \right ) \right ]
\spcend ,
\end{eqnarray}
which reduces to the usual symmetric \mexhat\ wavelet for
the special case where $\sigx=\sigy$.  The elliptical \mexhat\
wavelet is invariant under integer azimuthal rotations of $\pi$, 
thus the rotation angle \eulerc\ is always quoted in the range $[0,\pi)$.  

We define the eccentricity of an elliptical \mexhat\ wavelet as the
eccentricity of the ellipse defined by the first zero-crossing, given
by
\begin{equation}
\eccen = \sqrt{1 - \left ( \frac{\sigy}{\sigx} \right ) ^4}
\spcend .
\end{equation}
Elliptical \mexhat\ wavelets are subsequently parameterised by their
eccentricity; the standard deviation in each direction is set by
$\sigy=1$ and $\sigx = \sigy \sqrt[4]{1 - \eccen^2}$.
Elliptical \mexhat\ wavelets are illustrated in
\fig{\ref{fig:mother_wavelets}} (a) and (b) for eccentricities
$\eccen=0.00$ and $\eccen=0.95$ respectively.

We define the effective size on the sky of a spherical elliptical \mexhat\
wavelet for a particular dilation as the angular separation between
the first zero-crossings on the major axis of the ellipse, given by
\begin{equation}
\effsize^{\rm Mex}_1(\scale) = 4 \tan^{-1} \left ( \frac{\scale}{\sqrt{2}}  \right ) 
\approx 2 \sqrt{2} \: \scale
\spcend .
\end{equation}

%\pagebreak
\subsubsection{\Morlet\ wavelet}

The \morlet\ wavelet is defined by 
\begin{equation}
\wav_{\real^2}^{\rm Mor}(\bmath{x}; \bmath{k}) 
= \rm{Re} \left [
\exp \left ( \frac{\img \: \bmath{k} \cdot \bmath{x}}{\sqrt{2}} \right )
\exp \left ( \frac{-\| \bmath{x} \|^2}{2} \right )
\right ]
\spcend ,
\end{equation}
where $\bmath{k}$ is the wave vector of the wavelet 
(henceforth we consider only wave vectors of the form 
$\bmath{k} = \left(k_0, 0 \right)^{T}$).  We have scaled
the usual definition of the \morlet\ wavelet to achieve size
{consis\-tency} with the elliptical \mexhat\ wavelet.  The \morlet\
spherical wavelet is also invariant under integer azimuthal rotations of $\pi$, 
thus the rotation angle \eulerc\ is always quoted in the domain $[0,\pi)$.

The \morlet\ wavelet has two orthogonal scales: one defining the
overall size of the wavelet and the other defining the size of its
internal structure. 
The overall effective size on the sky of the spherical \morlet\ wavelet is
defined as the angular separation between opposite
$e^{-1}$ roll-off points of the 
%$e ^{ \frac{-\| \bmath{x} \|^2}{2} }$
exponential decay factor,
and is given by
\begin{equation}\effsize^{\rm Mor}_1(\scale) = 4 \tan^{-1} \left (
\frac{\scale}{\sqrt{2}}  \right ) 
\approx 2 \sqrt{2} \: \scale
\spcend .
\end{equation}
Notice that for a given dilation \scale, the spherical elliptical
\mexhat\ and \morlet\ wavelets have an equivalent overall effective
size on the sky.
The effective size on the sky of the internal structure of the
\morlet\ wavelet is defined as the angular separation between the
first zero-crossings in the direction of the wave vector
$\bmath{k}$, and is given by
\begin{equation}
\effsize^{\rm Mor}_2(\scale) = 4 \tan^{-1} \left ( 
\frac{\scale \pi}{4 k_0}  \right ) 
\approx \frac{\scale \pi}{k_0}
\spcend .
\end{equation}

% -----------------------------------------------------------------------------

\section{Non-Gaussianity analysis}
\label{sec:non_gaussianity}

Spherical wavelet analysis is applied to probe the \wmap\ 1-year data
for possible deviations from Gaussianity.  
We follow a similar strategy to \citet{vielva:2003}, however we extend
the analysis to directional spherical wavelets to probe orientated
structure in the \cmb.

% ---------------------------------------

\subsection{Data preprocessing}

We consider the same data set analysed by both \citet{komatsu:2003}
and \citet{vielva:2003} in their non-Gaussianity studies.
The observed \wmap\ maps for which the \cmb\ is the dominant signal
(two Q-band maps at 40.7\ghz, two V-band maps at 60.8\ghz\ and four W-band
maps at 93.5\ghz) are combined to give a single signal-to-noise
ratio enhanced map.  These maps, together with
receiver noise and beam properties, are available from the
Legacy Archive for Microwave Background Data Analysis (\lambdaarch)
website\footnote{http://cmbdata.gsfc.nasa.gov/}.  The maps are
provided in the \healpix\footnote{http://www.eso.org/science/healpix/}
\citep{gorski:1999} format at a resolution of $\nside=512$ (the number
of pixels in a \healpix\ map is given by $12 \,  \nside^2$).
The data processing pipeline specified by \citet{komatsu:2003} is
applied to produce a single co-added map for analysis. 
The co-added temperature at a given position on the sky \sa\ is given
by
\begin{equation}
\cmbtemp(\sa) = \frac
{\sum_{r=3}^{10}  w_r(\sa) \: \cmbtemp_r(\sa)}
{\sum_{r=3}^{10} w_r(\sa)}
\spcend ,
\label{eqn:wmap_combine}
\end{equation}
where $\cmbtemp(\sa)$ is a \cmb\ temperature map and
the $r$ index corresponds to the Q-, V- and W-band
receivers respectively (indices $r=1,2$ correspond to the K and Ka
receiver bands that are excluded from the analysis).  The noise
weights $w_r(\sa)$ are defined by
\begin{equation}
w_r(\sa) = \frac{N_r(\sa)}{{\sigma_{0_r}}^2}
\spcend ,
\end{equation}
where $N_r(\sa)$ specifies the number of observations at each point on the
sky for each receiver band and $\sigma_{0_r}$ is the receiver noise
dispersion.

%\begin{table*}
%\begin{minipage}{128mm}
%\centering
%\caption{Noise dispersion for each \wmap\ differencing assembly.}
%\label{tbl:rec_noise}
%%
%\begin{tabular}{lcccccccc} \hline
%Assembly $r$ & 3 & 4 & 5 & 6 & 7 & 8 & 9 & 10 \\ \hline
%Differencing assembly & Q1 & Q2 & V1 & V2 & W1 & W2 & W3 & W4 \\
%Noise dispersion $\sigma_{0_r}$ (m\kelvin) & 2.267 & 2.156 & 3.288 & 2.937 & 5.852 & 6.533 & 6.880 & 6.725 \\ \hline
%\end{tabular}
%%
%\end{minipage}
%\end{table*}

Foreground cleaned sky maps, where the Galactic foreground signal
(consisting of synchrotron, free-free, and dust emission) has been
removed, are directly available from the \lambdaarch\ website.  The Galactic
foreground signal is removed by using the \mbox{3-band},
\mbox{5-parameter} template fitting method described by 
\citet{bennett:2003}.  We use these foreground cleaned maps in
our analysis. 

An independent foreground analysis of the \wmap\ data is performed by
\citet{tegmark:2003}.  The Tegmark \etal\ map is also constructed from a linear
summation of observed \wmap\ maps, however the weights used vary
over both position on the sky and scale.  We also perform our analysis
on the Tegmark \etal\ map to ensure any detected deviations from
Gaussianity are not due to differences in the various foreground
removal techniques. 

Following the analysis of \citet{vielva:2003} we down-sample 
map resolutions to $\nside=256$, since the very small scales are
dominated by noise (and also to reduce computational requirements).
The conservative \kpzero\ exclusion mask provided by the \wmap\ team 
(appropriately downsampled to conserve point source exclusion regions
in the coarser resolution) is
applied to remove emissions due to the Galactic plane and known point
sources.
The final preprocessed co-added map (hereafter referred to as the
\wmap\ team map, or simply \wmap\ map) and the map produced by
\cite{tegmark:2003} (hereafter referred to as the Tegmark map) are 
illustrated in \mbox{\fig{\ref{fig:cmb_wmap}}}.

\newlength{\mapwidth}
\setlength{\mapwidth}{70mm}

\begin{figure*}
\begin{minipage}{145mm}
\centering
\mbox{
\subfigure[Co-added \wmap\ team map]{\includegraphics[width=\mapwidth]{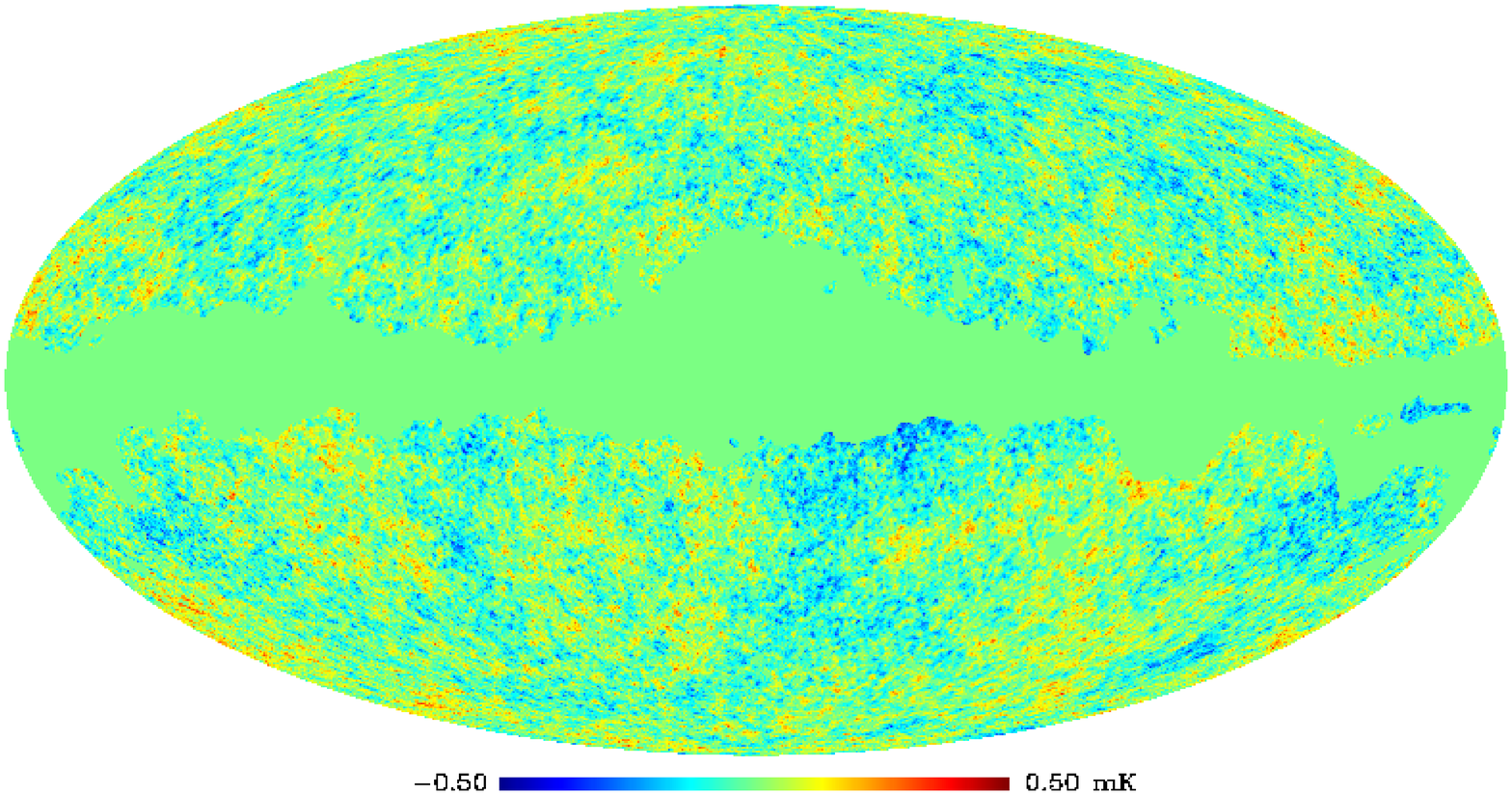}} \quad
\subfigure[Tegmark map]{\includegraphics[width=\mapwidth]{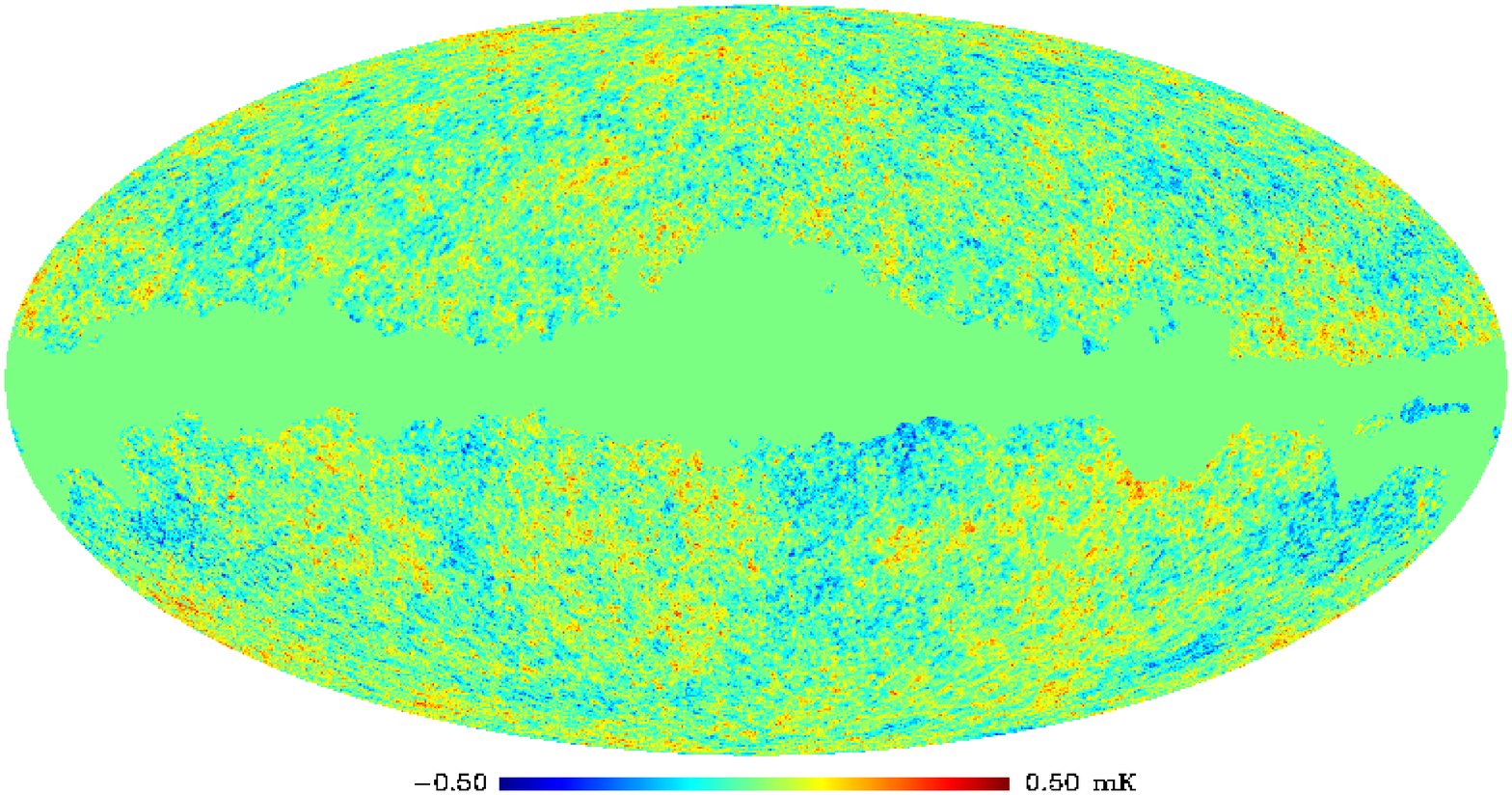}}
}
\caption{Preprocessed \wmap\ maps considered in the non-Gaussianity analysis.}
\label{fig:cmb_wmap}
\end{minipage}
\end{figure*}

% ---------------------------------------

\subsection{Monte Carlo simulations}

Monte Carlo simulations are performed to construct confidence bounds on
the test statistics used to probe for non-Gaussianity in the \wmap\
1-year data.  \ngsim\ Gaussian
\cmb\ realisations are produced from the theoretical power spectrum
fitted by the \wmap\ team.\footnote{The theoretical power
spectrum used is based on a \lcdmtext\ (\lcdm) model
using a power law for the primordial spectral index which best fits the
\wmap, Cosmic Background Imager ({CBI}) and Arcminute Cosmology
Bolometer Array Receiver ({ACBAR}) \cmb\ data, and is also directly
available from the \lambdaarch\ website.} 

To simulate the \wmap\ observing strategy each Gaussian \cmb\ realisation is
convolved with the beam transfer function of each of the Q-, V- and W-band
receivers.  White noise of dispersion
$\sigma_r(\sa) = \sigma_{0_r}/\sqrt{N_r(\sa)}$ is added to each band.
The resultant simulated  Q-, V- and W-band maps are combined
in the same manner used to construct the co-added map,
before down-sampling and \mbox{applying} the \kpzero\
mask, to give a final simulated Gaussian co-added map for
analysis.

The same Gaussian simulations are also used for comparison with
the Tegmark map.  Since the weights used to construct the Tegmark
map differ from those used to construct the \wmap\ team map, one
should strictly 
produce a second set of Gaussian simulations following the Tegmark
map construction method.  The weights for the Tegmark
map vary as a function of angular scale, and unfortunately are not quoted
explicitly. Nevertheless, for both the \wmap\ and Tegmark maps, the
weights sum to unity and the difference in the linear
combination of maps used by \citet{tegmark:2003} should not lead to
significant changes in the Gaussian confidence limits as compared with
those obtained using the Gaussian simulations produced to model the
\wmap\ map.

% ---------------------------------------

\subsection{Wavelet analysis}

The \cswt\ is a linear operation; hence the wavelet coefficients of a
Gaussian map will also follow a Gaussian distribution.  
One may therefore probe a full sky \cmb\ map for non-Gaussianity
simply by looking for deviations from Gaussianity in the distribution
of the spherical wavelet coefficients.  

The analysis consists of first taking the \cswt\ at a
range of scales and, for directional wavelets on the sphere, a range of
\eulerc\ {direc\-tions}.  The scales we consider (and the corresponding
\mbox{effective} size on the sky for both the \mexhat\ and \morlet\ wavelets)
are shown in \tbl{\ref{tbl:scales}}.  For directional wavelets we
consider five evenly spaced \eulerc\ orientations in the domain $[0,\pi)$. 

\begin{table*}
\begin{minipage}{145mm}
\centering
\caption{Wavelet scales considered in the non-Gaussianity analysis.
  The overall size on the sky $\effsize_1$ for a given
  scale are the same for both the \mexhat\ and \morlet\ wavelets.
  The size on the sky of the internal structure of the \morlet\
  wavelet $\effsize_2$ is also quoted.
}
\label{tbl:scales}
\begin{tabular}{lcccccccccccc} \hline
Scale & 1 & 2 & 3 & 4 & 5 & 6 & 7 & 8 & 9 & 10 & 11 & 12 \\ \hline
Dilation \scale  & 50\arcmin & 100\arcmin & 150\arcmin & 200\arcmin & 250\arcmin & 300\arcmin & 350\arcmin & 400\arcmin & 450\arcmin & 500\arcmin & 550\arcmin & 600\arcmin \\
Size on sky $\effsize_1$ & 141\arcmin & 282\arcmin & 424\arcmin & 565\arcmin & 706\arcmin & 847\arcmin & 988\arcmin & 1130\arcmin & 1270\arcmin & 1410\arcmin & 1550\arcmin & 1690\arcmin \\ 
Size on sky $\effsize_2$ & 15.7\arcmin & 31.4\arcmin & 47.1\arcmin & 62.8\arcmin & 78.5\arcmin & 94.2\arcmin & 110\arcmin & 126\arcmin & 141\arcmin & 157\arcmin & 173\arcmin & 188\arcmin \\ 
\hline
\end{tabular}
\end{minipage}
\end{table*}

Those wavelet coefficients distorted by the application of the
\kpzero\ mask are removed, as subsequently described, 
before test statistics are calculated from the 
wavelet coefficients.  An identical analysis is performed on
each Monte Carlo \cmb\ simulation in order to construct significance
measures for the test statistics.

\subsubsection{Coefficient exclusion masks}

The application of the \kpzero\ exclusion mask distorts coefficients
corresponding to wavelets that overlap with the mask exclusion
region. These contaminated wavelet coefficients must be
removed from any subsequent non-Gaussianity analysis.  An 
\emph{extended coefficient exclusion mask} is required to remove all
contaminated wavelet coefficients.

On small scales masked point sources introduce significant distortion
in wavelet coefficient maps and should not be neglected.  On larger
scales the masked Galactic plane introduces the most significant
distortion, as point source distortions are averaged over a large
wavelet support.
Our construction of an {ex\-tended} coefficient mask inherently accounts
for the dominant type of \mbox{distortion} on a particular scale. Firstly,
the \cswt\ of the original \kpzero\ 
mask is taken.  
Admissible spherical wavelets have zero mean \citep{antoine:2002}, hence
the only non-zero wavelet coefficients are those that are distorted by
the mask boundary. These distorted coefficients may be easily detected
and the coefficient exclusion mask extended accordingly. 
Coefficient exclusion masks are illustrated in
\fig{\ref{fig:cmask_mexhat000}} for the \mexhat\ $\eccen=0.00$
wavelet for a range of scales and in \fig{\ref{fig:cmask_morlet}}
for the \morlet\ wavelet for a given scale (the scale that a
significant non-Gaussianity detection is subsequently made) and a range
of orientations.  
As scale increases the dominant form of distortion may be seen in
\fig{\ref{fig:cmask_mexhat000}} to shift from point source to Galactic
plane.

\citet{vielva:2003} construct an extended coefficient 
mask simply by extending the Galactic plane region of the \kpzero\ mask by
$2.5\scale$ (the point source components of the original mask are not
extended).  
Several other definitions for coefficient exclusion masks
are analysed in detail by \citet{mw:2004}, none of which 
alter the results of subsequent non-Gaussianity analysis.
Although it is important to account correctly for the distortions
introduced by the \kpzero\ mask, the results of Gaussianity analysis
appear to be relatively insensitive to the particular mask chosen.

\newlength{\maskwidth}
\setlength{\maskwidth}{40mm}

\begin{figure*}
\begin{minipage}{150mm}
\centering
\mbox{
\subfigure[$\scale_1=50\arcmin$]
  {\includegraphics[width=\maskwidth]{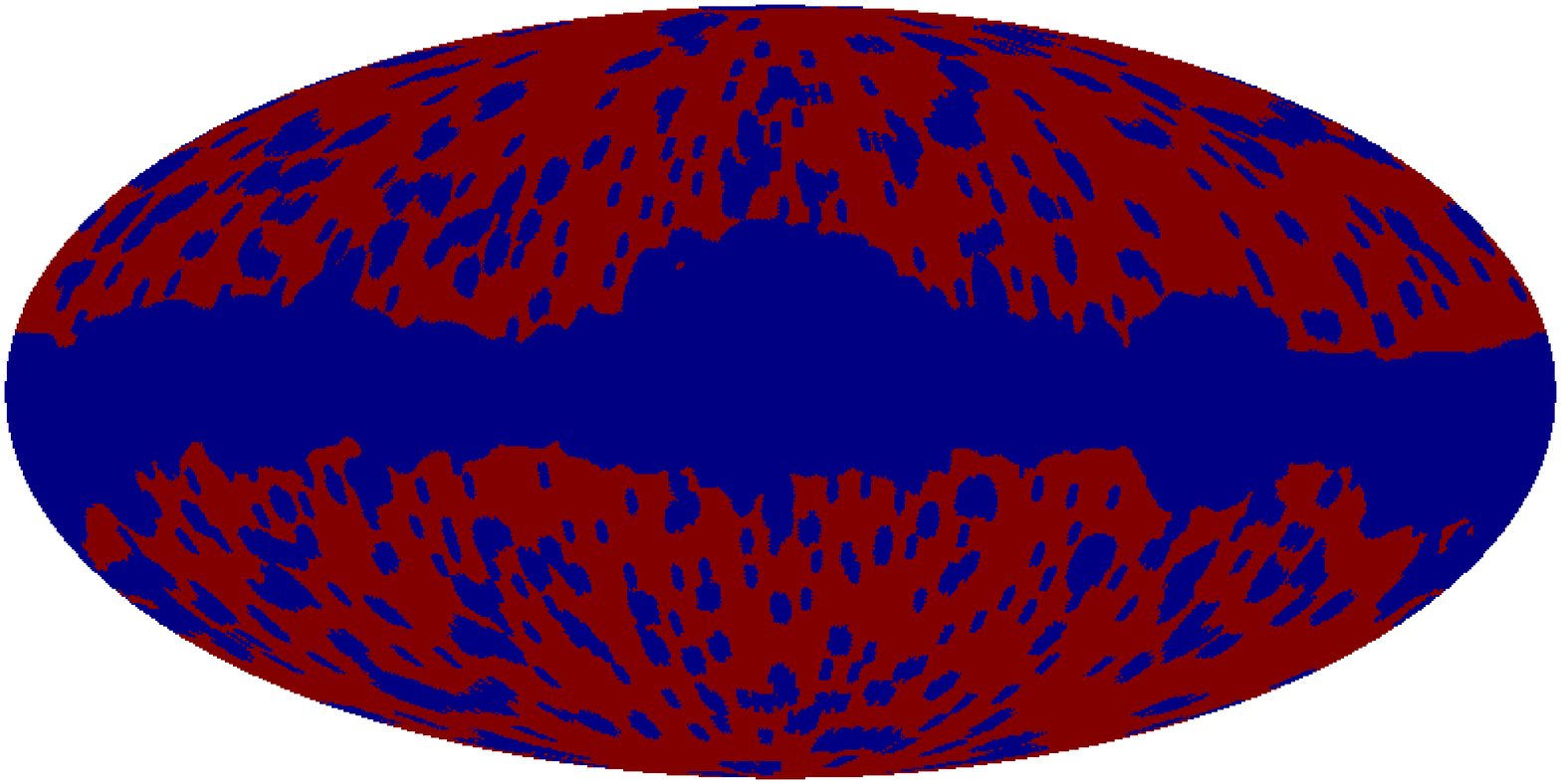}} \quad
\subfigure[$\scale_2=100\arcmin$]
  {\includegraphics[width=\maskwidth]{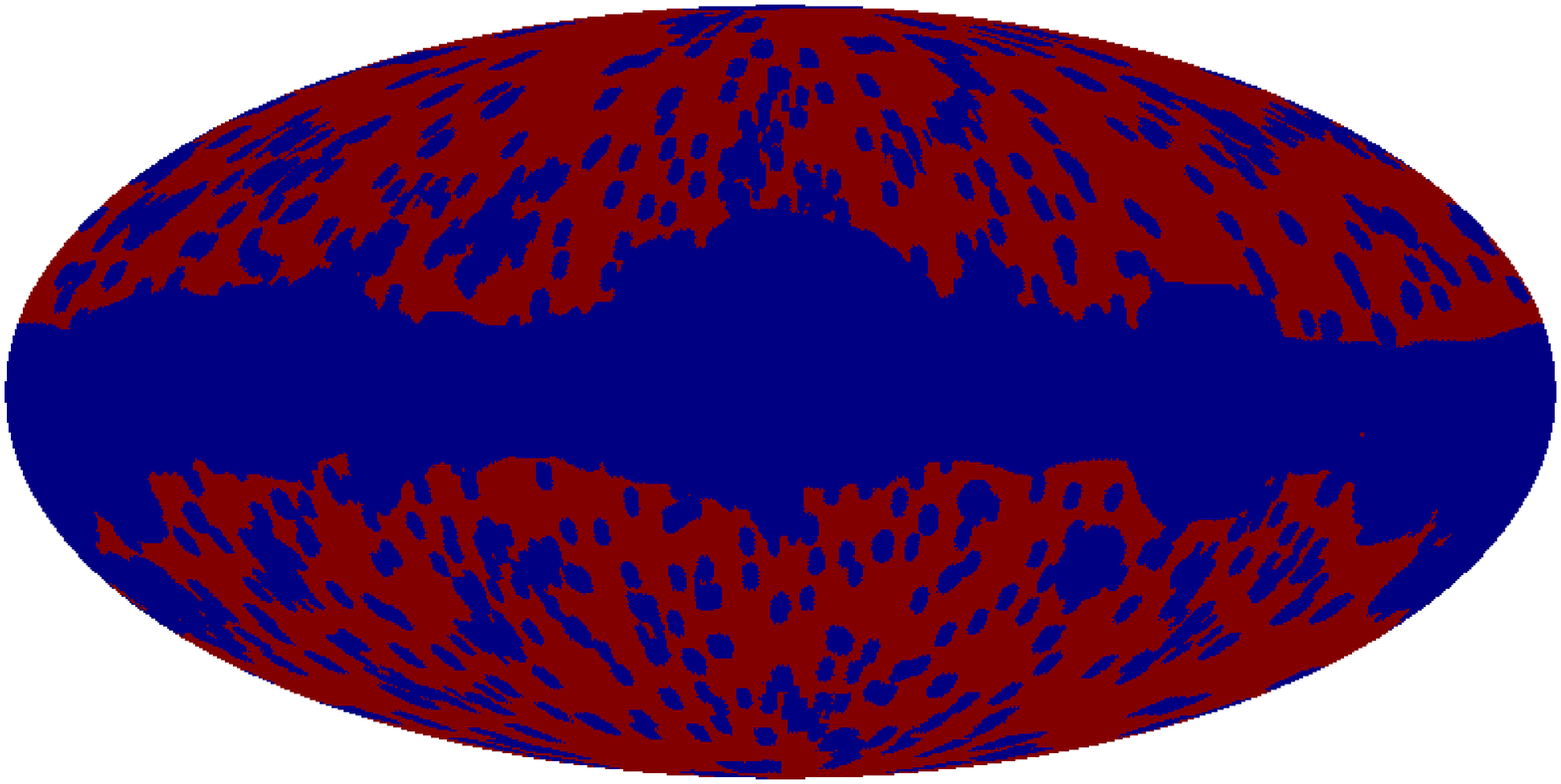}} \quad
\subfigure[$\scale_3=150\arcmin$]
  {\includegraphics[width=\maskwidth]{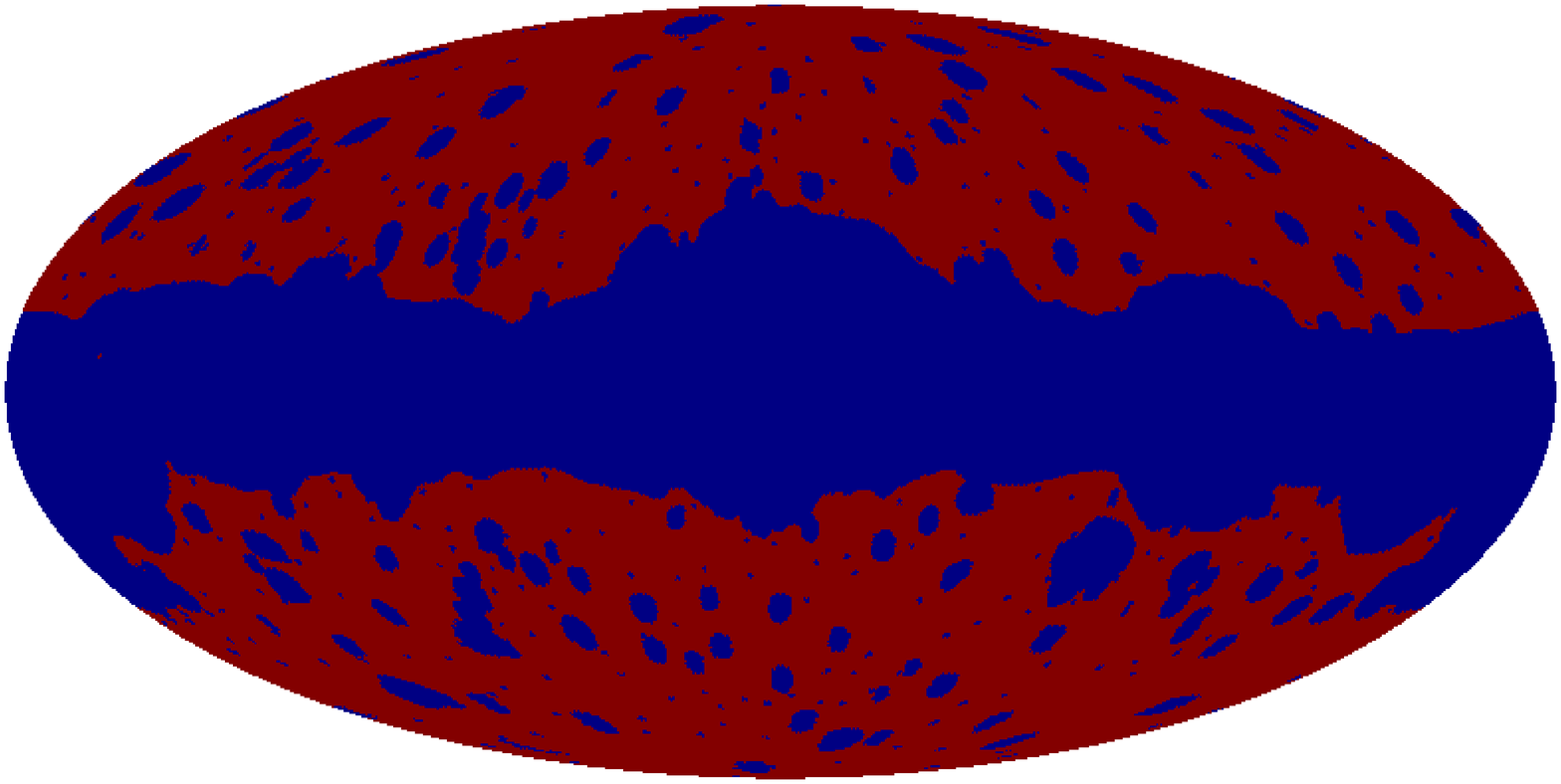}} 
}
\mbox{
\subfigure[$\scale_4=200\arcmin$]
  {\includegraphics[width=\maskwidth]{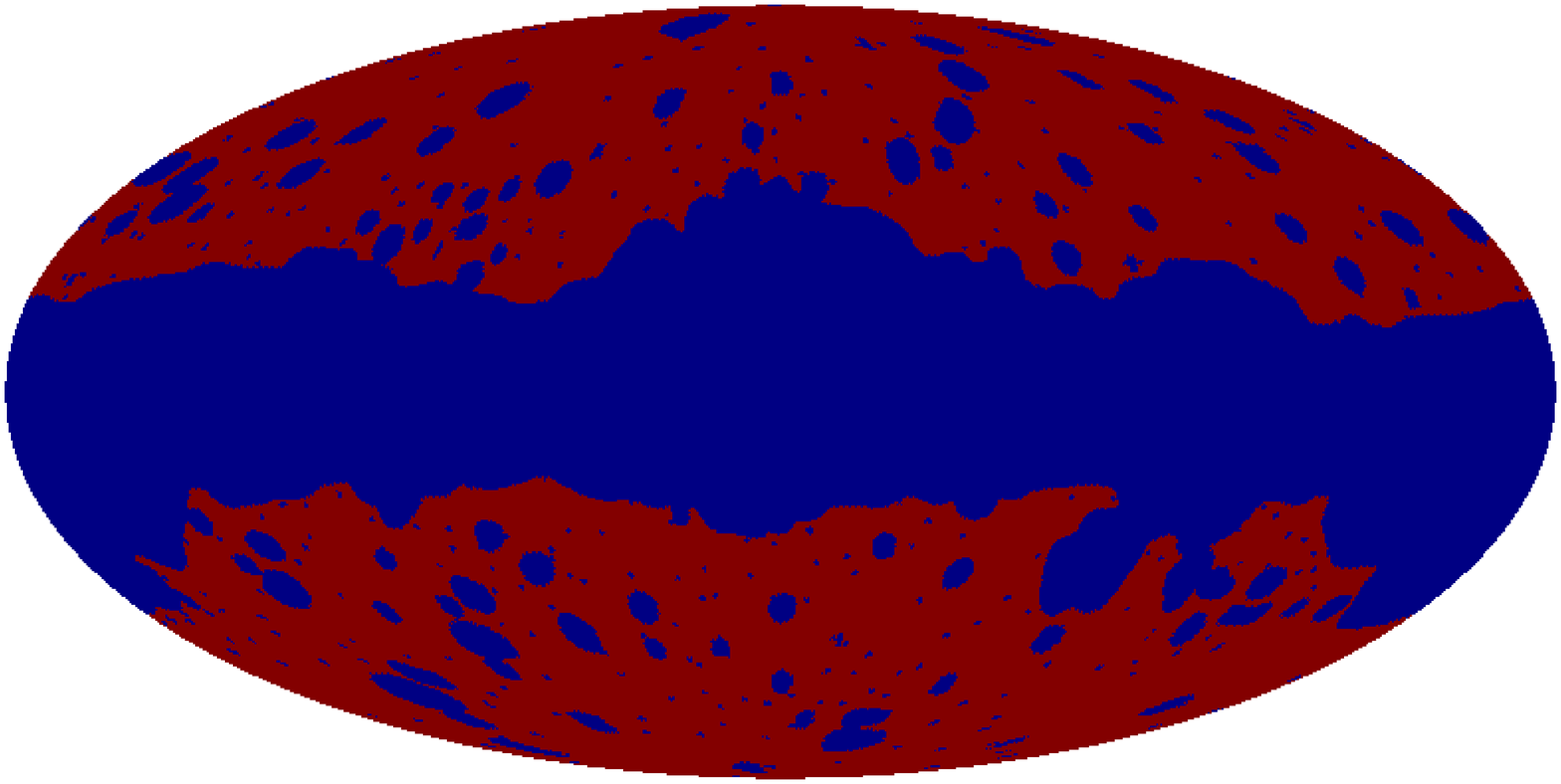}} \quad
\subfigure[$\scale_5=250\arcmin$]
  {\includegraphics[width=\maskwidth]{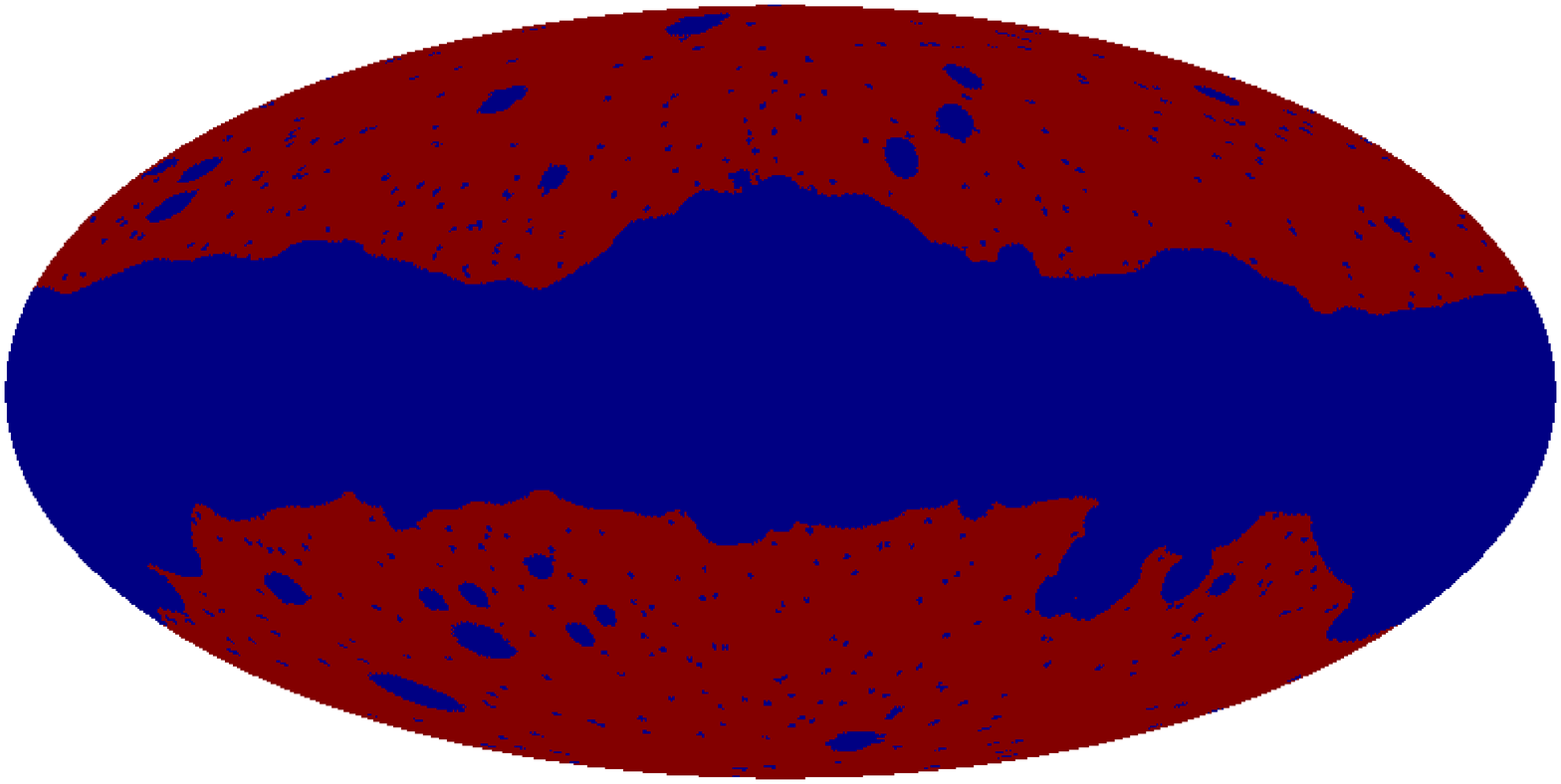}} \quad
\subfigure[$\scale_6=300\arcmin$]
  {\includegraphics[width=\maskwidth]{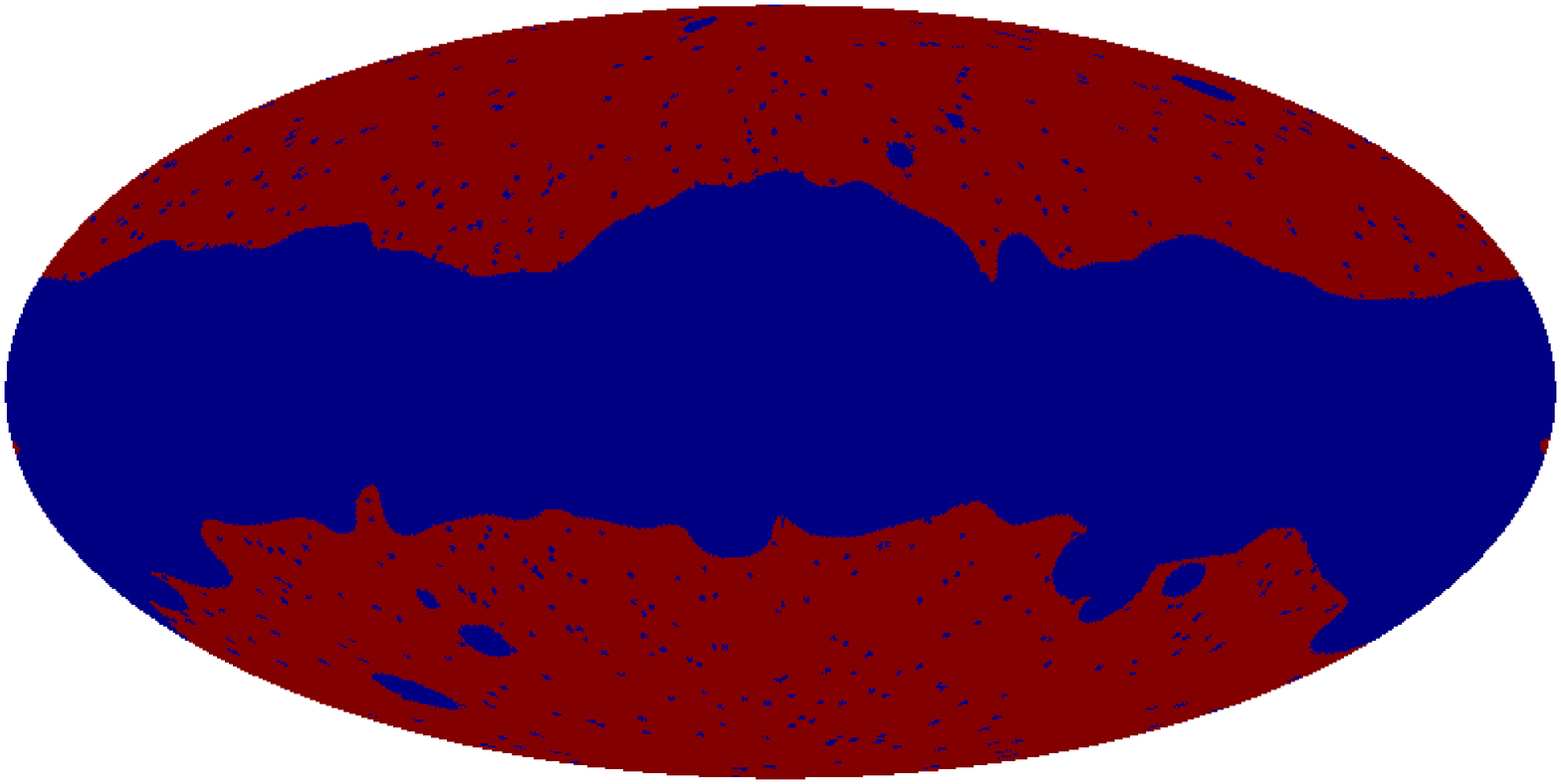}} 
}
\mbox{
\subfigure[$\scale_7=350\arcmin$]
  {\includegraphics[width=\maskwidth]{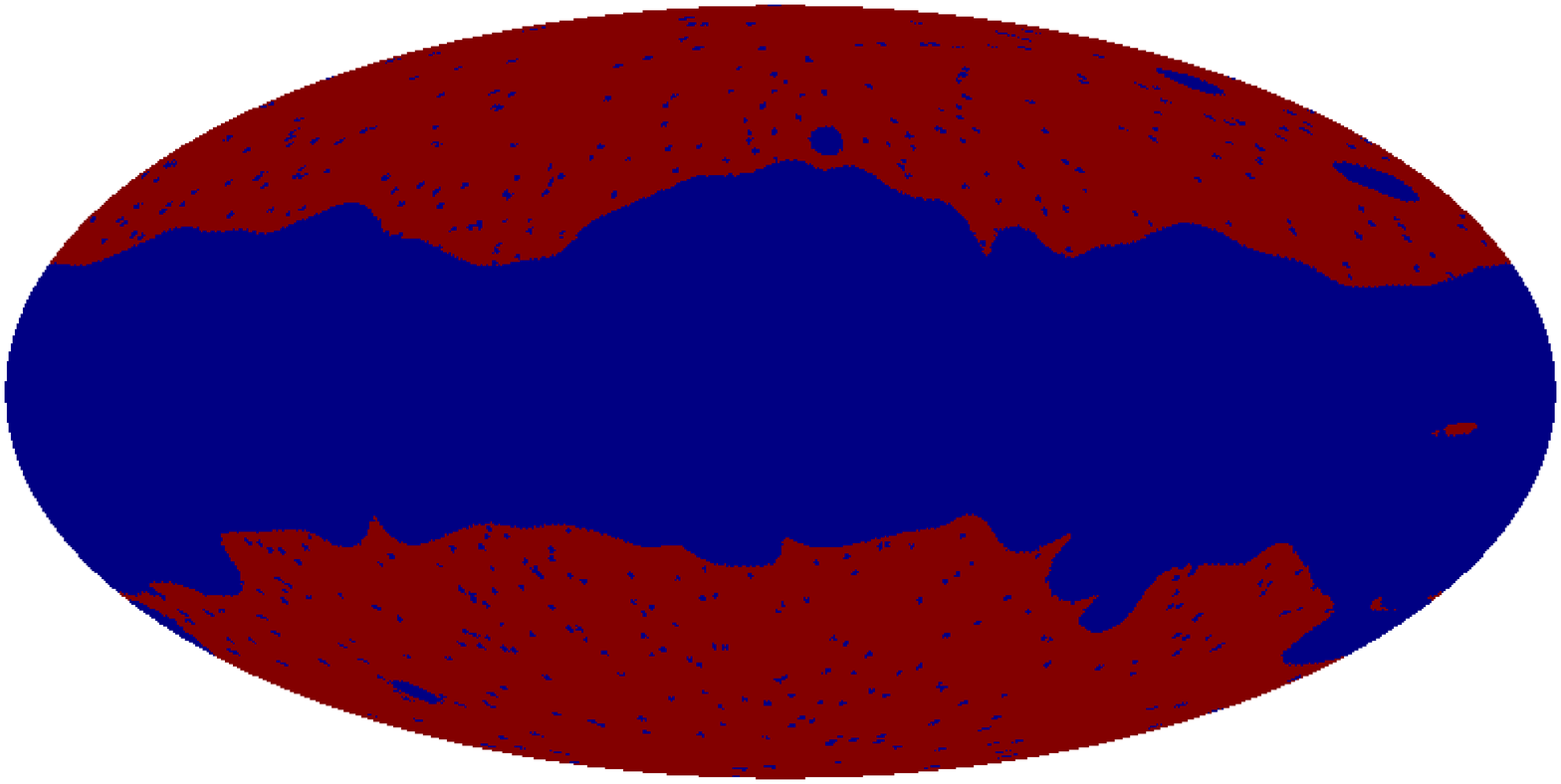}} \quad
\subfigure[$\scale_8=400\arcmin$]
  {\includegraphics[width=\maskwidth]{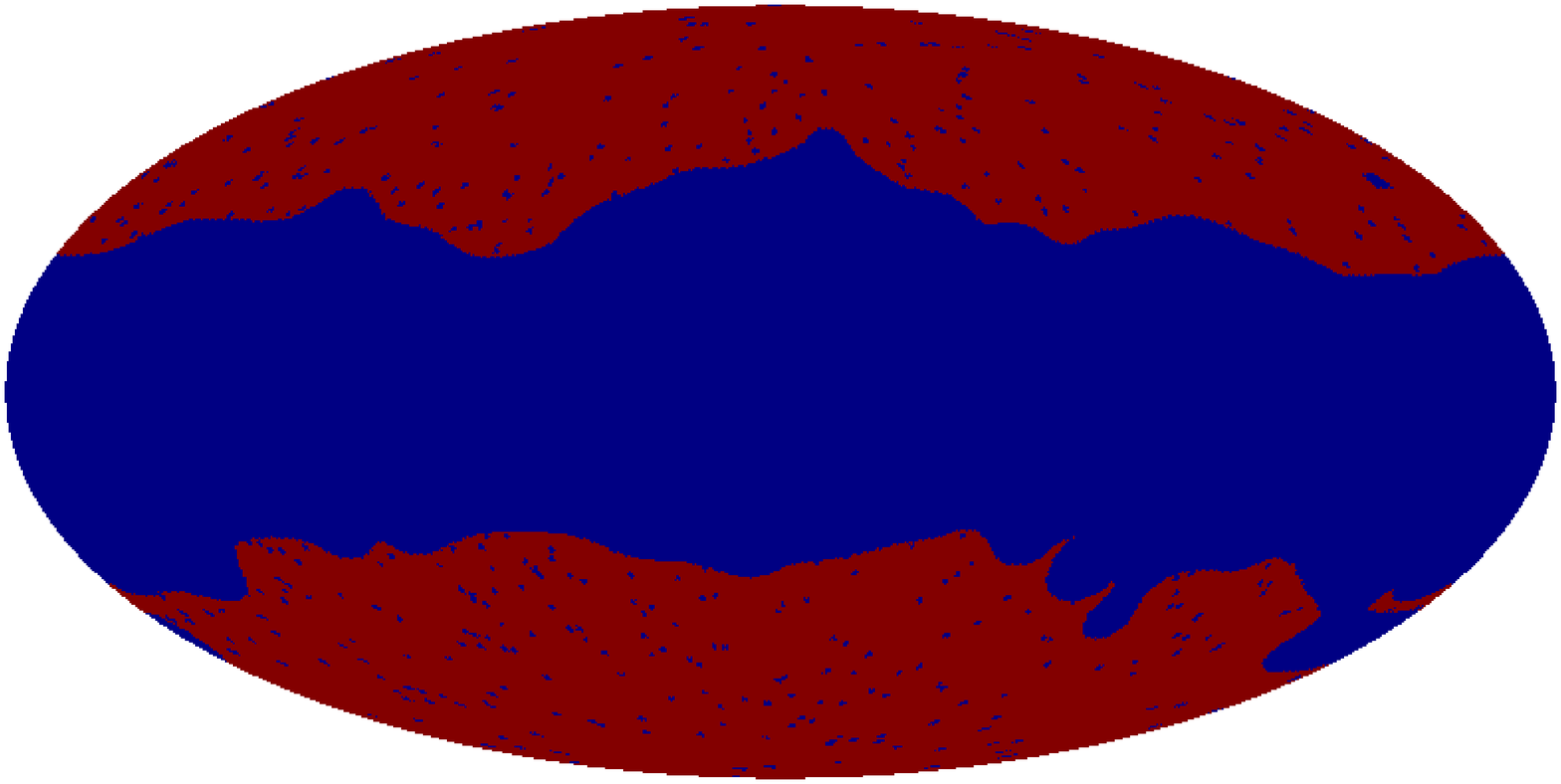}} \quad
\subfigure[$\scale_9=450\arcmin$]
  {\includegraphics[width=\maskwidth]{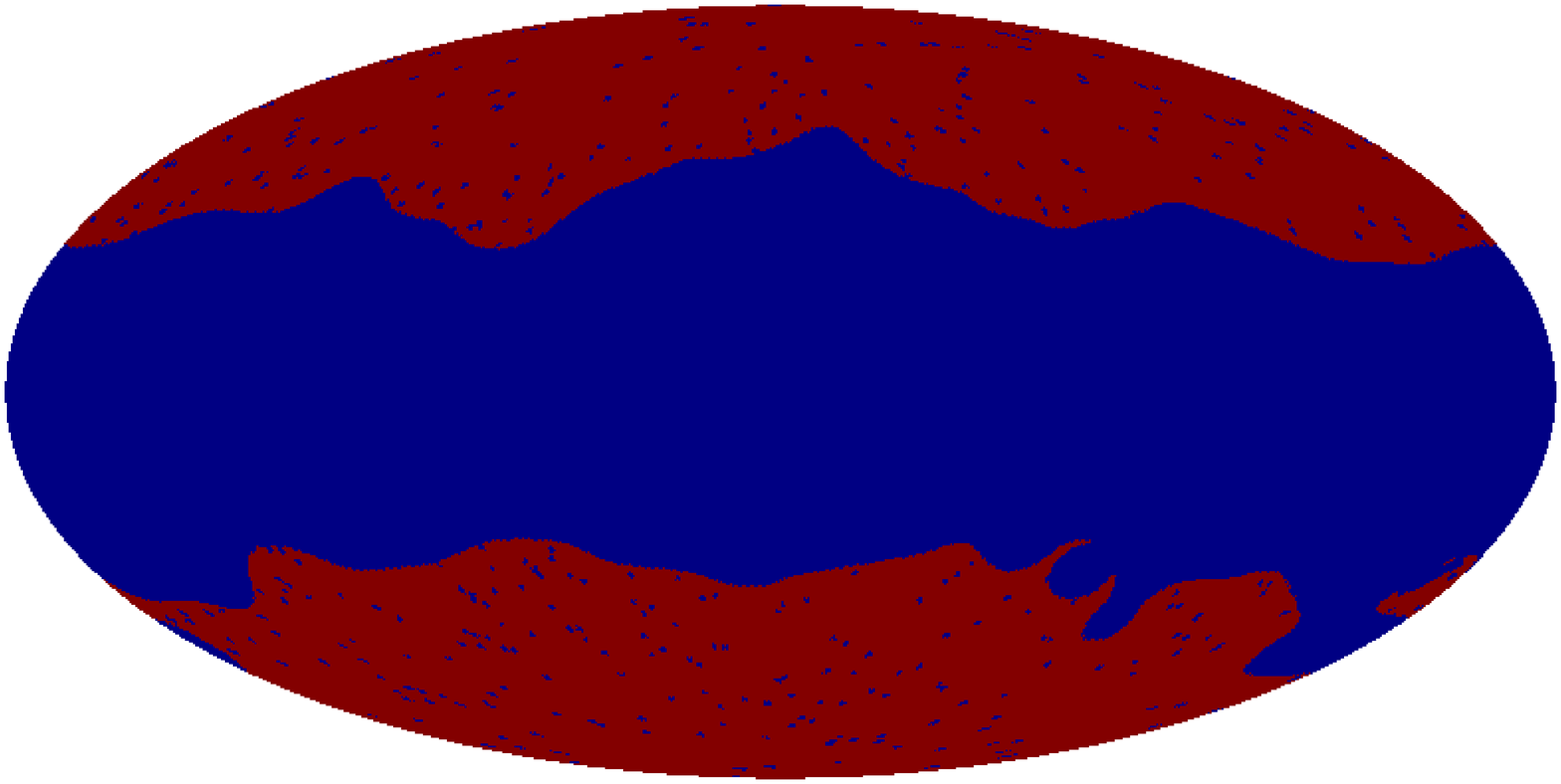}} 
}
\mbox{
\subfigure[$\scale_{10}=500\arcmin$]
  {\includegraphics[width=\maskwidth]{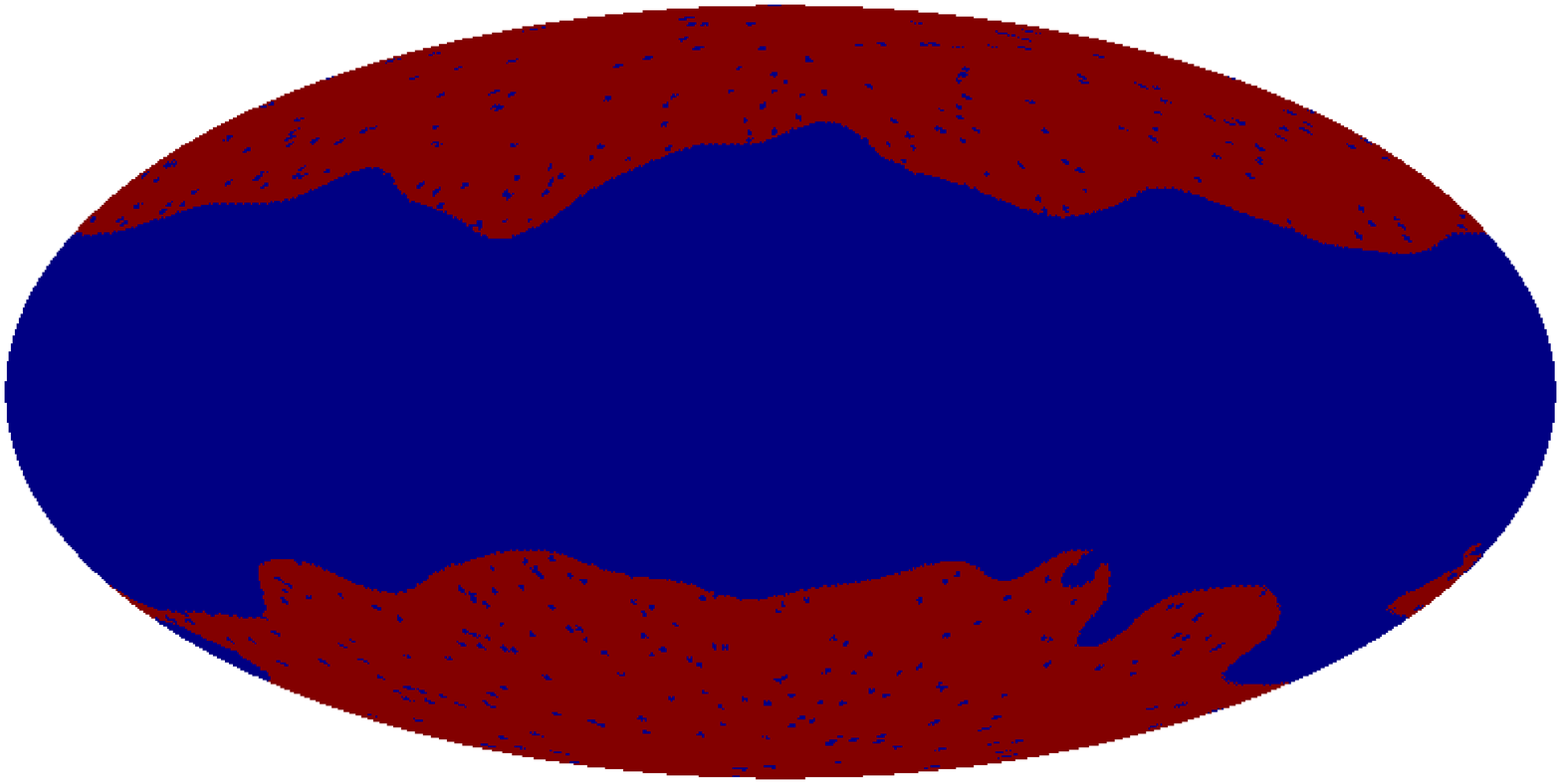}} \quad
\subfigure[$\scale_{11}=550\arcmin$]
  {\includegraphics[width=\maskwidth]{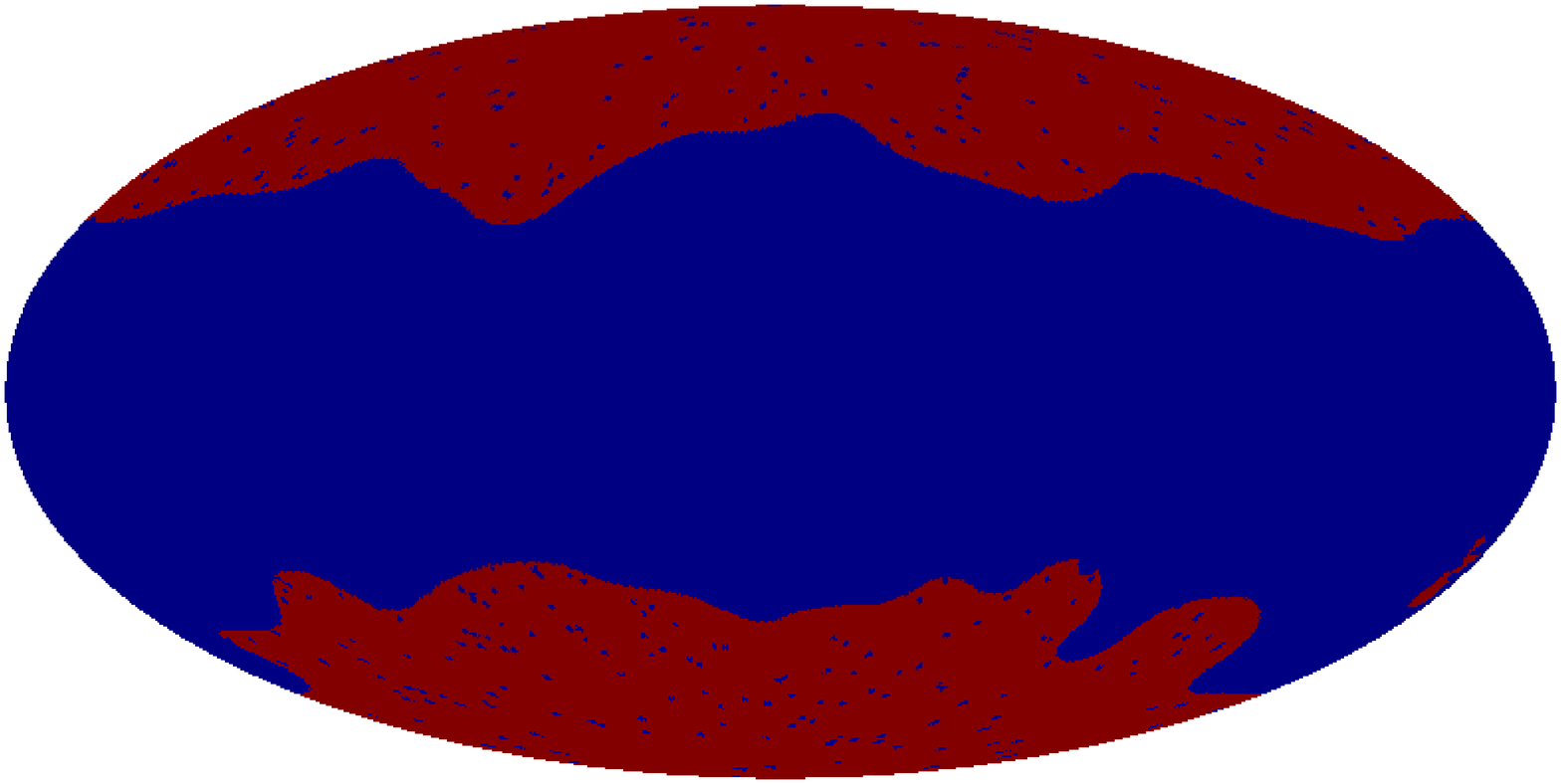}} \quad
\subfigure[$\scale_{12}=600\arcmin$]
  {\includegraphics[width=\maskwidth]{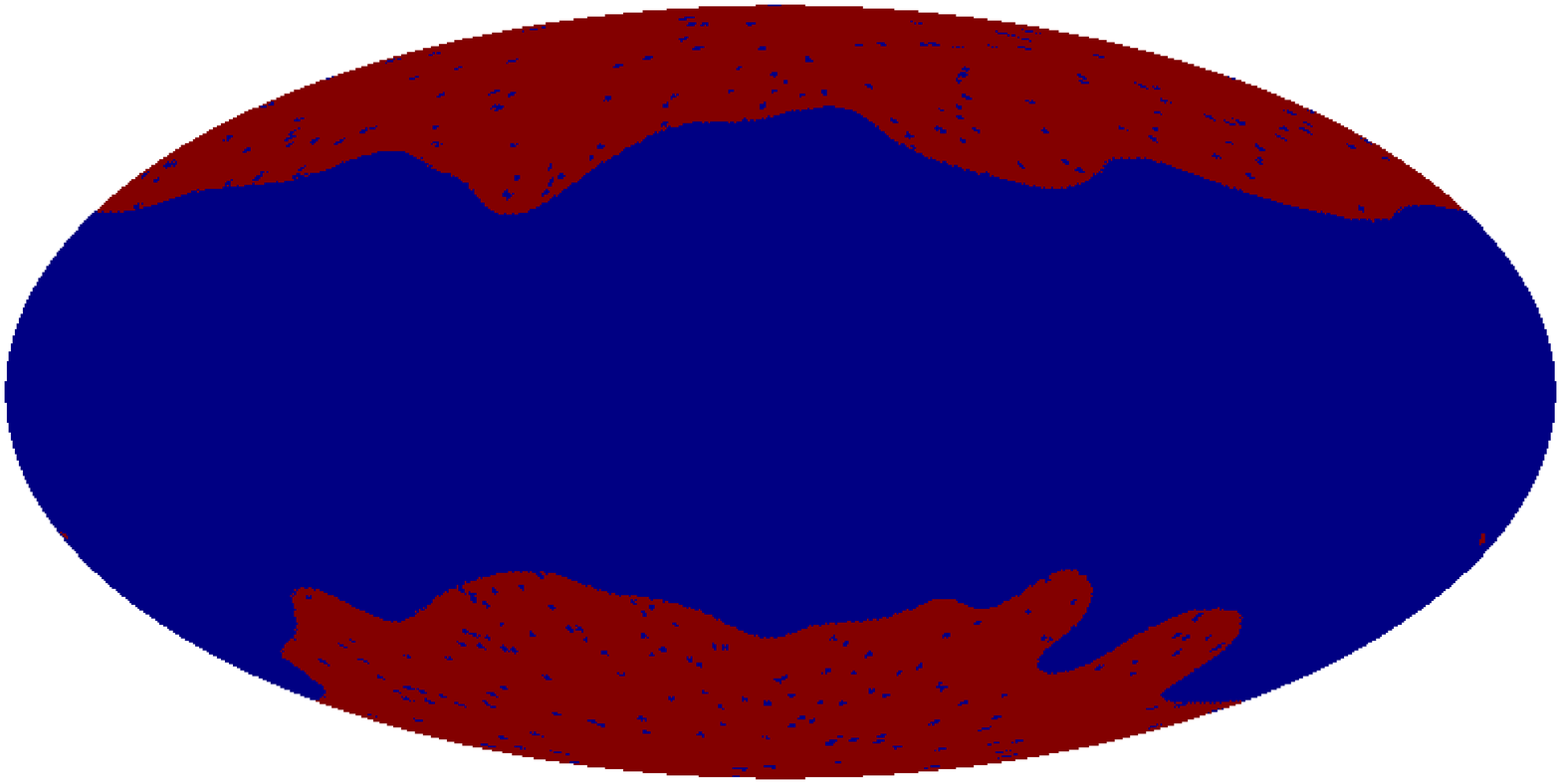}} 
}
\caption{Symmetric \mexhat\ $\eccen=0.00$ wavelet
  coefficient exclusion masks for each scale.}
\label{fig:cmask_mexhat000}
\end{minipage}
\end{figure*}

\begin{figure*}
\begin{minipage}{150mm}
\centering
\mbox{
\subfigure[$\eulerc=0^\circ$]
  {\includegraphics[width=\maskwidth]{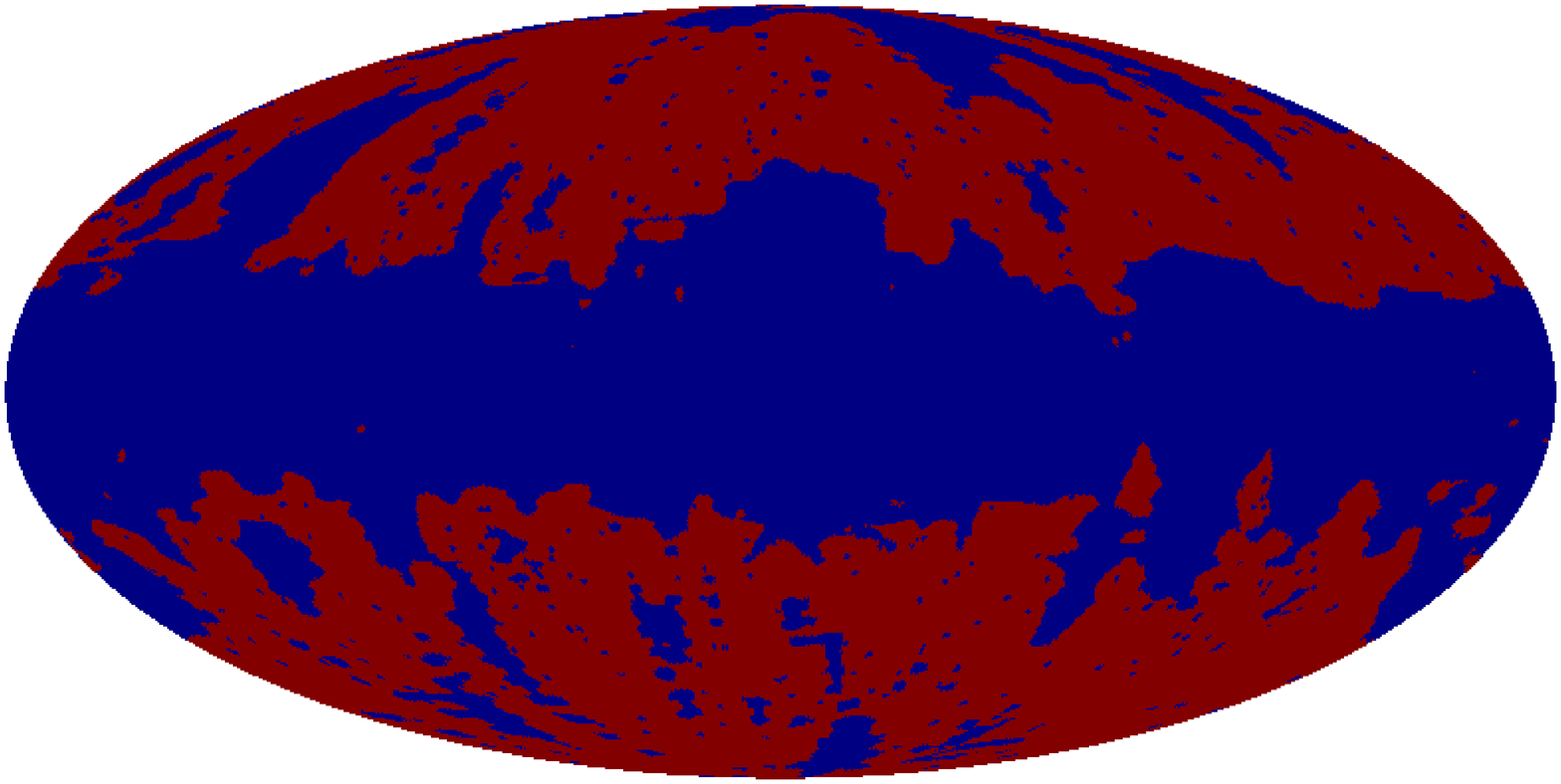}} \quad
\subfigure[$\eulerc=72^\circ$]
  {\includegraphics[width=\maskwidth]{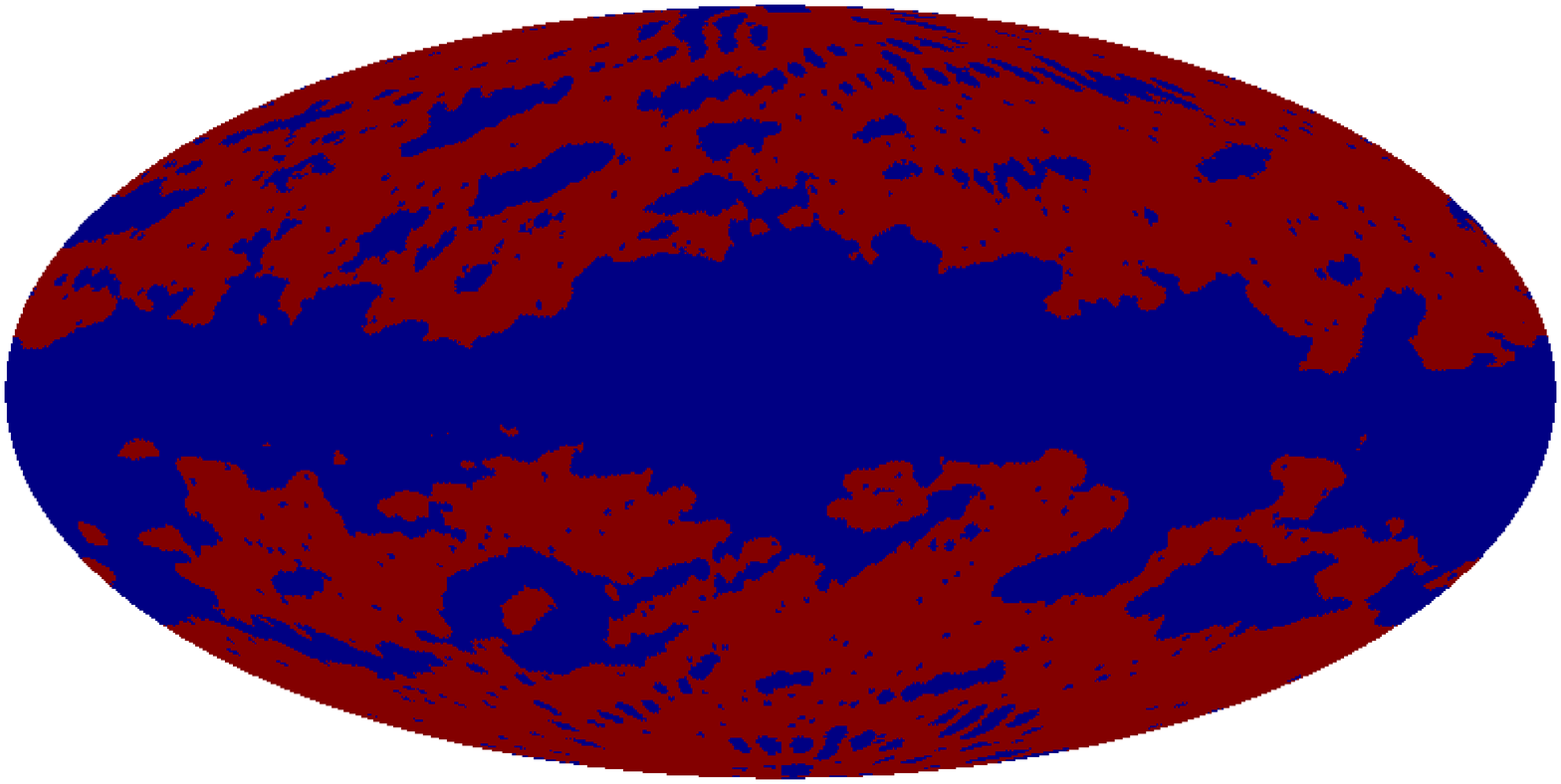}} \quad
\subfigure[$\eulerc=144^\circ$]
  {\includegraphics[width=\maskwidth]{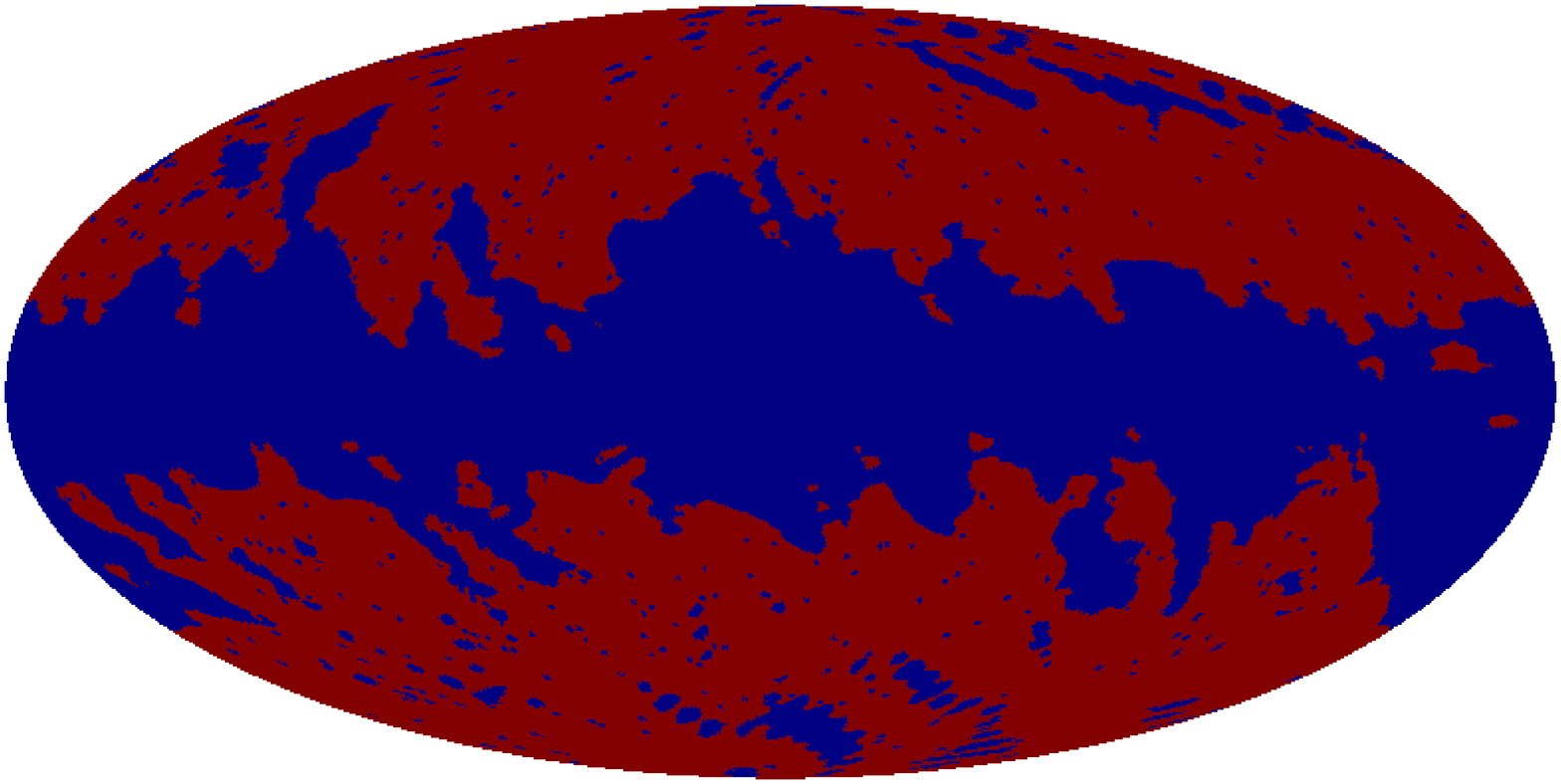}} 
}
\mbox{
\subfigure[$\eulerc=216^\circ$]
  {\includegraphics[width=\maskwidth]{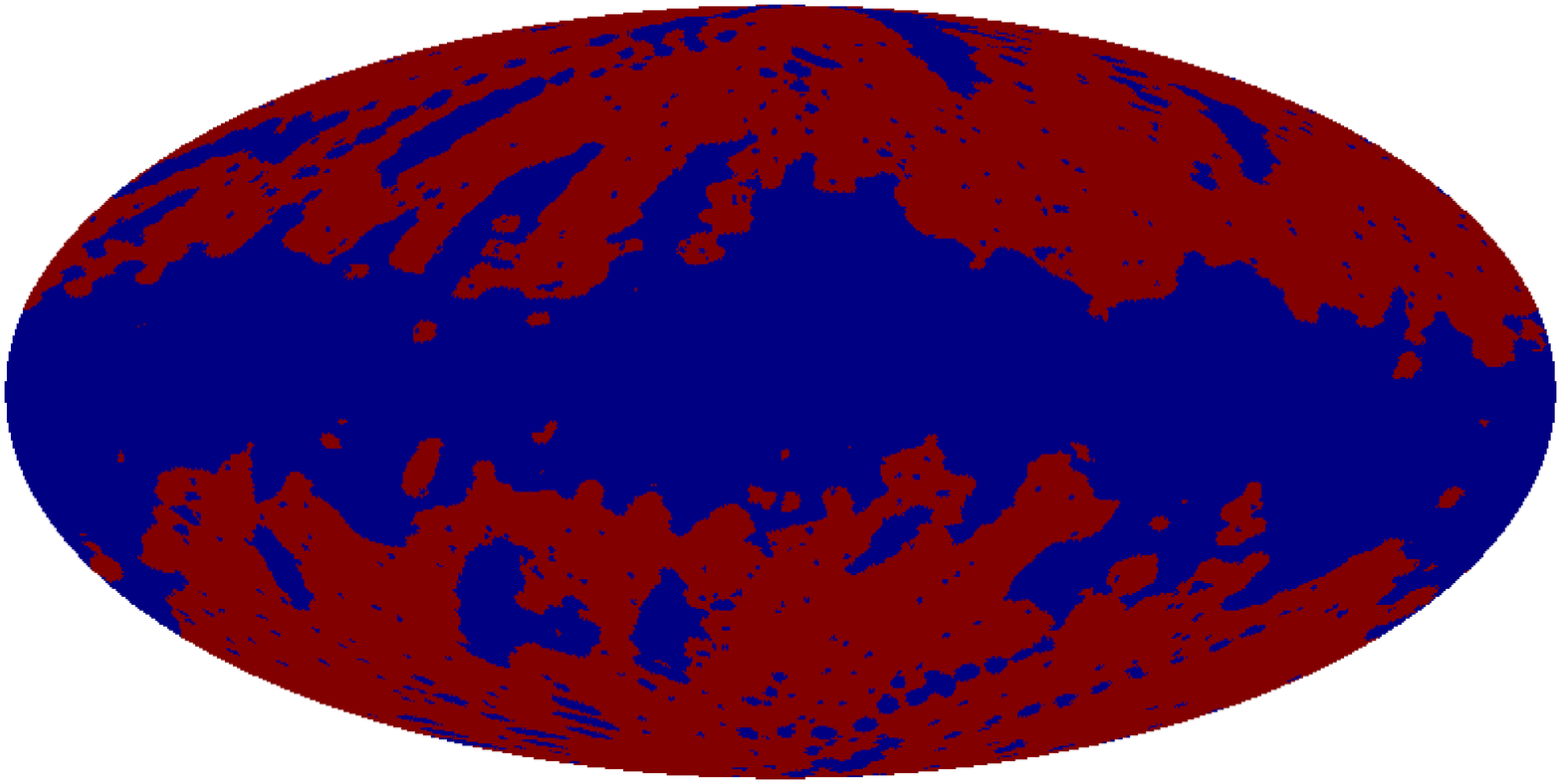}} \quad
\subfigure[$\eulerc=288^\circ$]
  {\includegraphics[width=\maskwidth]{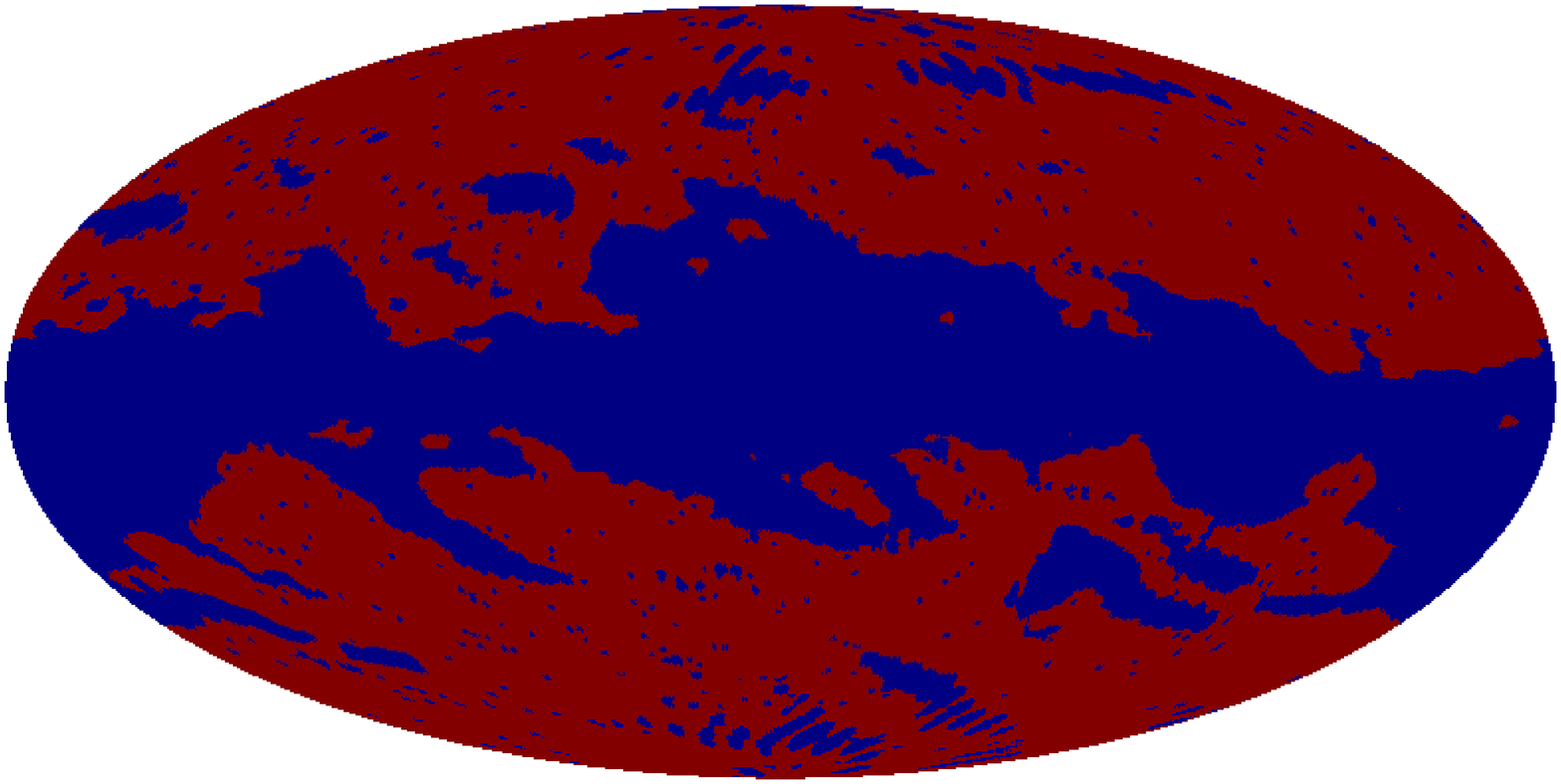}}
}
\caption{\Morlet\ wavelet coefficient exclusion masks at scale
  $\scale_{11}=550\arcmin$ for each orientation.}
\label{fig:cmask_morlet}
\end{minipage}
\end{figure*}

\subsubsection{Test statistics}

The third (skewness) and fourth (kurtosis) moments about the mean are
considered to test spherical wavelet coefficients for deviations
from Gaussianity.  These estimators describe the degree of
symmetry and the degree of peakedness in the underlying distribution
respectively.  Skewness is defined by
\begin{equation}
\skewness(\scale, \eulerc) = \frac{1}{\neff} \sum_{i=1}^{\neff} 
\frac{ \left [ \: \skywav_i(\scale, \eulerc) - \mean(\scale, \eulerc)
  \: \right ] ^3}
{\sigma^3(\scale, \eulerc)}
\end{equation}
\pagebreak
and excess kurtosis by
\begin{equation}
\kurtosis(\scale, \eulerc) = \frac{1}{\neff} \sum_{i=1}^{\neff} 
\frac{ \left [ \: \skywav_i(\scale, \eulerc) - \mean(\scale, \eulerc)
  \: \right ] ^4 }
{\sigma^4(\scale, \eulerc)}
\: - \: 3
\spcend ,
\end{equation}
where $\mu$ is the mean and $\sigma$ the dispersion of the
wavelet coefficients.
The $i$ index ranges over all wavelet coefficients not excluded by the
coefficient exclusion mask and indexes both \eulera\ and \eulerb\
components.  The number of spherical wavelet coefficients retained in
the analysis after the application of the coefficient exclusion mask is
given by \neff.

Skewness and excess kurtosis for a Gaussian distribution are both zero.
We look for deviations from zero in these test statistics to
indicate the existence of non-Gaussianity in the distribution of
spherical wavelet coefficients, and hence also in the
corresponding \cmb\ map.

% -----------------------------------------------------------------------------

\section{Results}
\label{sec:results}

To probe for non-Gaussianity in the \wmap\ 1-year data, the
analysis procedure described in \sectn{\ref{sec:non_gaussianity}} is
performed on both the \wmap\ team and Tegmark maps.
 The three spherical wavelets illustrated in
\fig{\ref{fig:mother_wavelets}} are considered, namely the
symmetric \mexhat\ $\eccen=0.00$ wavelet, the
elliptical \mexhat\ $\eccen=0.95$ wavelet and the \morlet\  
$\bmath{k}=\left( 10, 0 \right)^{T}$ wavelet.  The \mexhat\
$\eccen=0.00$ case has previously been analysed by
\citet{vielva:2003} (although some scales considered differ),
thereby providing a consistency check for the analysis.

% ---------------------------------------

\subsection{Wavelet coefficient statistics}

For a given wavelet, the skewness and kurtosis of wavelet \mbox{coefficients}
is calculated for each scale and orientation.  These \mbox{statistics} are
displayed in \fig{\ref{fig:stat_plot}}, with confidence intervals
{con\-structed} from the Monte Carlo simulations also shown.  For
\mbox{directional} wavelets, only the orientations corresponding to the
maximum deviations from Gaussianity are shown.

Our coefficient exclusion mask differs slightly from that applied by
\citet{vielva:2003}, thus for comparison purposes we also perform the
\mexhat\ $\eccen=0.00$ analysis without applying any extended
coefficient mask, as \citet{vielva:2003} also do initially.  
These results, although not shown, correspond
identically.  By applying different coefficient masks the shape of the
plots differ slightly, nevertheless the findings drawn remain the
same.  Deviations from Gaussianity are detected in the kurtosis
outside of the 99\% confidence region constructed from Monte-Carlo
simulations, on scales $\scale_5=250\arcmin$ and $\scale_6=300\arcmin$. 
Furthermore, a deviation outside the 99\% confidence region is detected
in the skewness at scale $\scale_2=100\arcmin$.  \citet{vielva:2003}
measure a similar skewness value at this scale, although this lies
directly on the boundary of their 99\% confidence region.

Deviations from Gaussianity are also detected in both skewness and
kurtosis using the elliptical \mexhat\ \mbox{$\eccen=0.95$} wavelet. 
In each case the observed deviations occur on a slightly larger scale
than those found using the symmetric \mexhat\ $\eccen=0.00$ wavelet.
This behaviour also appears typical for simulated Gaussian map
realisations.
Adjacent orientations exhibit similar results, although not at such 
large confidence levels (but still outside of the 99\% confidence
level). 

An extremely significant deviation from Gaussianity is observed in the
skewness of the \morlet\ wavelet coefficients at scale 
$\scale_{11}=550\arcmin$ and orientation $\eulerc=72^\circ$.  The kurtosis
measurement on the same scale and orientation also lies outside of
the 99\% confidence region. 

% ******************************************************************
% Note about plots:
% When update plots must change bounding box for statistic plots and
% histograms.
% Bounding box magic numbers: 74, 55                  
% ** top of plot cut off, increase one of numbers 75->74 ** XX - no
% **** Now changed to 69, 30 to account for axis labels ****
% --> Not any longer, now just use ps files.
% ******************************************************************

\newlength{\statplotwidth}
\setlength{\statplotwidth}{55mm}

\begin{figure*}
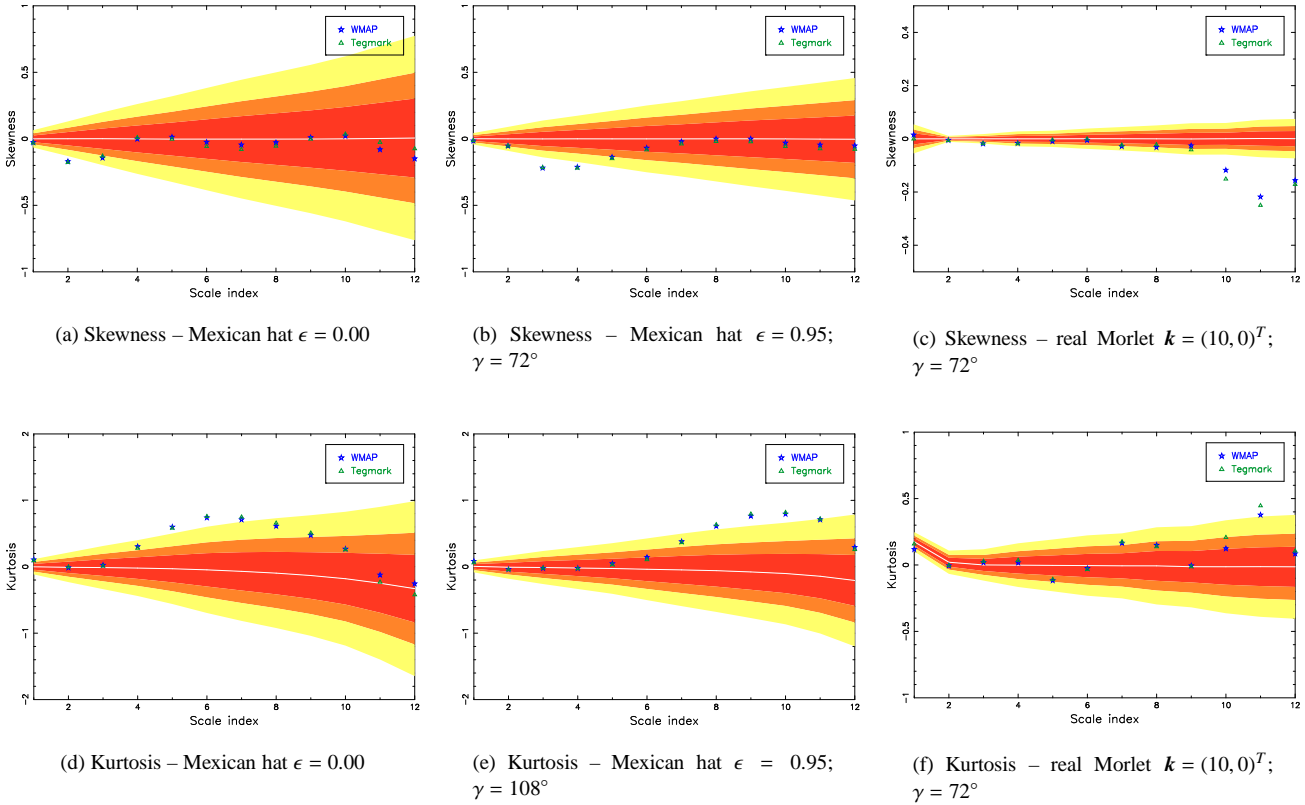

\begin{minipage}{175mm}
\centering
\mbox{
\subfigure[Skewness -- \mexhat\ $\eccen=0.00$]
  {\includegraphics[clip=,angle=-90,width=\statplotwidth]{figures/skewness_cmask_mexhat000_ig01.ps}} \quad
\subfigure[Skewness -- \mexhat\ \mbox{$\eccen=0.95$}; \mbox{$\eulerc=72^\circ$}]
  {\includegraphics[clip=,angle=-90,width=\statplotwidth]{figures/skewness_cmask_mexhat095_ig02.ps}} \quad
\subfigure[Skewness -- \morlet\ \mbox{$\bmath{k}=\left( 10, 0 \right)^{T}$}; \mbox{$\eulerc=72^\circ$}]
  {\includegraphics[clip=,angle=-90,width=\statplotwidth]{figures/skewness_cmask_morlet_ig02.ps}}
}
\mbox{
\subfigure[Kurtosis -- \mexhat\ $\eccen=0.00$]
  {\includegraphics[clip=,angle=-90,width=\statplotwidth]{figures/kurtosis_cmask_mexhat000_ig01.ps}} \quad
\subfigure[Kurtosis -- \mexhat\ $\eccen=0.95$; \mbox{$\eulerc=108^\circ$}]
  {\includegraphics[clip=,angle=-90,width=\statplotwidth]{figures/kurtosis_cmask_mexhat095_ig05.ps}} \quad
\subfigure[Kurtosis -- \morlet\ \mbox{$\bmath{k}=\left( 10, 0 \right)^{T}$}; \mbox{$\eulerc=72^\circ$}]
  {\includegraphics[clip=,angle=-90,width=\statplotwidth]{figures/kurtosis_cmask_morlet_ig02.ps}}
}
\caption{Spherical wavelet coefficient statistics for each wavelet.  Confidence regions
  obtained from \ngsim\ Monte Carlo simulations are shown for 68\% (red), 95\%
  (orange) and 99\% (yellow) levels, as is the mean (solid white
  line).
  Only the orientations corresponding to the most significant deviations
  from Gaussianity are shown for the \mexhat\ $\eccen=0.95$ and
  \morlet\ wavelet cases.}
\label{fig:stat_plot}
\end{minipage}
\end{figure*}

% ---------------------------------------

\subsection{Statistical significance of detections}

We now consider in more detail, 
the most significant deviation from Gaussianity obtained in each of the
panels in \fig{\ref{fig:stat_plot}}.  In particular, we
examine the distribution of each statistic, obtained from the 
Gaussian Monte Carlo simulations, and also perform a $\chi^2$ test for
each statistic.
Significance measures of the non-Gaussianity detections may then be
constructed from each test. 

\begin{figure*}
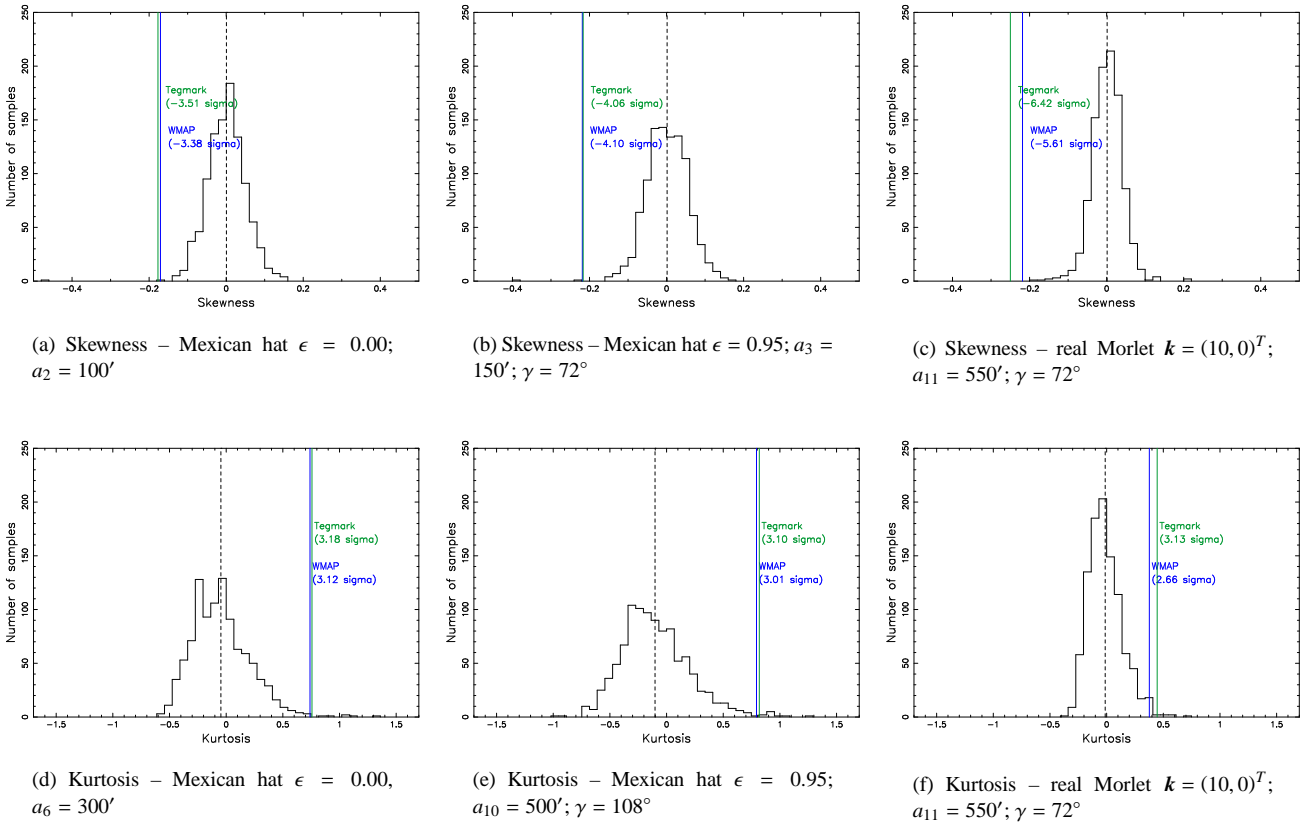

\begin{minipage}{175mm}
\centering
\mbox{
\subfigure[Skewness -- \mexhat\ $\eccen=0.00$; \mbox{$\scale_2=100\arcmin$}]
  {\includegraphics[clip=,angle=-90,width=\statplotwidth]{figures/hist_skew_mexhat000_ia02_ig01.ps}} \quad
\subfigure[Skewness -- \mexhat\ \mbox{$\eccen=0.95$}; $\scale_3=150\arcmin$; $\eulerc=72^\circ$]
  {\includegraphics[clip=,angle=-90,width=\statplotwidth]{figures/hist_skew_mexhat095_ia03_ig02.ps}} \quad
\subfigure[Skewness -- \morlet\ \mbox{$\bmath{k}=\left( 10, 0 \right)^{T}$}; $\scale_{11}=550\arcmin$; $\eulerc=72^\circ$]
  {\includegraphics[clip=,angle=-90,width=\statplotwidth]{figures/hist_skew_morlet_ia11_ig02.ps}}
}
\mbox{
\subfigure[Kurtosis -- \mexhat\ $\eccen=0.00$, \mbox{$\scale_6=300\arcmin$}]
  {\includegraphics[clip=,angle=-90,width=\statplotwidth]{figures/hist_kurt_mexhat000_ia06_ig01.ps}} \quad
\subfigure[Kurtosis -- \mexhat\ $\eccen=0.95$; \mbox{$\scale_{10}=500\arcmin$}; $\eulerc=108^\circ$]
  {\includegraphics[clip=,angle=-90,width=\statplotwidth]{figures/hist_kurt_mexhat095_ia10_ig05.ps}} \quad
\subfigure[Kurtosis -- \morlet\ \mbox{$\bmath{k}=\left( 10, 0 \right)^{T}$}; \mbox{$\scale_{11}=550\arcmin$}; $\eulerc=72^\circ$]
  {\includegraphics[clip=,angle=-90,width=\statplotwidth]{figures/hist_kurt_morlet_ia11_ig02.ps}}
}
\caption{Histograms of spherical wavelet coefficient statistic
  obtained from \ngsim\ Monte Carlo simulations. The mean is shown by
  the dashed vertical line.   The observed statistics for
  the \wmap\ and Tegmark maps are shown by the blue and green lines
  respectively.  The number of standard deviations these observations
  deviate from the mean is also displayed on each plot.
  Only those scales and orientations corresponding to the most
  significant deviations from Gaussianity are shown for each wavelet.}
\label{fig:hist}
\end{minipage}
\end{figure*}

\Fig{\ref{fig:hist}} shows histograms constructed from the
Monte Carlo simulations for those test statistics corresponding to the most
significant deviations from Gaussianity.  The measured statistic for
both the \wmap\ team and Tegmark maps is also shown on each plot,
with the number of standard deviations these observations
deviate from the mean.  In particular, we note the large deviations
shown in panel~(c), corresponding to \nstdmorskew\ and
\nstdmorskewteg\ standard deviations for the \morlet\ wavelet analysis
of the \wmap\ and Tegmark maps respectively.  

Having determined separately the confidence level of the largest
non-Gaussianity detection in each panel of \fig{\ref{fig:stat_plot}},
we now consider the statistical significance of our results for each
wavelet as a whole. Treating each wavelet separately,
we search through the Gaussian simulations to determine the number of
maps that have an equivalent or greater deviation in \emph{any} of the
test {statis\-tics} calculated from that map using the given wavelet.  That
is, if any skewness or kurtosis statistic%
\footnote{
Although we recognise the distinction between skewness and kurtosis, 
there is no reason to partition the set of test statistics into
skewness and kurtosis subsets.  The full set of test statistics must
be considered.}
calculated from the Gaussian
map -- on any scale or orientation -- deviates more than the maximum
deviation observed in the \wmap\ data for that wavelet,
then the map is flagged as
exhibiting a more significant deviation.
This is an extremely conservative means of constructing significance levels
for the observed test statistics.
Significance levels corresponding to the detections considered in
\fig{\ref{fig:hist}} are calculated and displayed in
\tbl{\ref{tbl:num_deviations}}.
Notice that although several individual test statistics 
fall outside of the 99\% confidence region, the true significance
level of the detection when all statistics are taken into account is
considerably lower. 
Using our conservative test, the significance of the
non-Gaussianity detection made by \citet{vielva:2003}, previously
quoted at greater than 99.9\% significance, drops to a significance
level of \clmexkurt\%.
Of particular interest is the non-Gaussian detection in the skewness
of \morlet\ wavelet coefficients on scale $\scale_{11}=550\arcmin$ and
orientation $\eulerc=72^\circ$.  This statistic deviates from the mean
by \nstdmorskew\ standard deviations for the \wmap\ map and by
\nstdmorskewteg\ standard deviations for the Tegmark map.  
The detection is made at an overall significance level of
\clmorskew\%.

\begin{table}
\centering
\caption{Deviation and significance levels of spherical wavelet
  coefficient statistics calculated from the \wmap\ map (similar
  results are obtained using the Tegmark map).  
  Standard deviations and significant levels 
  are calculated from \ngsim\ Monte Carlo Gaussian simulations.
  The table variables are defined as follows: the number of standard
  deviations the observation deviates from the mean is given by \nstd;
  the number of simulated Gaussian maps that exhibit an equivalent or greater deviation
  in \emph{any} test statistics calculated using the given wavelet is
  given by \ndev; the corresponding significance level of the
  non-Gaussianity detection is given by \conflevel.
  Only those scales and orientations corresponding to the most
  significant deviations from Gaussianity are shown for each wavelet.}
\label{tbl:num_deviations}
\subfigure[\Mexhat\ $\eccen=0.00$]
{
\begin{tabular}{lcc} \hline
& Skewness & Kurtosis \\
& ($\scale_2=100\arcmin$) & ($\scale_6=300\arcmin$) \\ \hline
\nstd & \nstdmexskewsgn & \nstdmexkurtsgn \\
\ndev & \nstatmexskew\ maps & \nstatmexkurt\ maps \\
\conflevel & \clmexskew\% & \clmexkurt\% \\ \hline
\end{tabular}
}
\subfigure[\Mexhat\ $\eccen=0.95$]
{
\begin{tabular}{lcc} \hline
& Skewness & Kurtosis \\
& ($\scale_3=150\arcmin$; $\eulerc=72^\circ$) & ($\scale_{10}=500\arcmin$; $\eulerc=108^\circ$)  \\ \hline
\nstd & \nstdmexepskewsgn & \nstdmexepkurtsgn \\
\ndev & \nstatmexepskew\ maps & \nstatmexepkurt\ maps \\
\conflevel & \clmexepskew\% & \clmexepkurt\% \\ \hline
\end{tabular}
}
\subfigure[\Morlet\ $\bmath{k}=\left( 10, 0 \right)^{T}$]
{
\begin{tabular}{lcc} \hline
& Skewness & Kurtosis \\
& ($\scale_{11}=550\arcmin$; $\eulerc=72^\circ$) & ($\scale_{11}=550\arcmin$; $\eulerc=72^\circ$) \\ \hline
\nstd & \nstdmorskewsgn & \nstdmorkurtsgn \\
\ndev & \nstatmorskew\ maps & \nstatmorkurt\ maps \\
\conflevel & \clmorskew\% & \clmorkurt\% \\ \hline
\end{tabular}
}
\end{table}

\begin{figure}
\centering
\subfigure[\Mexhat\ $\eccen=0.00$]{\includegraphics[width=57mm,angle=-90]{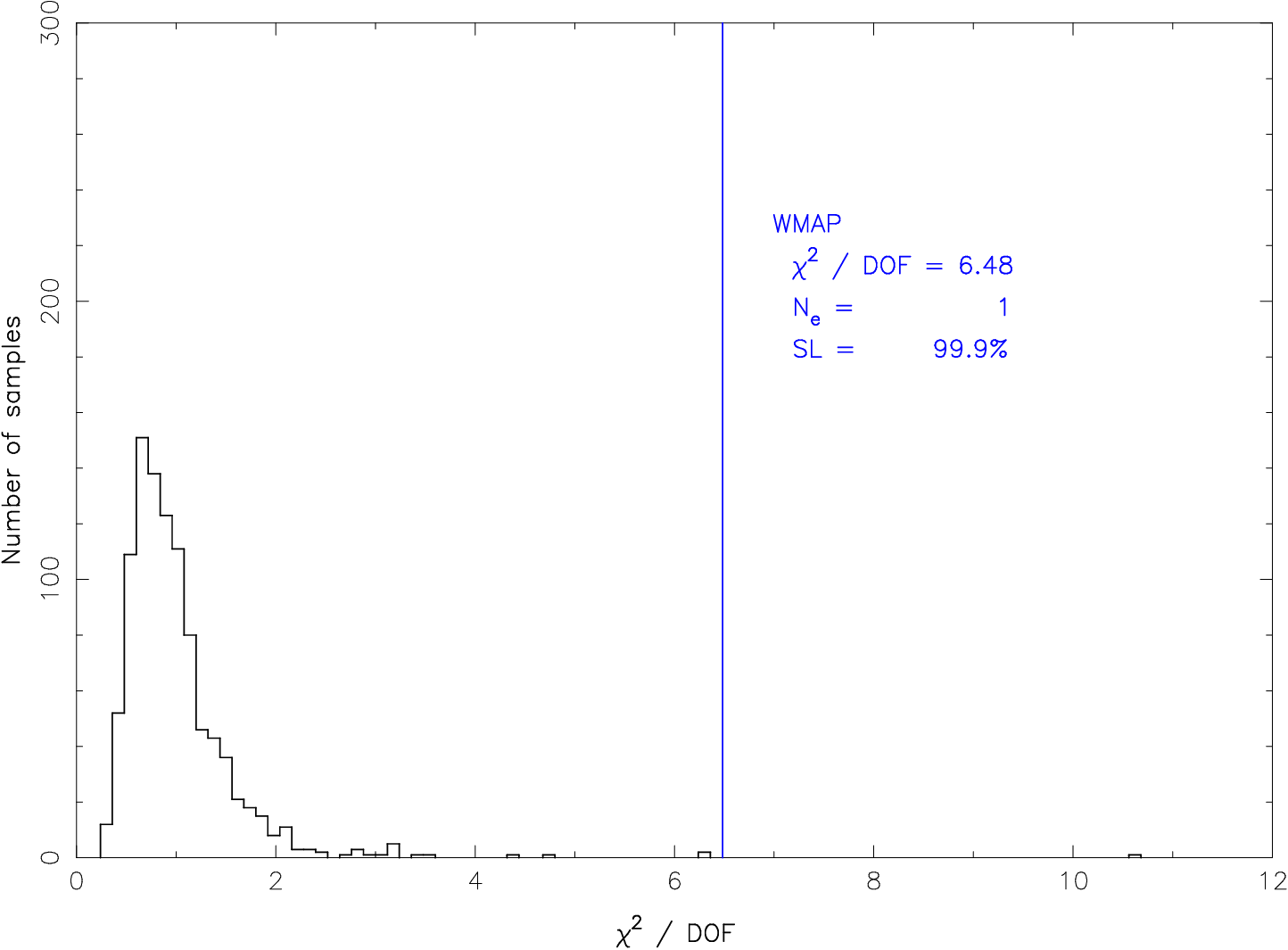}}
\subfigure[\Mexhat\ $\eccen=0.95$]{\includegraphics[width=57mm,angle=-90]{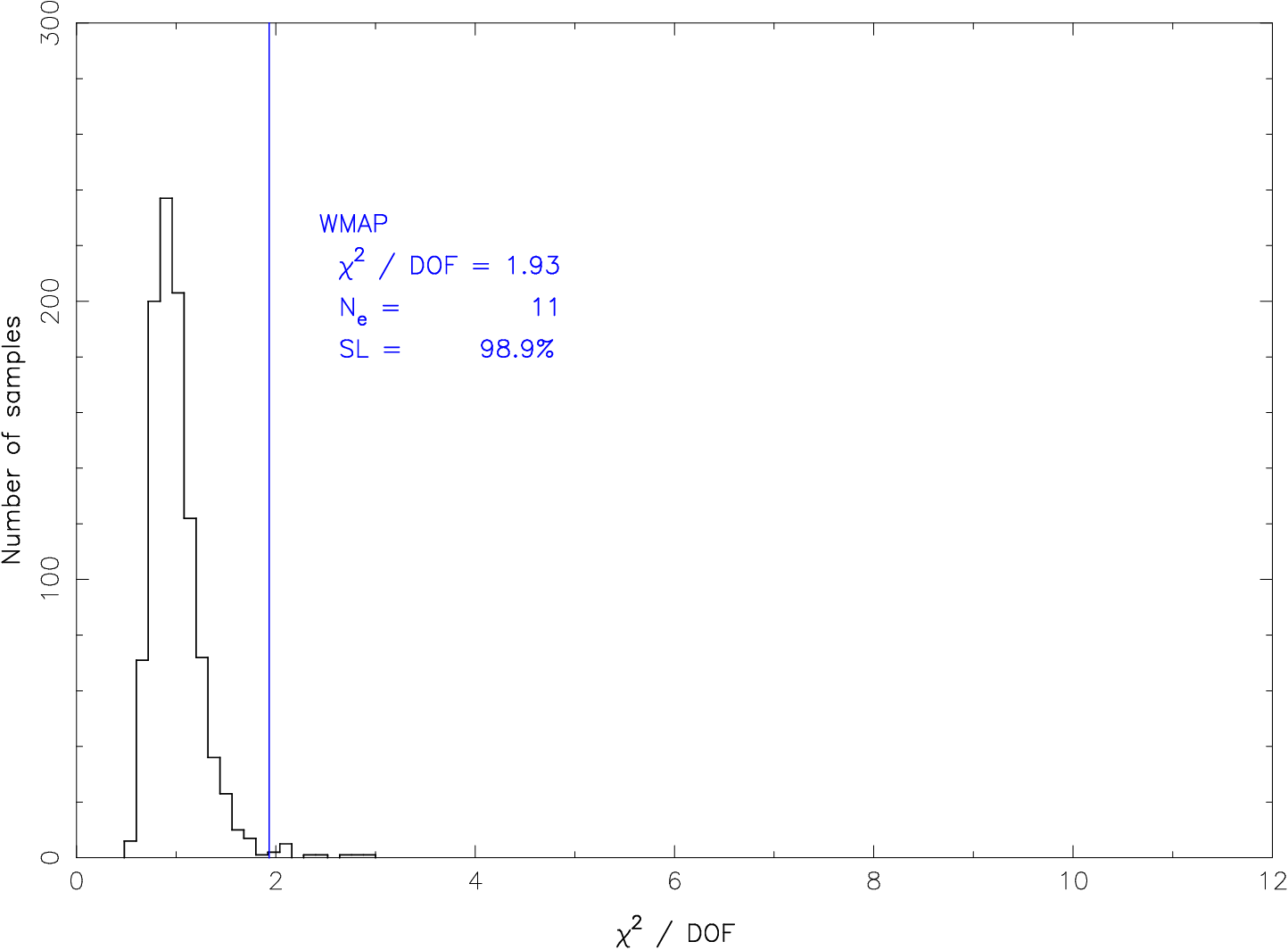}}
\subfigure[\Morlet\ $\bmath{k}=\left( 10, 0 \right)^{T}$]{\includegraphics[width=57mm,angle=-90]{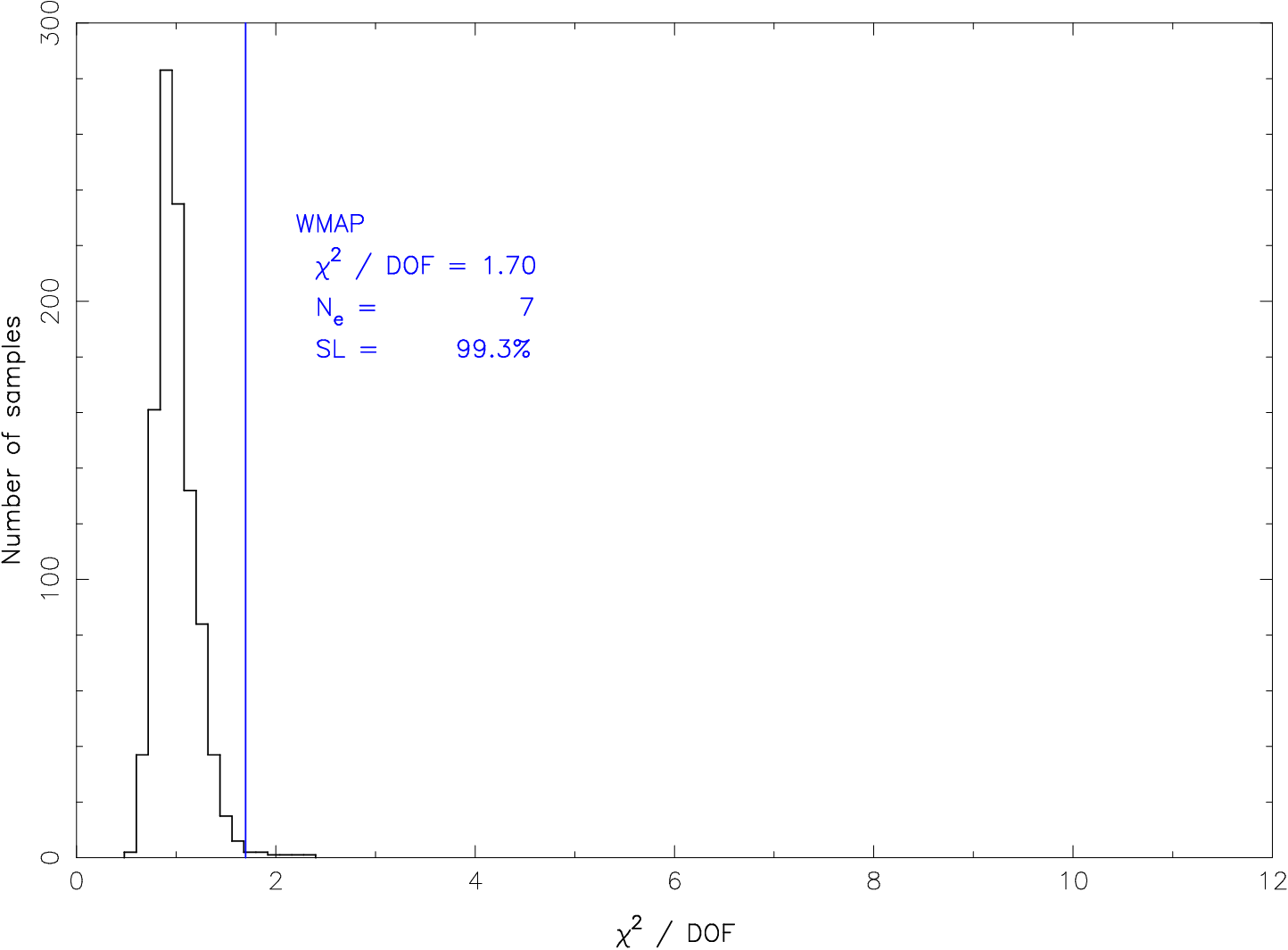}}
\caption{Histograms of normalised $\chi^2$ test
  statistics obtained from \ngsim\ Monte Carlo simulations.  The
  normalised $\chi^2$ value obtained from the \wmap\ map is indicated
  by the blue vertical line (similar results are obtained using the
  Tegmark map).  The number of simulated maps that exhibit a greater
  or equivalent $\chi^2$ value than the \wmap\ map is quoted ($N_e$),
  accompanied by the corresponding significance level (SL).}
\label{fig:chisqd}
\end{figure}

The preceding analysis is based on the marginal distributions of
individual statistics and makes a posterior selection of the \mbox{critical}
confidence limit from the most discrepant values obtained from the data.
A $\chi^2$ test provides an alternative
analysis and method of \mbox{constructing} significance measures.  This test
instead considers the set of test statistics for each wavelet as a
whole and hence is based on their joint distribution.
The posterior statistic selection problem is thus eliminated,
however including a large number of less useful test statistics has a
pronounced effect on down-weighting the overall significance of the
test.
The $\chi^2$ statistic is given by 
\begin{equation}
\chi^2 =
\sum_{i=1}^{N_{\rm stat}} \: \sum_{j=1}^{N_{\rm stat}} \:
( \tstat_i - \overline{\tstat}_i ) \:
( \cov^{-1} )_{ij} \:
( \tstat_j - \overline{\tstat}_j )
\spcend ,
\end{equation}
where $\tstat_i$ gives each test statistic.
For Gaussian distributed test statistics this should satisfy a
$\chi^2$ distribution.  Although our test statistics are not Gaussian
distributed, one may still use the $\chi^2$ test if one is willing to
estimate {signi\-fi\-cance} levels using Monte Carlo simulations.
The test statistics include both skewness and
kurtosis statistics for each scale and orientation, hence there are $24$
statistics for the symmetric \mexhat\ analysis and $120$ for
each of the directional wavelet analyses.  The mean value
for each test statistic $\overline{\tstat}$ and the covariance matrix
\cov\ is calculated from the Gaussian simulations.
$\chi^2$ values are calculated for the \wmap\ map, and also for all
simulated Gaussian realisations. 
The previously described approach for constructing significance levels
is applied to the $\chi^2$ statistics.
%The same approach to constructing significant levels as previously described,
%is now applied to the $\chi^2$ statistics.  
\Fig{\ref{fig:chisqd}} summarises the results
obtained.  All spherical wavelet analyses flag deviations from
Gaussianity of very high significance when all test statistics are
incorporated in this manner.  In particular, the detection made
using the symmetric \mexhat\ wavelet occurs at the 99.9\% significance
level and that made with the \morlet\ wavelet occurs at the 99.3\%
significance level.  In this case the superiority of the symmetric
\mexhat\ wavelet over the \morlet\ wavelet arises since the \morlet\
wavelet analysis contains a number of additional less useful
statistics (due to the additional orientations), that dilute the
overall results.

%\begin{figure*}
%\begin{minipage}{175mm}
%\centering
%\mbox{
%\subfigure[\Mexhat\ $\eccen=0.00$]{\includegraphics[angle=-90,width=\statplotwidth]{figures/nchi2_mexhat000_n1000_sTkT.ps}}
%\subfigure[\Mexhat\ $\eccen=0.95$]{\includegraphics[angle=-90,width=\statplotwidth]{figures/nchi2_mexhat095_n1000_sTkT.ps}}
%\subfigure[\Morlet\ $\bmath{k}=\left( 10, 0 \right)^{T}$]{\includegraphics[angle=-90,width=\statplotwidth]{figures/nchi2_morlet_n1000_sTkT.ps}}
%}
%\caption{Histograms of normalised $\chi^2$ test
%  statistics obtained from \ngsim\ Monte Carlo simulations.  The
%  normalised $\chi^2$ value obtained from the \wmap\ map is indicated
%  by the blue vertical line (similar results are obtained using the
%  Tegmark map).  The number of simulated maps that exhibit a greater
%  or equivalent $\chi^2$ value than the \wmap\ map is quoted ($N_e$),
%  accompanied by the corresponding significance level (SL).}
%\label{fig:chisqd}
%\end{minipage}
%\end{figure*}

Since the first analysis is based on marginal distributions, as
opposed to the joint distribution of statistics probed by the $\chi^2$
test, the former is more conservative.  Thus we quote the overall
significance of all detections of non-Gaussianity at the level
calculated by the first method (which, in all cases, is the lower of
the values calculated by the two methods).

% ---------------------------------------

\subsection{Localised deviations from Gaussianity}

Wavelet analysis inherently affords the spatial localisation of
\mbox{interesting} signal characteristics.  The most pronounced deviations from
Gaussianity in the \wmap\ 1-year data may therefore be {loca\-lised} on
the sky.   
In addition, directional wavelets also allow signal components to be
{loca\-lised} in orientation.  

The wavelet coefficients corresponding to the most {signi\-fi\-cant}
non-Gaussian detections for each wavelet are displayed in 
\fig{\ref{fig:coeff}}, accompanied by corresponding thresholded maps
to localise the most pronounced deviations from Gaussianity.
The regions displayed in \fig{\ref{fig:coeff}}~(b) that are detected
from the kurtosis \mexhat\ $\eccen=0.00$ analysis are in close
accordance with those regions found by \citet{vielva:2003}.
Additional similarities appear to \mbox{exist} between the regions detected 
from different thresholded wavelet coefficient maps, as
apparent in \fig{\ref{fig:coeff}}.  To quantify these \mbox{similarities},
the cross-correlation of all combinations of thresholded coefficient
maps is computed
(the cross-correlation is normalised to lie in the range
$[-1,1]$, where unity indicates a fully correlated map).
\Tbl{\ref{tbl:correlation}} shows the normalised cross-correlation
values obtained. 
Deviation regions shown in \fig{\ref{fig:coeff}}~(b) and
\fig{\ref{fig:coeff}}~(d), detected by the \mexhat\ $\eccen=0.00$ and 
$\eccen=0.95$ wavelets respectively, are highly correlated.
Interestingly, these coefficient maps are both flagged by excess
kurtosis measures.  Furthermore, the regions shown in 
\fig{\ref{fig:coeff}}~(a) and \fig{\ref{fig:coeff}}~(c), detected by
the  \mexhat\ $\eccen=0.00$ and $\eccen=0.95$ wavelets respectively, 
are moderately correlated.  
These coefficient maps are both flagged by excess
skewness measures. No other combinations of thresholded coefficient
maps exhibit any significant similarities.  In particular, the
deviation regions detected by the skewness \morlet\ analysis
(\fig{\ref{fig:coeff}}~(e)) do not correlate with any of the regions found
using the \mexhat\ wavelets.  This is expected since a different
wavelet that probes different structure is applied.  The
cross-correlation relationships exhibited here between detected
deviation regions for the \wmap\ map, also appear typical of simulated
Gaussian maps. 

To investigate the impact of these localised regions on the initial
non-Gaussianity detection, the corresponding coefficients
are removed from the calculation of skewness and kurtosis
test statistics.  
The non-Gaussian detections are substantially reduced for all of the
six most significant test statistics considered in 
\fig{\ref{fig:hist}}.
For the statistics considered in \fig{\ref{fig:hist}}~(c), (d) and (e)
the detection of non-Gaussianity is completely eliminated.  For the
remaining cases considered in \fig{\ref{fig:hist}}~(a), (b) and (f)
non-Gaussian detections are reduced in significance and lie between
the 95\% and 99\% confidence levels.  

Thus, the localised deviation regions identified do indeed 
appear to be the source of detected
non-Gaussianity.  Moreover, those detected regions shown in
\fig{\ref{fig:coeff}}~(a), (c) and (e) appear to \mbox{introduce} skewness
into the \wmap\ map, whereas those detected regions shown in
\fig{\ref{fig:coeff}}~(b) and (d) appear to introduce kurtosis.

\citet{cruz:2004}, in a continuation of the work of
\citet{vielva:2003}, consider localised regions in more
detail and find the large cold spot at 
\mbox{$(b=-57^\circ,\;  l=209^\circ)$} to be the main source of
non-Gaussianity detected by the symmetric \mexhat\ wavelet
analysis. A similar analysis may be performed using the 
wavelets we consider and the associated localised regions,
although we leave this for a further work.

\newlength{\coeffplotwidth}
\setlength{\coeffplotwidth}{52mm}

\begin{figure*}
\begin{minipage}{175mm}
\centering
\subfigure[\Mexhat\ $\eccen=0.00$; $\scale_2=100\arcmin$]
  {\includegraphics[width=\coeffplotwidth]{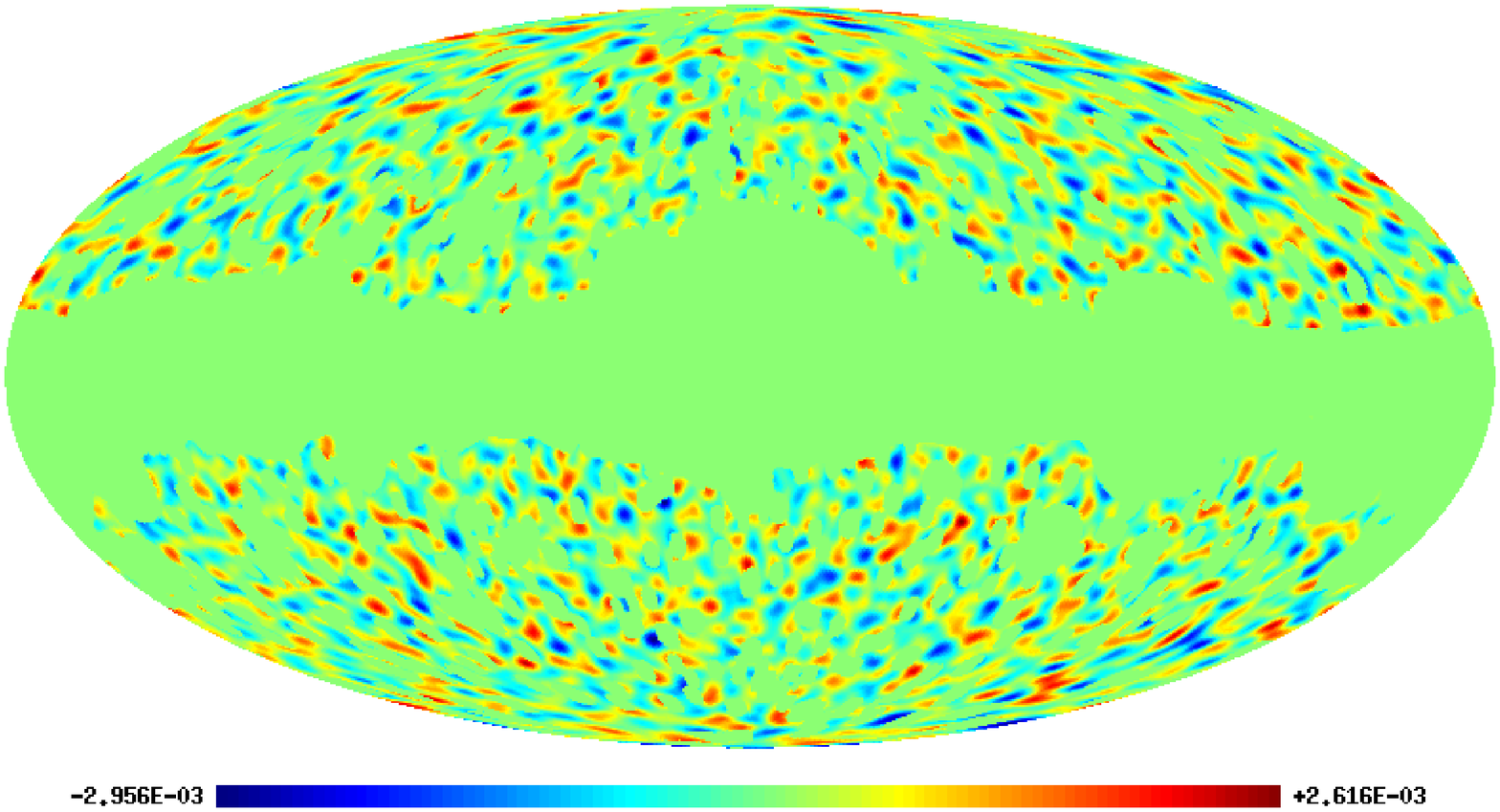} \hspace{5mm}
   \includegraphics[width=\coeffplotwidth]{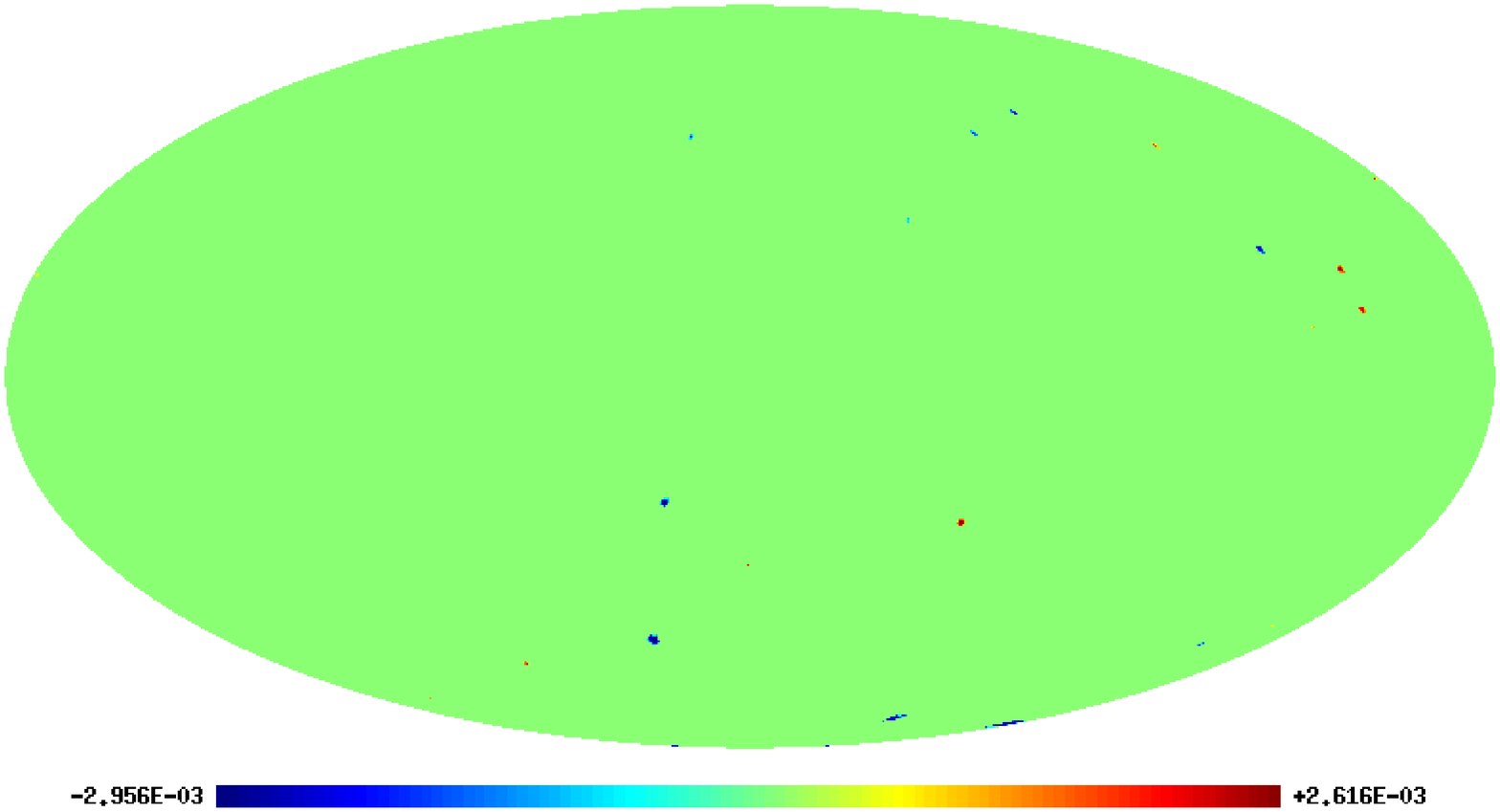} }
\subfigure[\Mexhat\ $\eccen=0.00$; $\scale_6=300\arcmin$]
  {\includegraphics[width=\coeffplotwidth]{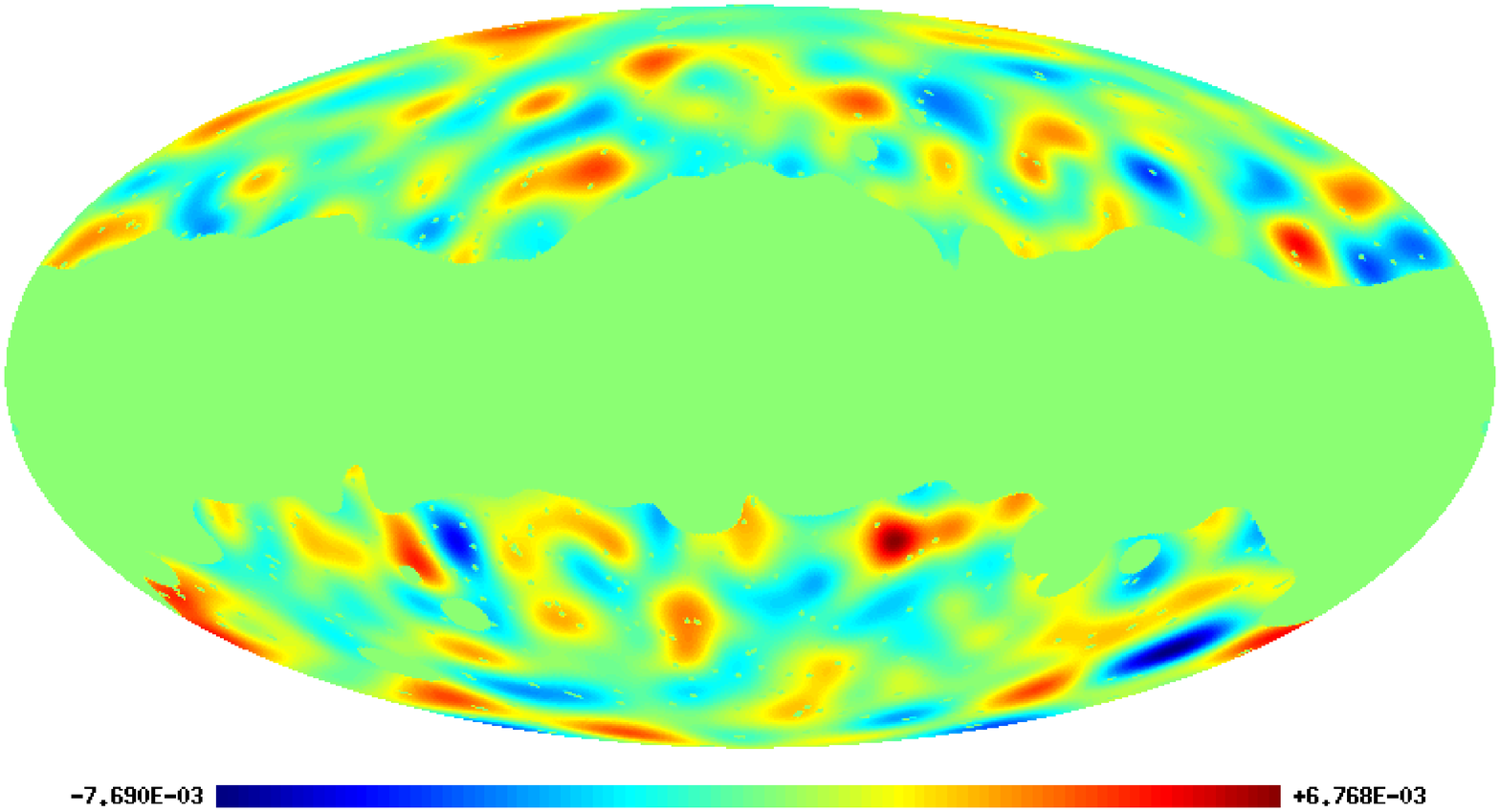} \hspace{5mm}
   \includegraphics[width=\coeffplotwidth]{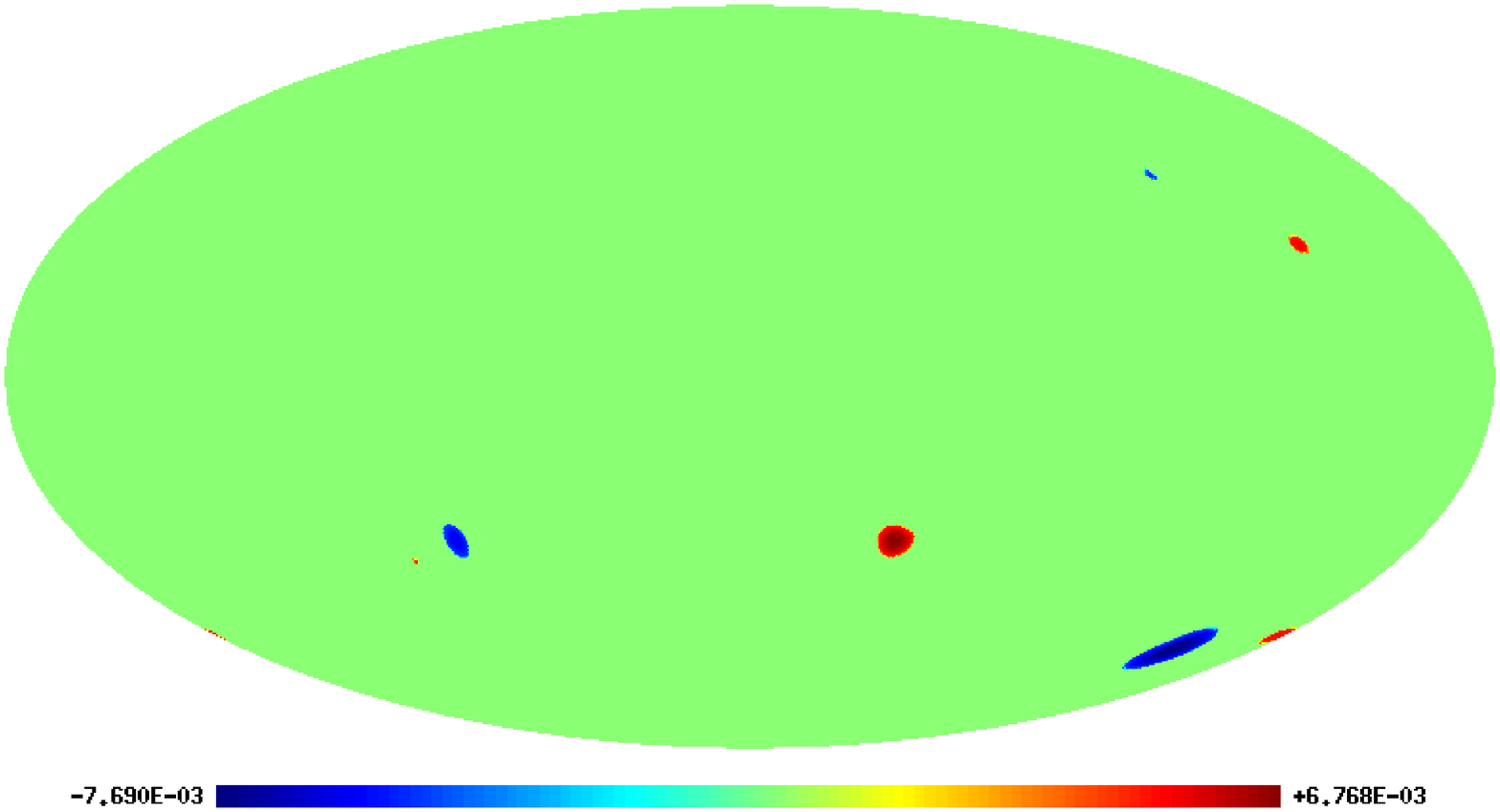} }
\subfigure[\Mexhat\ $\eccen=0.95$; $\scale_3=150\arcmin$; $\eulerc=72^\circ$]
  {\includegraphics[width=\coeffplotwidth]{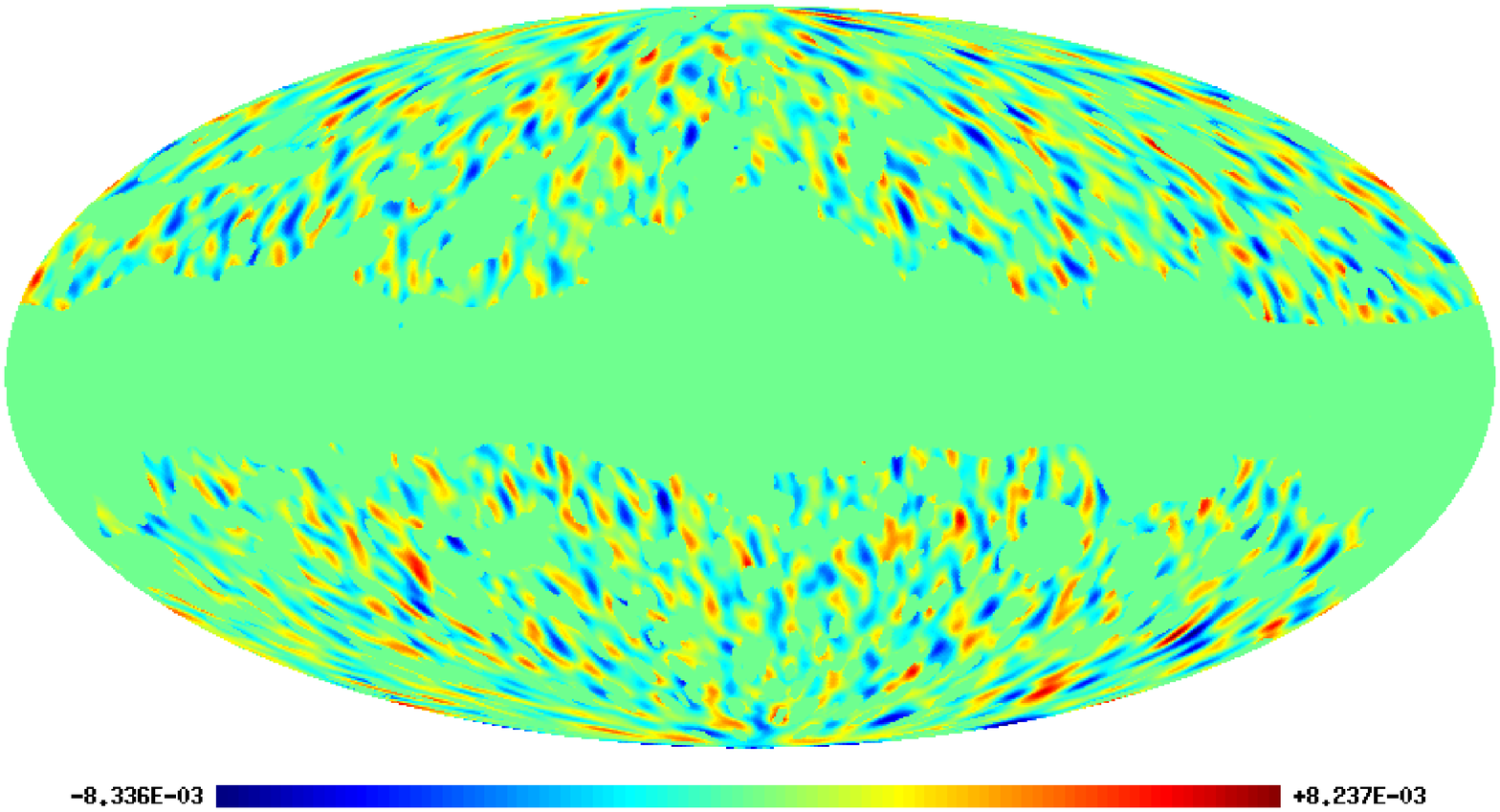} \hspace{5mm}
   \includegraphics[width=\coeffplotwidth]{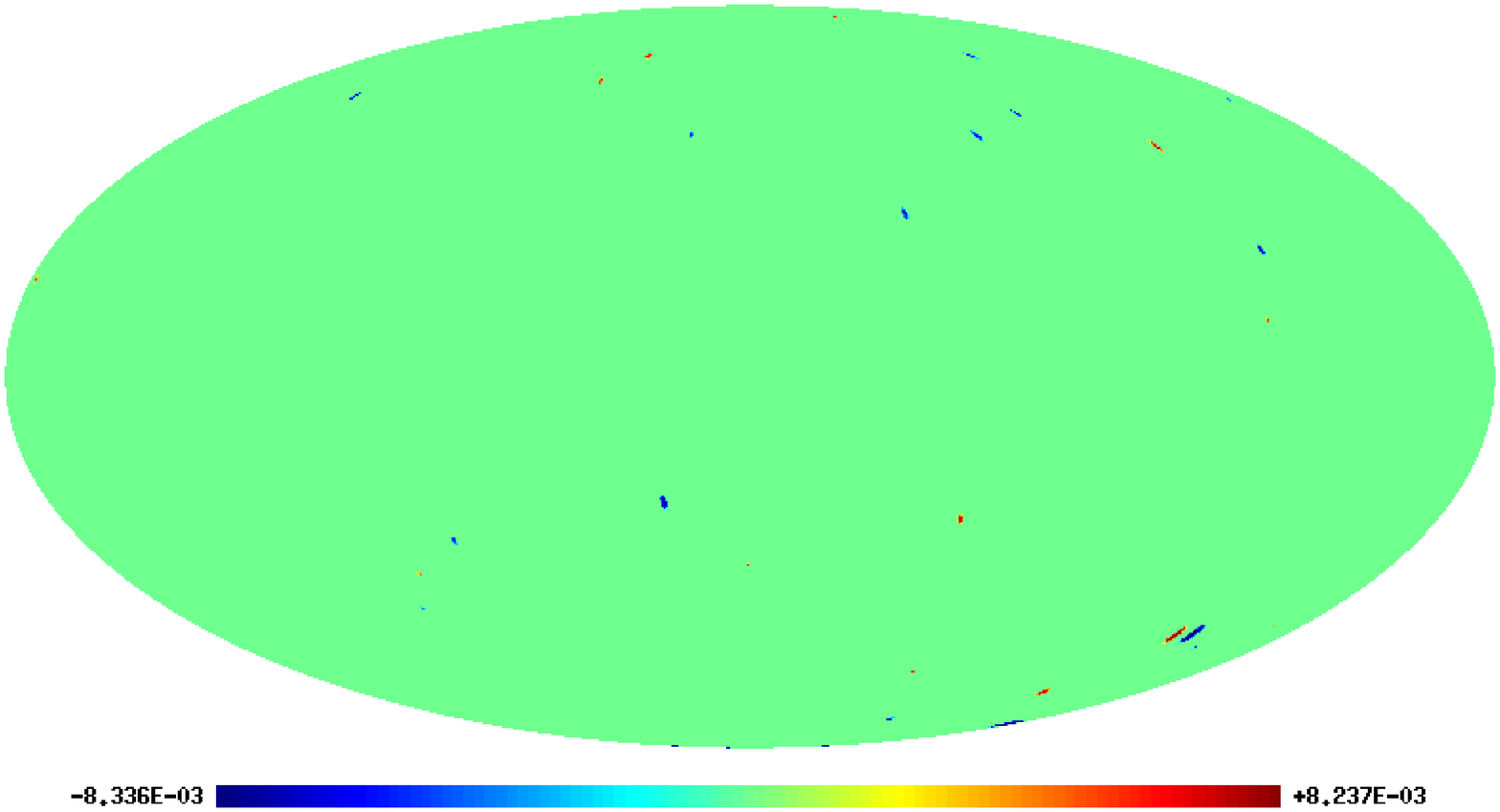} }
\subfigure[\Mexhat\ $\eccen=0.95$; $\scale_{10}=500\arcmin$; $\eulerc=108^\circ$]
  {\includegraphics[width=\coeffplotwidth]{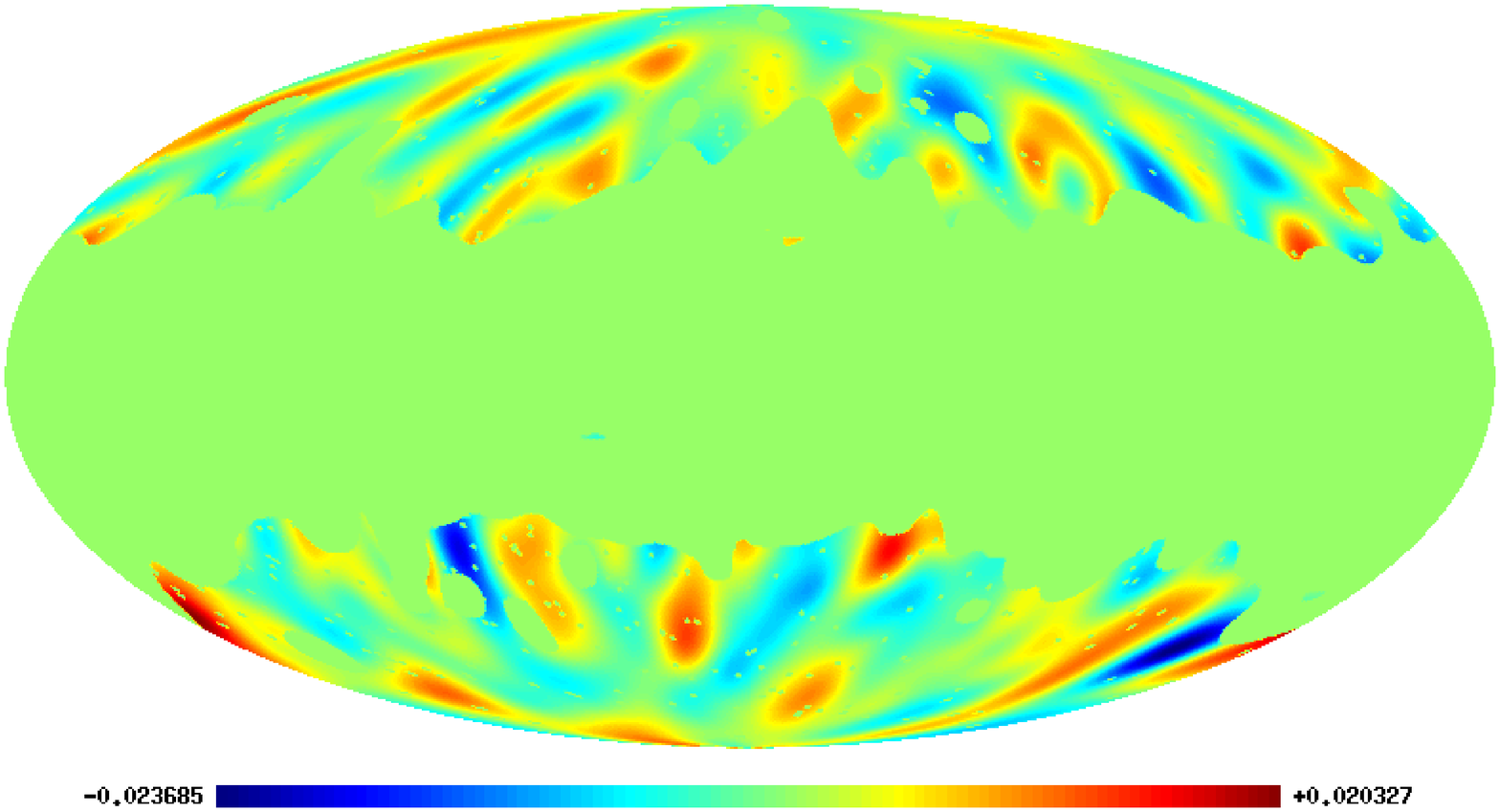} \hspace{5mm}
   \includegraphics[width=\coeffplotwidth]{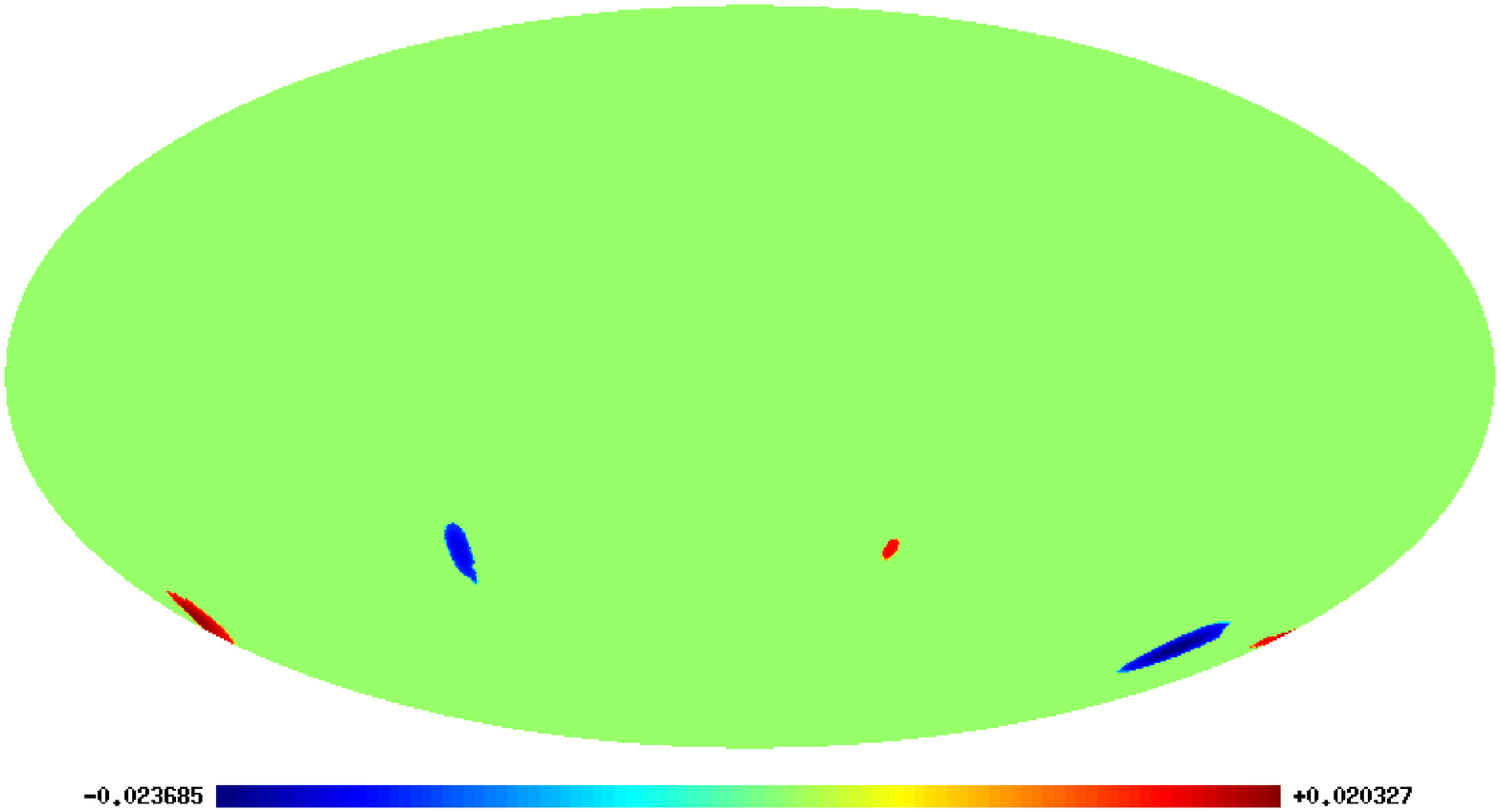} }
\subfigure[\Morlet\ $\bmath{k}=\left( 10, 0 \right)^{T}$; $\scale_{11}=550\arcmin$; $\eulerc=72^\circ$]
  {\includegraphics[width=\coeffplotwidth]{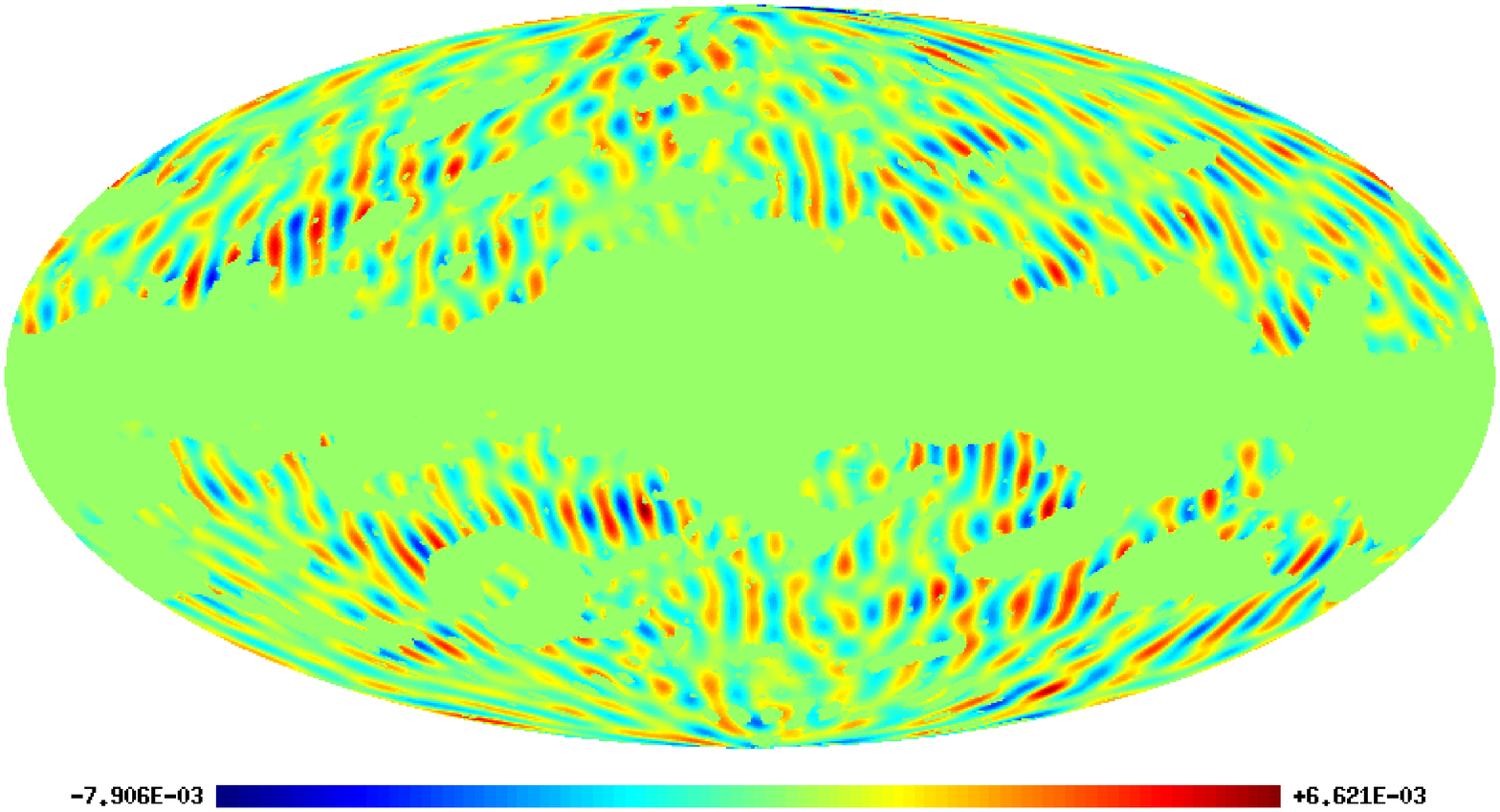} \hspace{5mm}
   \includegraphics[width=\coeffplotwidth]{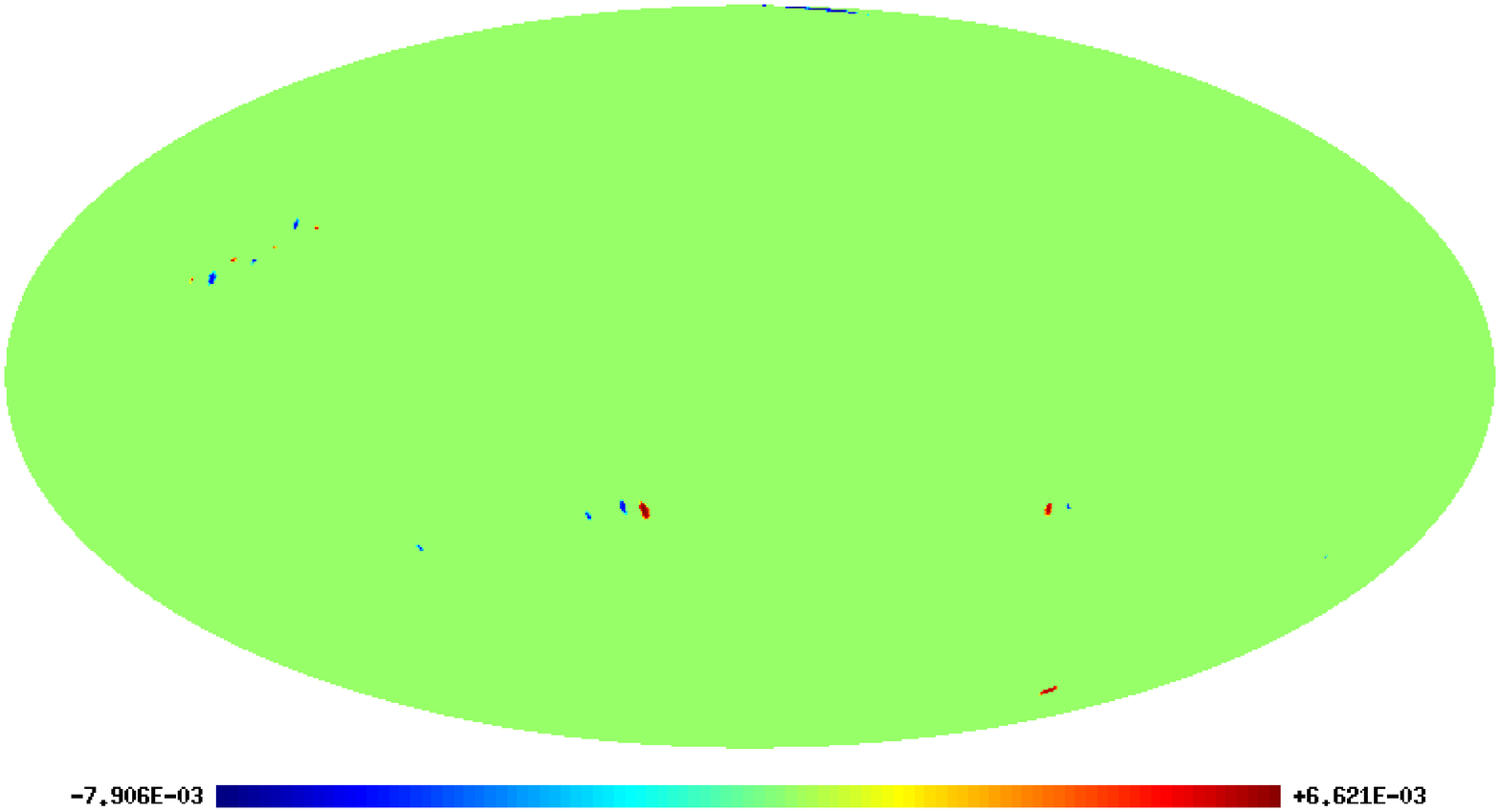} }
\caption{Spherical wavelet coefficient maps (left) and thresholded maps (right).  To
  localise most likely deviations from Gaussianity on the sky, the
  coefficient maps exhibiting strong non-Gaussianity are thresholded
  so that only those coefficients above $3\sigma$ (in absolute value)
  are shown.
  Due to the apparent similarity of the \wmap\ team and Tegmark
  maps, only coefficients for the analysis of the \wmap\ map
  are shown above.}
\label{fig:coeff}
\end{minipage}
\end{figure*}

\begin{table}
\centering
\caption{Normalised cross-correlation of thresholded spherical wavelet
  coefficient maps indicating the similarity between the localised
  most likely deviations from Gaussianity flagged by the most
  significant skewness and kurtosis observations for each wavelet.
  Notice that the regions detected from the skewness flagged maps of
  the symmetric and elliptical \mexhat\ wavelets are moderately
  correlated, while the regions detected from the kurtosis flagged
  maps of the symmetric and elliptical \mexhat\ wavelets are
  strongly correlated.  The regions flagged by the \morlet\ wavelet
  analysis are not correlated with any of the other regions detected
  by a \mexhat\ wavelet analysis, as expected since a different
  wavelet that probes different structure is applied.
  (Note that the lettered key corresponds to the thresholded coefficient
  maps contained in the panels of \fig{\ref{fig:coeff}}.)
}
\label{tbl:correlation}
\begin{tabular}{llccccc} \hline
\multicolumn{2}{l}{Thresholded coefficient map}%
  & (a)    & (b)    & (c)    & (d)    & (e)    \\ \hline
(a) & \Mexhat\ $\eccen=0.00$  %
    & 1.00 & 0.00 & 0.46 & 0.00 & 0.00 \\
  & $\scale_2=100\arcmin$ \\
(b) & \Mexhat\ $\eccen=0.00$ %
    & -    & 1.00 & 0.04 & 0.70 & 0.00 \\
  & $\scale_6=300\arcmin$ \\
(c) & \Mexhat\ $\eccen=0.95$ %
    & -    & -    & 1.00 & 0.05 & 0.01 \\
  & $\scale_3=150\arcmin$; $\eulerc=72^\circ$ \\
(d) & \Mexhat\ $\eccen=0.95$ %
    & -    & -    & -    & 1.00 & 0.00 \\
  & $\scale_{10}=500\arcmin$; $\eulerc=108^\circ$ \\
(e) & \Morlet\ $\bmath{k} = \left(10, 0 \right)^{T}$ %
    & -    & -    & -    & -    & 1.00 \\
  & $\scale_{11}=550\arcmin$; $\eulerc=72^\circ$ \\ \hline
\end{tabular}
\end{table}

% ---------------------------------------

\subsection{Preliminary noise analysis}

Naturally, 
one may wish to consider
%the next step is to consider 
possible sources of the non-Gaussianity  detected. 
We briefly consider the deviation regions detected to see if they
correspond to regions on the sky that have higher noise dispersion
than typical.  We leave the analysis of residual foregrounds or
further systematics, or whether the features detected  
do indeed exist in the \cmb, for a further work.

\begin{figure}
\centerline{\includegraphics[width=60mm,angle=-90]{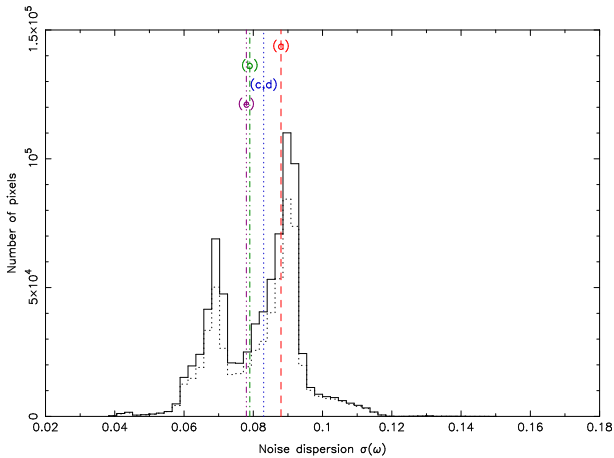}}
\caption{\wmap\ sky noise dispersion histogram, with the mean
  noise dispersion level obtained in detected deviation regions also
  shown.  The solid histogram shown corresponds to the full sky
  noise dispersion map, whereas the dotted histogram corresponds to
  the \kpzero\ masked noise dispersion map.  The dashed vertical lines
  indicate the mean noise dispersion level in the detected deviation
  regions illustrated in \fig{\ref{fig:coeff}}.  None of the detected
  deviation regions correspond to noise of higher dispersion than typical.
  The lettered key corresponds to the thresholded coefficient
  maps contained in the panels of \fig{\ref{fig:coeff}}: 
  (a) \Mexhat\ $\eccen=0.00$, $\scale_2=100\arcmin$; 
  (b) \Mexhat\ $\eccen=0.00$, $\scale_6=300\arcmin$;
  (c) \Mexhat\ $\eccen=0.95$, $\scale_3=150\arcmin$, $\eulerc=72^\circ$;
  \mbox{(d) \Mexhat} $\eccen=0.95$, $\scale_{10}=500\arcmin$, $\eulerc=108^\circ$;
  (e) \Morlet\ $\bmath{k} = \left(10, 0 \right)^{T}$, $\scale_{11}=550\arcmin$, $\eulerc=72^\circ$.}
\label{fig:noise_hist}
\end{figure}

The noise dispersion map for each \wmap\ band is combined, 
\\ 
according to 
\begin{equation}
\sigma(\sa) = \sqrt{ \frac{\sum_{r=3}^{10}
\: w_r{}^2(\sa) \: \sigma_r{}^2(\sa) }
{ \left [ \: \sum_{r=3}^{10} w_r(\sa) \: \right ]^2}
} \:
\spcend ,
\end{equation}
to produce a single noise dispersion sky map for the \wmap\ map, and
equivalently for the Gaussian realised maps. A histogram of this map,
and a \kpzero\ masked version of the map, is illustrated in
\fig{\ref{fig:noise_hist}} to investigate the noise dispersion
distribution for the \wmap\ observing strategy.
Also plotted is the mean noise dispersion level in the detected
deviation regions for each thresholded coefficient map of
\fig{\ref{fig:coeff}}.  All mean noise dispersion levels of deviation
regions lie within the central region of the full noise distribution.
Furthermore, full noise dispersion histograms for
the deviation regions were also produced to ensure outliers did not
exist.  No outliers were observed in any deviation regions.
These additional five plots are not shown
to avoid clutter and since no pertinent additional findings may be
drawn from them.
It is therefore apparent that the deviation regions
detected do not correspond to regions with greater noise dispersion
than typical.  

% -----------------------------------------------------------------------------

\section{Conclusions}
\label{sec:conclusions}

A directional spherical wavelet analysis, facilitated by our fast
\cswt, has been applied to the \wmap\ 1-year data to probe for
deviations from Gaussianity. 
Directional spherical wavelets allow one to probe orientated structure 
inherent in the data.  Non-Gaussianity has been
detected by a number of test statistics for a range of wavelets.

We have reproduced the results obtained by \citet{vielva:2003} using
the symmetric \mexhat\ $\eccen=0.00$ wavelet,
thereby confirming their findings, whilst also providing a consistency
check for our analysis.  
Deviations in the skewness and kurtosis of wavelet
coefficients on scale $\scale_2=100\arcmin$ and $\scale_6=300\arcmin$
were detected, although using our more conservative test we make these
detections at the \clmexskew\% and \clmexkurt\% overall significance levels
respectively (lower than the 99.9\% significance level quoted by
\citet{vielva:2003} for their kurtosis detection).

Similar detections of non-Gaussianity were made using the elliptical
\mexhat\ $\eccen=0.95$ wavelet, although on slightly larger scales.
In particular, a deviation from Gaussianity was detected in the
skewness of the \mexhat\ $\eccen=0.95$ wavelet coefficients on scale
$\scale_3=150\arcmin$ and orientation $\eulerc=72^\circ$ at the
\clmexepskew\% significance level.
Although a detection was observed in the kurtosis outside of the 99\%
confidence region on scale $\scale_{10}=500\arcmin$ and orientation
$\eulerc=108^\circ$, the full statistical analysis of Monte Carlo
simulations gave a significance of only \clmexepkurt\% for this
detection. 

The most interesting result, however, is the deviation from
Gaussianity observed in the \morlet\ wavelet skewness {mea\-sure\-ment} on scale
$\scale_{11}=550\arcmin$ and orientation $\eulerc=72^\circ$.  This
wavelet scale corresponds to an effective size on the sky of
$\sim26^\circ$ ($\sim3^\circ$ for the internal structure of the
\morlet\ wavelet), or equivalently a spherical harmonic
scale of $\lsh\sim7$ ($\lsh\sim63$). The detection
deviates from the mean of \ngsim\ Gaussian Monte Carlo
simulations by \nstdmorskew\ standard
deviations for the \wmap\ map and by \nstdmorskewteg\ standard
deviations for the Tegmark map.  Only \nstatmorskew\ of \ngsim\ 
Gaussian simulated maps
exhibited a deviation this large in \emph{any} \morlet\ test statistic,
hence the detection is conservatively made at \clmorskew\%
significance.  

Significance levels were also calculated from $\chi^2$ tests for each
spherical wavelet.  This approach avoids the posterior selection
of particular statistics, but rather considers the set of test
statistics in aggregate.  By considering the joint distribution of
test statistics in this manner the analysis results may be diluted by
including a large number of less powerful test statistics.
Deviations from
Gaussianity at significance levels of 99.9\% and 99.3\% were found
using the symmetric \mexhat\ and \morlet\ wavelet respectively.
In this case the directional \morlet\ analysis is more severely
affected by a larger number of less useful test statistics,
nevertheless both deviations from Gaussianity are made at very high
significance.  We quote the overall significance of our findings,
however, at the lower significance levels found using the previous
most conservative test.

Deviations from Gaussianity corresponding to the most significant
detections for each wavelet were localised on the sky. By
removing the coefficients corresponding to these regions from the
initial analysis, all significant non-Gaussianity detections were
\mbox{eliminated.} 
These localised regions therefore appear to be the source of detected
non-Gaussianity.  Moreover, those regions that introduce skewness in
the \wmap\ map may be localised, as may those regions that introduce
kurtosis.  Preliminary noise analysis indicates that these detected
deviation regions do \emph{not} correspond to regions that have higher
noise dispersion than typical.
Further analysis is required, however, to ascertain whether
these regions correspond to the localised introduction of secondary
non-Gaussianity or systematics, or whether in fact the non-Gaussianity
detected in the \wmap\ 1-year data is due to intrinsic primordial
fluctuations in the \cmb. 

An interesting first step in deducing
whether the non-Gaussian signal discovered is of cosmological
origin would be to repeat our analysis on the 4-year \cobe\
data. Although it has been shown that these data contain some
systematic effects that lead to non-Gaussianity \citep{mm:2004}, it is
likely that these systematics are not shared by \wmap.
Provided the most significant detection in the \wmap\ data is
predominantly due to the global structure of the \morlet\ wavelet at
an angular scale of $\sim26^\circ$, the angular resolution of the
\cobe\ data should be sufficient to observe it if it is
astrophysical in origin. Clearly, it will also be of great interest to
investigate whether the non-Gaussianity detections reported here are
still present in the 2-year \wmap\ data.

% -----------------------------------------------------------------------------

\section*{Acknowledgements}

JDM would like to thank the Association of Commonwealth
Universities and the Cambridge Commonwealth Trust for the 
support of a Commonwealth (Cambridge) Scholarship.
DJM is supported by PPARC.
Some of the results in this paper have been derived using the
\healpix\ package \citep{gorski:1999}.
We acknowledge the use of the Legacy Archive for Microwave Background
Data Analysis (\lambdaarch).  Support for \lambdaarch\ is provided by
the NASA Office of Space Science.

% -----------------------------------------------------------------------------

% -----------------------------------------------------------------------------

\appendix
\section{A fast directional \cswt}
\label{sec:fast_cswt}

The \cswt\ at a particular scale is essentially a spherical
{con\-volu\-tion}; hence it is possible to apply the fast spherical
convolution algorithm proposed by \citet{wandelt:2001} to evaluate
the wavelet transform.  The harmonic representation of the
\cswt\ is first presented, followed by a discretisation and fast
implementation.  Without loss of generality we consider a single
dilation only.

% ---------------------------------------

\subsection{Harmonic formulation}
\label{sec:harmonic}

There does not exist any finite point set on the sphere that is
invariant under rotations (due to geometrical properties of the
sphere), hence it is more natural, and in fact more accurate for
numerical purposes, to recast the \cswt\ defined by
\eqn{\ref{eqn:cswt}} in harmonic space. 

Both the wavelet and signal are represented in terms of a spherical
harmonic expansion, defined for an arbitrary function $f \in
L^2(\sphere)$ by
\begin{equation}
f(\sa) = \sum_{\lsh=0}^{\infty} \sum_{m=-\lsh}^{\lsh}
\shcoeff{f}_{\lsh m} \sh_{\lsh m}(\sa)
\spcend ,
\label{eqn:sh_expansion}
\end{equation}
where the spherical harmonic coefficients are given by the usual
projection of the signal onto each spherical harmonic basis function
$\sh_{\lsh m}(\sa)$,
\begin{equation}
\shcoeff{f}_{\lsh m} 
= \int_{\sphere}
f(\sa) \: {\sh_{\lsh m}}^\conj(\sa) \: \dx \mu(\sa)
\spcend .
\end{equation}
In practice one requires that at least one of the functions, usually the
wavelet, has a finite band limit so that negligible power is present in
those coefficients above a certain $\lsh _{\rm max}$.  For all practical
purposes, the outer summation of \eqn{\ref{eqn:sh_expansion}} may then
be truncated to $\lsh _{\rm max}$.

Substituting the spherical harmonic expansions of the wavelet and signal
into the \cswt\ of \eqn{\ref{eqn:cswt}} and noting the orthogonality of
the spherical harmonics, yields the harmonic representation 
\begin{equation}
\skywav(\eulera, \eulerb, \eulerc) =
\sum_{\lsh =0}^{\lsh _{\rm max}} \:
\sum_{m=-\lsh }^{\lsh } \:
\sum_{m\p=-\lsh }^{\lsh }
\left [
{\dmatbig_{mm\p}^{\lsh }(\eulers) \,
\shcoeff{\wav}_{\lsh m\p}} \right ]^\conj \:
\shcoeff{\sky}_{\lsh m}
 .
\label{eqn:harmonic:1}
\end{equation}
The additional summation and $\dmatbig_{mm\p}^\lsh$ Wigner rotation
matrices that are introduced
characterise the rotation of a spherical harmonic, noting that a rotated
spherical harmonic may simply be represented by a sum of rotated 
harmonics of the same $\lsh$ by \citep{inui:1996}
\begin{equation}
(\rot_{\eulers}\sh_{\lsh m})(\sa) = 
\sum_{m\p=-\lsh}^{l} 
\dmatbig_{mm\p}^{\lsh}(\eulers) \: \sh_{\lsh m\p}(\sa)
\spcend .
\end{equation}
The Wigner rotation matrices may be decomposed as
\begin{equation}
\dmatbig_{mm\p}^{\lsh}(\eulers)
= e^{-\img m\eulera} \:
\dmatsmall_{mm\p}^\lsh(\eulerb) \:
e^{-\img m\p\eulerc}
\spcend ,
\label{eqn:d_decomp}
\end{equation}
where the real polar \dmatsmall-matrix is defined by, for example,
\cite{brink:1993}.
The relationship shown in \eqn{\ref{eqn:d_decomp}} is exploited by factoring
the rotation
$\rot_{\eulers}$ into two separate rotations, both of which only contain
a constant $\pm \pi/2$ polar rotation:
\begin{equation}
\rot_{\eulers}
= \rot_{\eulera-\pi/2, \; -\pi/2, \; \eulerb} \:\:
\rot_{0, \; \pi/2, \; \eulerc+\pi/2}
\spcend .
\label{eqn:rot_factor}
\end{equation}
By factoring the rotation in this manner and applying the decomposition
described by \eqn{\ref{eqn:d_decomp}}, \eqn{\ref{eqn:harmonic:1}} 
can be written as
\begin{eqnarray}
\lefteqn{
\skywav(\eulers) =
\sum_{\lsh=0}^{\lsh_{\rm max}} \:
\sum_{m=-\lsh}^{\lsh} \:
\sum_{m\p=-\lsh}^{\lsh} \:
\sum_{m\pp=-\min(m_{\rm max},\lsh)}^{\min(m_{\rm max},\lsh)}
\dmatsmall_{m\p m}^\lsh(\pi/2) \:
\dmatsmall_{m\p m\pp}^\lsh(\pi/2)
} \hspace{10mm}\nonumber \\
& & 
\times \:\:
\shcoeff{\wav}_{\lsh m\pp}{}^\conj \:
\shcoeff{\sky}_{\lsh m} \:
e^{\img [m(\eulera-\pi/2) + m\p\eulerb + 
m\pp(\eulerc+\pi/2) ]}
\spcend ,
\label{eqn:harmonic:2}
\end{eqnarray}
where the symmetry relationship
\mbox{$\dmatsmall_{m m\p}^{\lsh}(-\eulerb)=\dmatsmall_{m\p m}^{\lsh}(\eulerb)$}
has been applied.  In many cases it is likely that the wavelet will have
minimal azimuthal structure compared to the signal under analysis, in
which case it may also have a lower effective azimuthal band limit
$m_{\rm max} \ll \lsh_{\rm max}$.

The harmonic formulation presented replaces the continuous integral
of \eqn{\ref{eqn:cswt}} by finite summations, although evaluating these
summations directly would be no more efficient that approximating the
initial integral using simple quadrature.  Rotations are
elegantly represented in harmonic space, however, and the
approximation and interpolation required in any real space
discretisation is avoided.  Moreover, \eqn{\ref{eqn:harmonic:2}}
is represented in such a way that the presence of complex exponentials
may be exploited such that fast Fourier transforms (\fft s) may be
applied to evaluate rapidly the three summations simultaneously.

% ---------------------------------------

\subsection{Fast implementation}

\newlength{\eqnindent}
\setlength{\eqnindent}{7mm}

Azimuthal rotations may be applied with far less computational expense
than polar rotations since they appear within complex exponentials
in \eqn{\ref{eqn:harmonic:2}}.  Although the \dmatsmall-matrices can be
evaluated reasonably quickly and reliably using recursion formulae
(\eg\ those given by \citealt{risbo:1996}), the
basis for the fast implementation is to avoid these polar rotations as
much as possible and use \fft s to evaluate rapidly all of the azimuthal
rotations simultaneously. 
This is the motivation for factoring the rotation by
\eqn{\ref{eqn:rot_factor}} so that all Euler angles occur as azimuthal
rotations.

The discretisation of each Euler angle may in general be {arbi\-trary}.
However, to exploit standard \fft\ routines uniform sampling is adopted.
The uniformly sampled spherical wavelet coefficient samples are
defined by%
\footnote{Whilst \eulera\ and \eulerc\ both cover the range $0$ to
$2\pi$, evaluating \eulerb\ over the same range is redundant, covering
the $SO(3)$ manifold exactly twice.  Nonetheless, the use of the fast
\fft-based algorithm requires this range.}
\begin{equation}
\skywav_{\ind_\eulera, \ind_\eulerb, \ind_\eulerc} =
\skywav
\left (
\frac{2\pi \ind_\eulera}{\num_\eulera},
\frac{2\pi \ind_\eulerb}{\num_\eulerb},
\frac{2\pi \ind_\eulerc}{\num_\eulerc}
\right )
\spcend .
\end{equation}
Discretising \eqn{\ref{eqn:harmonic:2}} in this manner and performing
a little algebra we may recast it in a form amenable to fast Fourier
techniques:
\begin{eqnarray}
\lefteqn{
\skywav_{\ind_\eulera, \ind_\eulerb, \ind_\eulerc} =
e^{-\img 2\pi ( \ind_\eulera \lsh_{\rm max} / \num_\eulera +
\ind_\eulerb \lsh_{\rm max} / \num_\eulerb +
\ind_\eulerc m_{\rm max} / \num_\eulerc)}  
} \hspace{\eqnindent} \nonumber \\
& & \times \:
\sum_{j=0}^{\num_\eulera-1} \:
\sum_{j\p=0}^{\num_\eulerb-1} \:
\sum_{j\pp=0}^{\num_\eulerc-1}
\cswtfftterm_{j, j\p, j\pp} \:
e^{ \img 2\pi (j\ind_\eulera/\num_\eulera 
+ j\p \ind_\eulerb/\num_\eulerb
+ j\pp \ind_\eulerc/\num_\eulerc)}
\spcend ,
\label{eqn:cswt_fast}
\end{eqnarray}
where the second line is simply the unnormalised \dthree\ inverse
discrete Fourier transform (\dft) of 
\begin{eqnarray}
\lefteqn{\cswtfftterm_{m+\lsh_{\rm max}, m\p+\lsh_{\rm max}, m\pp+m_{\rm max}} = 
e^{i(m\pp-m)\pi/2}} \hspace{\eqnindent} \nonumber \\
& & \times \:
\sum_{\lsh=\max(\mid m \mid, \mid m\p \mid, \mid m\pp \mid)}^{\lsh_{\rm max}}
\dmatsmall_{m\p m}^\lsh(\pi/2) \:
\dmatsmall_{m\p m\pp}^\lsh(\pi/2) \:
\shcoeff{\wav}_{\lsh m\pp}{}^\conj \:
\shcoeff{\sky}_{\lsh m}
\spcend  ,
\label{eqn:cswt_fast_term}
\end{eqnarray}
where the shifted indices show the conversion between the harmonic and
Fourier conventions.  The number of samples for each Euler angle is
\mbox{$\num_\eulera = 2 \, \lsh_{\rm max}+1$},
\mbox{$\num_\eulerb = 2 \, \lsh_{\rm max}+1$} and 
\mbox{$\num_\eulerc = 2 \, m_{\rm max}+1$}, enforced by uniform
sampling and the standard Fourier relationship.

The \cswt\ may be performed very rapidly in spherical harmonic
space by constructing the \cswtfftterm-matrix of
\eqn{\ref{eqn:cswt_fast_term}} from spherical harmonic coefficients and
precomputed \mbox{\dmatsmall-matrices},
followed by the application of a \fft\ to 
evaluate rapidly all three Euler angles of the discretised \cswt\
simultaneously, before applying a final modulating complex exponential
factor.  Memory and computational requirements may be reduced by a
further factor of two for real signals by exploiting the conjugate
symmetry relationship
$\cswtfftterm_{-m,-m\p,-m\pp}=\cswtfftterm_{m,m\p,m\pp}^\conj$.

The computational cost of the fast \cswt\ is dominated by the
calculation of the \cswtfftterm-matrix, which is of order 
$\complexity(\lsh_{\rm max}^{\: 3} m_{\rm max})$.
A direct quadrature approximation of the \cswt\ integral is of order 
$\complexity( \num_\eulera{}^2  \num_\eulerb{}^2 \num_\eulerc)$.
The harmonic and real space size parameters are of the same order,
that is \mbox{$\complexity(\lsh_{\rm max}) \sim \complexity(\num_\eulera) \sim
\complexity(\num_\eulerb)$} and 
\mbox{ $\complexity(m_{\rm max}) \sim \complexity(\num_\eulerc)$},
hence the fast algorithm provides a saving of 
$\complexity(\lsh_{\rm max})$.  
We give a more detailed comparison of the complexity of various \cswt\
implementations and typical execution times in \citet{mcewen:2004}.

% -----------------------------------------------------------------------------

\label{lastpage}


\begin{thebibliography}{}
  
  \bibitem[\protect\citeauthoryear{Antoine \& Vandergheynst}{1998}]{antoine:1998}
    Antoine J. -P. and Vandergheynst P., 1998,
    Journal of Mathematical Physics, 39, 8, 3987--4008
 
  \bibitem[\protect\citeauthoryear{Antoine J.-P., Demanet L., Jacques
    L.}{Antoine \etal}{2002}]{antoine:2002} Antoine J. -P., Demanet
    L., Jacques L., 2002, Applied Computational Harmonic
    Analysis, 13, 3, 177--200

  \bibitem[\protect\citeauthoryear{Barreiro \& Hobson}{2001}]{bh:2001}
    Barreiro R. B., Hobson M. P., 2001, \mnras, 327, 813

  \bibitem[\protect\citeauthoryear{Barreiro R. B., Hobson
    M. P., Lasenby A. N., Banday A. J., G\'{o}rski K. M. and Hinshaw
    G.}{Barreiro \etal}{2000}]{barreiro:2000} Barreiro R. B., Hobson
    M. P., Lasenby A. N., \mbox{Banday A. J.}, G\'{o}rski K. M. and
    Hinshaw G., 2000, \mnras, 318, 475

  \bibitem[\protect\citeauthoryear{Bennett \etal}{2003}]{bennett:2003}
    Bennett C. L. \etal, 2003,
    ApJS, 148, 97

  \bibitem[\protect\citeauthoryear{Brink \& Satchler}{1993}]{brink:1993}
    Brink D. M., Satchler G. R., 1993,
    Angular Momentum 3rd Ed., Clarendon Press, Oxford

  \bibitem[\protect\citeauthoryear{Cabella \etal}{2004}]{cabella:2004}
   Cabella P., Liguori M., Hansen F. K., Marinucci D., Matarrese S.,
   Moscardini L., Vittorio N., 2004,
   MNRAS, submitted (astro-ph/0406026)

  \bibitem[\protect\citeauthoryear{Cay\'{o}n , Sanz,
    Martinez-Gonz\'{a}lez, Banday, Arg\"{u}eso, Gallegos, G\'{o}rski
    \& Hinshaw}{Cay\'{o}n \etal}{2001}]{cayon:2001} Cay\'{o}n L., Sanz
    J. L., Martinez-Gonz\'{a}lez E., Banday A. J., Arg\"{u}eso F.,
    Gallegos J. E., G\'{o}rski K. M., Hinshaw G., 2001,
    \mnras, 326, 1243

  \bibitem[\protect\citeauthoryear{Chiang \etal}{2003}]{chiang:2003}
   Chiang L. -Y., Naselsky P. D., Verkhodanov O. V., 2003, ApJ, 590, 65

  \bibitem[\protect\citeauthoryear{Colley \& Gott}{2003}]{cg:2003}
    Colley W. N., Gott J. R., 2003, \mnras, 344, 686

  \bibitem[\protect\citeauthoryear{Coles \etal}{2004}]{coles:2004}
   Coles P., Dineen P., Earl J., Wright D., 2004, \mnras, 350, 989

  \bibitem[\protect\citeauthoryear{Copi, Huterer \& Starkman}
    {Copi \etal}{2004}]{copi:2004}
    Copi C. J., Huterer D., Starkman G. D., 2004, Phys. Rev. D, in press
    (asto-ph/0310511)

  \bibitem[\protect\citeauthoryear{Cruz, Martinez-Gonz\'{a}lez,
    Vielva \& Cay\'{o}n} {Cruz \etal}{2004}]{cruz:2004}
    Cruz M., Martinez-Gonz\'{a}lez E., Vielva P., Cay\'{o}n L.,
    2004, \mnras, in press (astro-ph/0405341) 

  \bibitem[\protect\citeauthoryear{Eriksen \etal}{2004}]{eriksen:2004}
    Eriksen, H. K., Novikov, D. I., Lilje, P. B, Banday, A. J.,
    G\'{o}rski K. M.,
    2004, ApJ, in press (astro-ph/0401276) 

  \bibitem[\protect\citeauthoryear{Gaztanaga \& Wagg}{2003}]{gw:2003}
    Gaztanaga E., Wagg J., 2003, Phys. Rev. D., 68, 21302

  \bibitem[\protect\citeauthoryear{G\'{o}rski K. M., Hivon
    E. \&Wandelt B. D.}{G\'{o}rski \etal}{1999}]{gorski:1999}
    G\'{o}rski K. M., Hivon E., Wandelt B. D., 1999, preprint
    (\astroph/9812350)

  \bibitem[\protect\citeauthoryear{Hansen \etal}{2004}]{hansen:2004}
    Hansen F.K., Cabella P., Marinucci D., Vittorio N., 2004,
    ApJL, submitted (astro-ph/0402396)

  \bibitem[\protect\citeauthoryear{Hobson, Jones \& Lasenby}{Hobson
    \etal}{1999}]{hobson:1999} Hobson M. P., Jones A. W., Lasenby
    A. N., 1999, \mnras, 309, 125

%  \bibitem[\protect\citeauthoryear{Holschneider}{1996}]{holschneider:1996}
%    Holschneider M., 1996,
%    \textit{Journal of Mathematical Physics}, 37, 4156--4165.

  \bibitem[\protect\citeauthoryear{Inui T., Tanabe Y. \& Onodera
    Y.}{Inui \etal}{1996}]{inui:1996} Inui T., Tanabe Y., Onodera Y.,
    1990, Group Theory and its Application in Physics,
    Springer Verlag

  \bibitem[\protect\citeauthoryear{Komatsu \etal}{2003}]{komatsu:2003}
    Komatsu E. \etal, 2003,
    ApJS, 148, 119

  \bibitem[\protect\citeauthoryear{Land \& Magueijo}{2004}]{lm:2004}
    Land K, Magueijo J., 2004, \mnras, submitted (astro-ph/0405519)

  \bibitem[\protect\citeauthoryear{Larson \& Wandelt}{2004}]{larson:2004}
    Larson D. L., Wandelt B. D., 2004, ApJ, 613, 85

  \bibitem[\protect\citeauthoryear{McEwen, Hobson, Lasenby \&
    Mortlock}{McEwen \etal}{2004}]{mcewen:2004} McEwen J. D., Hobson
    M. P., Lasenby A. N., \mbox{Mortlock D. J.}, 2004, XXXIXth
    Rencontres de Moriond (astro-ph/0409288)

  \bibitem[\protect\citeauthoryear{Magueijo \& Medeiros}{2004}]{mm:2004}
    Magueijo J., Medeiros J., 2004, \mnras, 351, L1

  \bibitem[\protect\citeauthoryear{Martinez-Gonz\'{a}lez, Gallegos,
    Arg\"{u}eso, Cay\'{o}n \& Sanz}{Martinez-Gonz\'{a}lez
    \etal}{2002}]{martinez:2002} Martinez-Gonz\'{a}lez E., Gallegos
    J. E., Arg\"{u}eso F., \mbox{Cay\'{o}n L.,} Sanz J. L., 2002,
    \mnras, 336, 22

  \bibitem[\protect\citeauthoryear{Mukherjee \& Wang}{2004}]{mw:2004}
    Mukherjee P., Wang Y., 2004, ApJ, submitted (astro-ph/0402602)

  \bibitem[\protect\citeauthoryear{Mukherjee \etal}{2000}]{mhl:2000}
    Mukherjee P., Hobson M. P., Lasenby A. N., 2000, \mnras, 318, 1157

  \bibitem[\protect\citeauthoryear{de Oliveira-Costa, Tegmark, Zaldarriaga 
    \& Hamilton}{de Oliveira-Costa  \etal}{2004}]{oliveira:2004}
    de Oliveira-Costa A., Tegmark M., Zaldarriaga M., Hamilton A., 2004,
    Phys. Rev. D, 69, 63516

  \bibitem[\protect\citeauthoryear{Pando \etal}{1998}]{pando:1998}
    Pando J., Valls-Gabaud D., Fang L. -Z., 1998, Phys. Rev. Lett., 81, 4568

%  \bibitem[\protect\citeauthoryear{Rees \& Sciama}{1968}]{rees:1968}
%    Rees M. J., Sciama D. W., 1968, Nature, 217, 511

  \bibitem[\protect\citeauthoryear{Risbo}{1996}]{risbo:1996}
    Risbo T., 1996,
    Journal of Geodesy, 70, 383

%  \bibitem[\protect\citeauthoryear{Sachs \& Wolfe}{1967}]{sachs:1967}
%    Sachs R. K., Wolfe A. M., 1967, ApJ, 147, 73

  \bibitem[\protect\citeauthoryear{Slosar \& Seljak}{2004}]{slosar:2004}
    Slosar A., Seljak U., 2004, Phys. Rev. D, submitted, (astro-ph/0404567)

%  \bibitem[\protect\citeauthoryear{Sunyaev \& Zel'dovich}{1972}]{sunyaev:1972}
%    Sunyaev R. A., Zel'dovich Y. B., 1972, Comm. Astrophys. Space Sci., 4, 173

  \bibitem[\protect\citeauthoryear{Tegmark M., de Oliveira-Costa A. \&
     Hamilton A. J. S.}{Tegmark \etal}{2003}]{tegmark:2003} Tegmark
     M., de Oliveira-Costa A., Hamilton A. J. S., 2003,
     Phys. Rev. D, 68, 12

  \bibitem[\protect\citeauthoryear{Vielva, Martinez-Gonz\'{a}lez,
    Barreiro, Sanz \& Cay\'{o}n}{Vielva \etal}{2003}]{vielva:2003}
    Vielva P., Martinez-Gonz\'{a}lez E., Barreiro R. B., \mbox{Sanz
    J. L.}, Cay\'{o}n L., 2003, ApJ, 609, 22
    
  \bibitem[\protect\citeauthoryear{Wandelt \& G\'{o}rski}{2001}]{wandelt:2001}
    Wandelt B. D., G\'{o}rski K. M., 2001,
    Phys. Rev. D, 63, 123002, 1--6

%  \bibitem[\protect\citeauthoryear{Zel'dovich \& Sunyaev}{1968}]{sunyaev:1968}
%    Zel'dovich Y. B., Sunyaev R. A., 1969, Astrophys. Space Sci., 4, 301

\end{thebibliography}
\end{document}